\documentclass[12pt,onecolumn,oneside,a4paper,peerreview]{IEEEtran}

\usepackage{cite}
\usepackage{graphicx}
\usepackage{amsmath}
\usepackage{times}
\usepackage{latexsym}
\usepackage{graphicx}
\usepackage{bm}
\usepackage{amssymb}
\usepackage[center]{caption2}
\usepackage{stfloats}
\usepackage{cases}
\usepackage{url}
\usepackage{array}
\usepackage{setspace}
\usepackage{fancyhdr}
\usepackage{color}
\usepackage{subfigure}
\usepackage{chemarrow}
\usepackage{extarrows}

\usepackage{algorithm}
\usepackage{algorithmic}
\usepackage{multirow}

\renewcommand{\QED}{\QEDopen}
\newtheorem{theorem}{Theorem}
\newtheorem{lemma}{Lemma}
\newtheorem{corollary}{Corollary}

\def\proof{\noindent\hspace{2em}{\itshape Proof: }}
\def\endproof{\hspace*{\fill}~$\square$\par\endtrivlist\unskip}

\begin{document}
\title{Multipair Two-way Half-Duplex Relaying with Massive Arrays and Imperfect CSI}
\author{Chuili Kong, \emph{Student Member}, \emph{IEEE}, Caijun Zhong, \emph{Senior Member}, \emph{IEEE}, Michail Matthaiou, \emph{Senior Member}, \emph{IEEE}, Emil Bj{\"o}rnson, \emph{Member}, \emph{IEEE}, and Zhaoyang Zhang, \emph{Member}, \emph{IEEE}
\thanks{C. Kong, C. Zhong and Z. Zhang are with the Institute of Information and Communication Engineering, Zhejiang University, China. C. Zhong is also affiliated with the National Mobile Communications Research Laboratory, Southeast University, Nanjing, China (email: caijunzhong@zju.edu.cn).}
\thanks{M. Matthaiou is with the School of Electronics, Electrical Engineering and Computer Science, Queen's University Belfast, Belfast, BT3 9DT, U.K. (email:
m.matthaiou@qub.ac.uk).}
\thanks{E. Bj{\"o}rnson is with the Department of Electrical Engineering (ISY), Link{\"{o}}ping University, Link{\"{o}}ping, SE-581 83, Sweden (email: emil.bjornson@liu.se).}}

\maketitle
\begin{abstract}
We consider a two-way half-duplex relaying system where multiple pairs of single antenna users exchange information assisted by a multi-antenna relay. Taking into account the practical constraint of imperfect channel estimation, we study the achievable sum spectral efficiency of the amplify-and-forward (AF) and decode-and-forward (DF) protocols, assuming that the relay employs simple maximum ratio processing. We derive an exact closed-form expression for the sum spectral efficiency of the AF protocol and a large-scale approximation for the sum spectral efficiency of the DF protocol when the number of relay antennas, $M$, becomes sufficiently large. In addition, we study how the transmit power scales with $M$ to maintain a desired quality-of-service. In particular, our results show that by using a large number of relay antennas, the transmit powers of the user, relay, and pilot symbol can be scaled down proportionally to $1/M^\alpha$, $1/M^\beta$, and $1/M^\gamma$ for certain $\alpha$, $\beta$, and $\gamma$, respectively. This elegant power scaling law reveals a fundamental tradeoff between the transmit powers of the user/relay and pilot symbol. Finally, capitalizing on the new expressions for the sum spectral efficiency, novel power allocation schemes are designed to further improve the sum spectral efficiency.
\end{abstract}

\newpage
\section{Introduction}\label{section:1}
Relaying is a low-complexity and cost-effective means to extend network coverage and provide spatial diversity, which has attracted a great deal of research attention from both academia and industry \cite{H.Q.Ngo1,H.A.Suraweera,G.Kramer,Q.Wang}. Thus far, most practical relaying systems are assumed to operate in the half-duplex mode where the relay does not transmit and receive signals simultaneously. Yet, such half-duplex mechanism incurs a 50\% spectral efficiency loss. To reduce this loss in spectral efficiency, two-way relaying was proposed in \cite{R.Zhang,K.-J.Lee,G.Amarasuriya,R.Vaze}, where the two communicating nodes perform bidirectional simultaneous data transmission.

Multipair two-way relaying is a sophisticated generalization of single pair two-way relaying, where multiple pairs of users simultaneously establish a communication link with the aid of a single shared relay \cite{S.Jin,H.Cui,M.Tao}, hence substantially boosting the system spectral efficiency. The major challenge is to properly handle the inter-pair interference from co-existing communication pairs. Thus far, a number of advanced techniques have been introduced to mitigate inter-pair interference, such as dirty-paper coding \cite{S.Sima} and interference alignment \cite{R.S.Ganesan}. Unfortunately, the practical implementation of these techniques is in general very complex. On the other hand, the massive multiple-input multiple-output (MIMO) paradigm has demonstrated superior interference suppression capabilities, with very simple and low-complexity linear processing \cite{T.L.Marzetta,E.G.Larsson,F.Rusek}. Therefore, deploying large-scale antenna arrays in two-way relaying systems appears to be an extremely promising solution for inter-pair interference mitigation.

Some initial works have studied the fundamental performance of such systems \cite{S.Jin,H.Cui,H.Q.Ngo3}. In particular, for the amplify-and-forward (AF) protocol, \cite{H.Cui} investigated the achievable rates and power scaling laws of maximum ratio (MR) and zero-forcing (ZF) processing schemes. Moreover, \cite{S.Jin} derived a closed-form approximation of the ergodic rate of the MR scheme, and addressed the optimal user pair selection problem. However, one major limitation of the above works is that perfect channel state information (CSI) is assumed. Since obtaining perfect CSI is very challenging in the context of massive MIMO systems, it is important to look into the realistic scenario with imperfect CSI. An early attempt was made in \cite{H.Q.Ngo3}, where the authors studied the sum rate performance of training-based AF systems utilizing the ZF scheme. Although the derived results are useful for understanding the impact of imperfect CSI on the system performance, a number of important questions remain to be addressed. For instance, the fundamental tradeoff between the transmit powers of pilot, user, and the relay, and the performance analysis of alternative relay processing schemes, e.g., the decode-and-forward (DF) protocol, remain largely overlooked.

Motivated by this, in the current work, we consider a multipair two-way relaying system taking into account the channel estimation error, and present an in-depth analysis of the achievable spectral efficiency and power scaling law of MR processing for both the AF and DF protocols. Specifically, the main contributions of this paper are summarized as follows:
\begin{itemize}
  \item We investigate a multipair two-way relaying system that employs MR processing with imperfect CSI, and then derive an exact spectral efficiency expression in closed-form for the AF protocol and a large-scale approximation of the spectral efficiency for the DF protocol when $M \rightarrow \infty$, where $M$ is the number of relay antennas. Based on our analysis and results, a comprehensive performance comparison of the two protocols is conducted.
  \item We characterize the power scaling laws for both the AF and DF protocols, which generalize the results presented in \cite{S.Jin,H.Cui,H.Q.Ngo1}. It turns out that there exists a fundamental trade-off between the transmit powers of each user, pilot symbol and the relay; in other words, the same spectral efficiency can be achieved with different combination of power scaling parameters, which permits great flexibility in the design of practical systems. In addition, our theoretical findings suggest that it is worthwhile to spend more power on the pilot sequence to improve the CSI accuracy instead of just increasing the users' and relay power.
  \item Finally, to improve the sum spectral efficiency, we study the power allocation problem for both the AF and DF protocols subject to a sum power constraint. Near optimum solutions are obtained by solving a sequence of geometric programming (GP) problems. Our numerical results suggest that the proposed power allocation strategies improve the sum spectral efficiency by 34.8\% and 89.2\% when $M = 300$, for the AF and DF protocols, respectively, indicating that the spectral efficiency enhancement is more prominent for the DF protocol. In addition, for the special case where all users have the same transmit power, it turns out that the users should decrease their transmit power when the number of user pairs becomes large. On the other hand, the users should increase their transmit power when the number of relay antennas increases or when the channel estimation accuracy improves.
\end{itemize}

The remainder of the paper is organized as follows: Section \ref{section:2} introduces the multipair two-way half-duplex relaying system model under consideration. Section \ref{section:3} presents an exact spectral efficiency in closed-form for the AF protocol, and two large-scale approximations of the spectral efficiency for both protocols, with imperfect CSI, while Section \ref{section:4} studies the power scaling laws of different system configurations. The power allocation problem is discussed in Section \ref{section:5}. Finally, Section \ref{section:6} provides some concluding remarks.

{\it Notation}: We use bold upper case letters to denote matrices, bold lower case letters to denote vectors and lower case letters to denote scalars. Moreover, $(\cdot)^{H}$, $(\cdot)^{*}$, $(\cdot)^{T}$, and $(\cdot)^{-1}$ represent the conjugate transpose operator, the conjugate operator, the transpose operator, and the matrix inverse, respectively. Also, $|| \cdot ||$ is the Euclidian norm, $|| \cdot ||_{\text F}$ denotes the Frobenius norm, $| \cdot |$ is the absolute value, and ${\left[ {\bf{A}} \right]_{mn}}$ gives the $(m,n)$-th entry of $\bf{A}$. In addition, ${\bf x} \thicksim {{\cal CN} ({\bf 0},{\bf \Sigma})}$ denotes a circularly symmetric complex Gaussian random vector ${\bf x}$ with zero mean and variance matrix ${\bf \Sigma}$, while ${{\bf{I}}_k}$ is the identity matrix of size $k$. Finally, the statistical expectation operator is represented by ${\tt E}\{\cdot\}$, the variance operator is ${\text{Var}} \left(\cdot\right)$, the trace is denoted by ${\text{tr}}\left(\cdot\right)$, and the notation $\overset{a.s.}{\rightarrow}$ means almost sure convergence.
\section{System Model}\label{section:2}
\begin{figure}[!ht]
    \centering
    \includegraphics[scale=0.5]{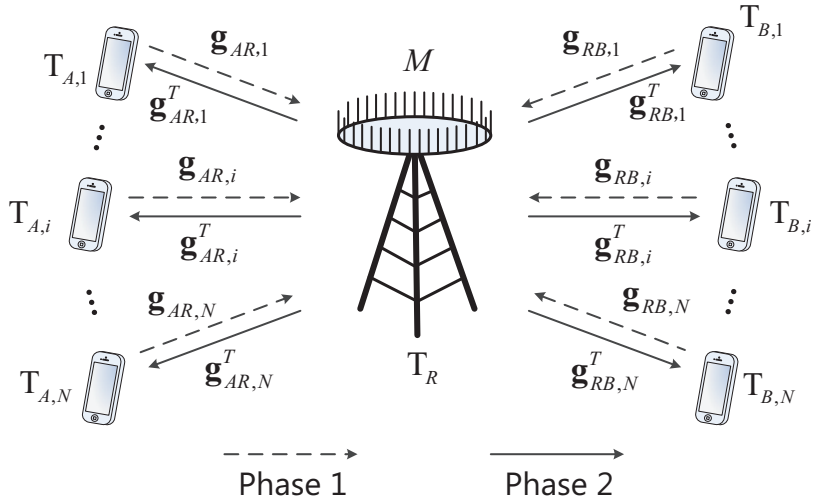}
    \caption{Illustration of the multipair two-way relaying system.}\label{fig:system_model}
  \end{figure}

Consider a multipair two-way relaying system as illustrated in Fig. \ref{fig:system_model}, where $N$ pairs of single-antenna users, denoted as ${\text T}_{A,i}$ and ${\text T}_{B,i}$, $i = 1,\ldots,N$, intend to exchange information with each other, under the assistance of a shared relay ${\text T}_R$ equipped with $M$ antennas. We assume that the direct links between ${\text T}_{A,i}$ and ${\text T}_{B,i}$ do not exist due to large obstacles or severe shadowing\cite{M.Tao}. Also, the relay operates in the half-duplex mode, i.e., it cannot transmit and receive simultaneously.

It is assumed that the system works under the time division duplex protocol and channel reciprocity holds. As such, the uplink and downlink channels between ${\text T}_{A,i}$ and ${\text T}_R$ can be denoted as ${\bf g}_{AR,i}$ and ${\bf g}^{\text T}_{AR,i}$, respectively. Similarly, the channels between ${\text T}_{B,i}$ and ${\text T}_R$ are denoted as ${\bf g}_{RB,i}$ and ${\bf g}^{\text T}_{RB,i}$, $i = 1,\ldots,N$, respectively, which account for both small-scale fading and large-scale fading. More precisely, ${\bf g}_{AR,i} \thicksim {\cal {CN}}\left({\bf 0}, \beta_{AR,i}{\bf I}_M\right)$ and ${\bf g}_{RB,i} \thicksim {\cal {CN}}\left({\bf 0}, \beta_{RB,i}{\bf I}_M\right)$. This model is known as uncorrelated Rayleigh fading, and ${\beta_{AR,i}}$ and ${\beta_{RB,i}}$ model the large-scale path-loss effect, which are assumed to be constant over many coherence intervals and known a priori. For notational convenience, the channel vectors can be collected together in a matrix form as ${\bf G}_{AR} \triangleq [{\bf g}_{AR,1}, \ldots, {\bf g}_{AR,N}]\in {\mathbb C}^{M \times N}$ and ${\bf G}_{RB} \triangleq [{\bf g}_{RB,1}, \ldots, {\bf g}_{RB,N}]\in {\mathbb C}^{M \times N}$.

For the considered multipair two-way relaying system, the entire information transmission process consists of two separate phases. In the first phase, the $N$ user pairs ${\text T}_{A,i}$ and ${\text T}_{B,i}$ simultaneously transmit their respective signals to ${\text T}_R$. Thus, the received signal at ${\text T}_R$ is given by
\begin{align}
  {\bf y}_r = \sum\limits_{i = 1}^{N} \left( \sqrt{p_{A,i}} {\bf g}_{AR,i} x_{A,i} + \sqrt{p_{B,i}} {\bf g}_{RB,i} x_{B,i} \right) + {\bf n}_R,
\end{align}
where $x_{A,i}$ and $x_{B,i}$ are Gaussian signals with zero mean and unit power transmitted by the \emph{i}-th user pair, $p_{A,i}$ and $p_{B,i}$ are the average transmit power of ${\text T}_{A,i}$ and ${\text T}_{B,i}$, respectively, and ${\bf n}_R$ is a vector of additive white Gaussian noise (AWGN) at ${\text T}_R$, whose elements are identically and independently distributed (i.i.d.) ${\cal {CN}}\left(0,1\right)$. Note that to keep the notation clean and without loss of generality, we take the noise variance to be $1$ here, and also in the subsequent sections. With this convention, $p_{A,i}$ and $p_{B,i}$ can be interpreted as the normalized transmit signal-to-noise (SNR).

In the second phase, the relay broadcasts a transformed version of the received signal to all users. Depending on the adopted relaying protocol, two separate scenarios are studied.

\subsubsection{AF protocol}
With the AF protocol, the relay first implements a type of linear processing on the received signal, and then forwards it to all users. Thus, the transmit signal from ${\text T}_R$ can be written as
\begin{align}\label{eq:y_t}
  {\bf y}_t^\text{AF} = {\rho_\text{AF}}{\bf F} {\bf y}_r,
\end{align}
where ${\bf F} \in {\mathbb C}^{M \times M}$ is the linear processing matrix to be specified shortly, and $\rho_\text{AF}$ is a normalization coefficient, which is chosen to satisfy the long-term total transmit power constraint at the relay, namely, ${\tt E}\left\{ ||{\bf y}_t^\text{AF}||^2 \right\} = p_r$, so that $p_r$ is the average transmit power of the relay.

Therefore, the received signals at ${\text T}_{X,i}$ (where $X \in \left\{A,B\right\}$) is given by
\begin{align}\label{eq:z_A}
  z_{X,i}^\text{AF} = {\bf g}^T_{XR,i}{\bf y}_t^\text{AF} + n_{X,i},
\end{align}
where $n_{X,i} \thicksim {\cal{CN}}(0,1)$ represents the AWGN at ${\text T}_{X,i}$. Please note, to simplify notation, we introduce ${\bf g}_{BR,i}$ which is defined as ${\bf g}_{BR,i} \triangleq {\bf g}_{RB,i}$ due to the channel reciprocity.

\subsubsection{DF protocol}
With the DF protocol, the relay first decodes the $2N$ symbols, and then re-encodes and forwards the signals to all users. To reduce the complexity, single user linear detection is considered at the relay. As such, the transformed signal after linear processing can be expressed as
\begin{align}
  {\bf r}^\text{DF} = {\bf W}^T {\bf y}_r,
\end{align}
where ${\bf W}^T \in {\mathbb C}^{2N \times M}$ is the linear receiver matrix, which will be specified shortly.

After decoding the $2N$ symbols ${\bf x}$ from ${\bf r}^\text{DF}$, the remaining task is broadcasting, but before this, a linear precoding matrix ${\bf J} \in {\mathbb C}^{M \times 2N}$ is applied to the re-generated symbols. As such, the transmit signal of ${\text T}_R$ is given by
\begin{align}
  {\bf y}_t^\text{DF} = {\rho_\text{DF}} {\bf J} {\bf x},
\end{align}
where ${\bf x} = \left[{\bf x}_A^T, {\bf x}_B^T\right]^T$ with ${\bf x}_A = [x_{A,1}, \ldots, x_{A,N}]^T$ and ${\bf x}_B = [x_{B,1}, \ldots, x_{B,N}]^T$, and $\rho_\text{DF}$ is the normalization coefficient, which is determined by the average power constraint at the relay, i.e., ${\tt E}\left\{ ||{\bf y}_t^\text{DF}||^2 \right\} = p_r$. Hence, the signals received at ${\text T}_{X,i}$ can be expressed as
\begin{align}
  {z}_{X,i}^\text{DF} = {\bf g}_{XR,i}^T{\bf y}_t^\text{DF} + {n}_{X,i}.
\end{align}
\subsection{Channel Estimation}
In practice, the channels ${\bf G}_{AR}$ and ${\bf G}_{RB}$ are not known and need to be estimated at the relay in every coherence interval. The typical way of doing this is to utilize pilots \cite{T.L.Marzetta}. To this end, during each coherence interval of length $\tau_c$ (in symbols), $\tau_p$ symbols are used for channel training. In this case, ${\text T}_{A,i}$ and ${\text T}_{B,i}$ transmit simultaneously their mutually orthogonal pilot sequences to ${\text T}_R$, for $i = 1,\ldots,N$. Thus, the received pilot matrix at the relay is given by
\begin{align}
  {\bf Y}_p = \sqrt{\tau_p p_p} {\bf G}_{AR} {\bf \Phi}_A^{T} + \sqrt{\tau_p p_p} {\bf G}_{RB} {\bf\Phi}_B^{T} + {\bf N}_p,
\end{align}
where $p_p$ is the transmit power of each pilot symbol, ${\bf N}_p$ is AWGN matrix including i.i.d. ${\cal{CN}}(0,1)$ elements, while the \emph{i}-th columns of ${\bf \Phi}_A \in {\mathbb C}^{\tau_p \times N}$ and ${\bf \Phi}_B \in {\mathbb C}^{\tau_p \times N}$ are the pilot sequences transmitted from ${\text T}_{A,i}$ and ${\text T}_{B,i}$, respectively. Since all pilot sequences are assumed to be mutually orthogonal, $\tau_p \geq 2N$ is required, and we have that ${\bf \Phi}_A^T {\bf \Phi}_A^* = {\bf I}_N$, ${\bf \Phi}_B^T {\bf \Phi}_B^* = {\bf I}_N$, and ${\bf \Phi}_A^T {\bf \Phi}_B^* = {\bf 0}_N$.

As in \cite{H.Q.Ngo1,H.Q.Ngo2,H.Q.Ngo3}, we assume that ${\text T}_R$ uses the minimum mean-square-error (MMSE) estimator to estimate ${\bf G}_{AR}$ and ${\bf G}_{RB}$. As such, the estimated channels of ${\bf G}_{AR}$ and ${\bf G}_{RB}$ are given by
\begin{align}\label{eq:G:hat}
  {\hat {\bf G}}_{AR} &= \frac{1}{\sqrt{\tau_p p_p}} {\bf Y}_p {\bf \Phi}_A^* \tilde{{\bf D}}_{AR} = \left( {\bf G}_{AR} + \frac{1}{\sqrt{\tau_p p_p}} {\bf N}_A \right) {\tilde{\bf D}}_{AR},\\
  {\hat {\bf G}}_{RB} &= \frac{1}{\sqrt{\tau_p p_p}} {\bf Y}_p {\bf \Phi}_B^* \tilde{{\bf D}}_{RB}= \left( {\bf G}_{RB} + \frac{1}{\sqrt{\tau_p p_p}} {\bf N}_B \right) {\tilde{\bf D}}_{RB},
\end{align}
respectively, where ${\tilde{\bf D}}_{AR} \triangleq \left( \frac{1}{\tau_p p_p}{\bf D}_{AR}^{-1} + {\bf I}_N \right)^{-1}$ with $\left[{\bf D}_{AR}\right]_{ii} = \beta_{AR,i}$ (${\bf D}_{AR}$ is an $N \times N$ diagonal matrix), ${\tilde{\bf D}}_{RB} \triangleq \left( \frac{1}{\tau_p p_p}{\bf D}_{RB}^{-1} + {\bf I}_N \right)^{-1}$ with $\left[{\bf D}_{RB}\right]_{ii} = \beta_{RB,i}$ (${\bf D}_{RB}$ is an $N \times N$ diagonal matrix), ${\bf N}_A \triangleq {\bf N}_p {\bf \Phi}_A^*$, and ${\bf N}_B \triangleq {\bf N}_p {\bf \Phi}_B^*$. Since ${\bf \Phi}_A^T {\bf \Phi}_A^* = {\bf I}_N$ and ${\bf \Phi}_B^T {\bf \Phi}_B^* = {\bf I}_N$, the elements of ${\bf N}_A$ and ${\bf N}_B$ are i.i.d. ${\cal{CN}}(0,1)$ random variables. Let ${\bf E}_{AR}$ and ${\bf E}_{RB}$ be the estimation error matrices of ${\bf G}_{AR}$ and ${\bf G}_{RB}$, respectively. Then, ${\bf G}_{AR}$ and ${\bf G}_{RB}$ can be decomposed as
\begin{align}\label{eq:G:matrix}
  {\bf G}_{AR} &= {\hat{\bf G}}_{AR} + {\bf E}_{AR},\\
  {\bf G}_{RB} &= {\hat{\bf G}}_{RB} + {\bf E}_{RB},
\end{align}
respectively. Due to the orthogonality property of MMSE estimators and the fact that ${\hat{\bf G}}_{AR}$, ${\bf E}_{AR}$, ${\hat{\bf G}}_{RB}$, and ${\bf E}_{RB}$ are complex Gaussian distributed, these matrices are independent of each other. By rewriting \eqref{eq:G:matrix} in vector form, we have
\begin{align}\label{eq:g:AR:vector}
  {\bf g}_{AR,i} &= {\hat{\bf g}}_{AR,i} + {\bf e}_{AR,i},\\
  {\bf g}_{RB,i} &= {\hat{\bf g}}_{RB,i} + {\bf e}_{RB,i},
\end{align}
where ${\hat{\bf g}}_{AR,i}$, ${\bf e}_{AR,i}$, ${\hat{\bf g}}_{RB,i}$, and ${\bf e}_{RB,i}$ are the \emph{i}-th columns of ${\hat{\bf G}}_{AR}$, ${\bf E}_{AR}$, ${\hat{\bf G}}_{RB}$, and ${\bf E}_{RB}$, respectively, which are mutually independent. Then, from \eqref{eq:G:hat}, the elements of ${\hat{\bf g}}_{AR,i}$, ${\bf e}_{AR,i}$ are Gaussian random variables with zero mean, variance $\sigma_{AR,i}^2$ and ${\tilde \sigma}_{AR,i}^2$, respectively, where $\sigma_{AR,i}^2 \triangleq \frac{\tau_p p_p\beta_{AR,i}^2}{1+\tau_p p_p\beta_{AR,i}}$ and ${\tilde \sigma}_{AR,i}^2 \triangleq \frac{\beta_{AR,i}}{1+\tau_p p_p\beta_{AR,i}}$. Similarly, the elements of ${\hat{\bf g}}_{RB,i}$, and ${\bf e}_{RB,i}$ are complex Gaussian random variables with zero mean, variance $\sigma_{RB,i}^2$ and ${\tilde \sigma}_{RB,i}^2$, respectively, where $\sigma_{RB,i}^2 \triangleq \frac{\tau_p p_p\beta_{RB,i}^2}{1+\tau_p p_p\beta_{RB,i}}$ and ${\tilde \sigma}_{RB,i}^2 \triangleq \frac{\beta_{RB,i}}{1+\tau_p p_p\beta_{RB,i}}$.
\subsection{Linear Processing Matrices}
The relay ${\text T}_R$ treats the channel estimates as the true channels and utilizes them to perform linear processing. For both the AF and DF protocols, the MR linear processing method is used.\footnote{Note that MR is a very attractive linear processing technique in the context of massive MIMO systems due to its low complexity. Most importantly, it can be implemented in a distributed manner \cite{T.L.Marzetta,E.G.Larsson}.}
\subsubsection{AF protocol}
With MR, the processing matrix ${\bf F} \in {\mathbb C}^{M \times M}$ is given by \cite{S.Jin}
\begin{align}
  {\bf F} = {\bf B}^*{\bf A}^H,
\end{align}
where ${\bf A} \triangleq \left[ {\hat{\bf G}}_{AR}, {\hat{\bf G}}_{RB} \right]$, ${\bf B} \triangleq \left[ {\hat{\bf G}}_{RB}, {\hat{\bf G}}_{AR} \right]$. Recall that $\rho_\text{AF}$ satisfies the long-term total transmit power constraint at the relay, and after some simple algebraic manipulations, we have
\begin{align}
  \rho_\text{AF} = \sqrt{\frac{p_r}{ \sum\limits_{i = 1}^N \left( p_{A,i} {\tt E}\left\{ ||{\bf F}{\bf g}_{AR,i}||^2 \right\} +  p_{B,i} {\tt E}\left\{ ||{\bf F}{\bf g}_{RB,i}||^2 \right\} \right) + {\tt E}\left\{ ||{\bf F}||_{\text F}^2  \right\}  }}.
\end{align}
\subsubsection{DF protocol}
With MR, the processing matrix ${\bf W}^T \in {\mathbb C}^{2N \times M}$ and ${\bf J} \in {\mathbb C}^{M \times 2N}$ are given by
\begin{align}
  {\bf W}^T &= \left[ {\bf {\hat G}}_{AR}, {\bf {\hat G}}_{RB} \right]^H,\\
  {\bf J} &=  \left[ {\bf {\hat G}}_{RB}, {\bf {\hat G}}_{AR} \right]^*,
\end{align}
respectively, while $\rho_\text{DF}$ is given by
\begin{align}
  \rho_\text{DF} = \sqrt{\frac{p_r}{{\tt E}\left\{||{\bf J}||_{\text F}^2\right\}}} = \sqrt{\frac{p_r}{M \sum\limits_{n = 1}^{N} \left(\sigma_{AR,n}^2 + \sigma_{RB,n}^2 \right)}}.
\end{align}
\section{Spectral efficiency}\label{section:3}
In this section, we investigate the spectral efficiency (in bit/s/Hz) of the two-way half-duplex relaying system. In particular, for the AF protocol, an exact spectral efficiency expression in closed-form is derived for arbitrary $M$. Furthermore, two large-scale approximations of the spectral efficiency for both protocols are deduced when $M \rightarrow \infty$.
\subsection{AF protocol}
Without loss of generality, we focus on the characterization of the achievable spectral efficiency of user ${\text T}_{A,i}$. With the AF protocol, when ${\text T}_{A,i}$ receives the superimposed signal from ${\text T}_R$, it first makes an attempt to subtract its own transmitted message according to its available CSI (known as self-interference cancellation). In the current work, we consider the realistic case where the users do not have access to instantaneous CSI, hence ${\text T}_{A,i}$ uses only statistical CSI to cancel the self-interference. Therefore, after cancelling partial self-interference, i.e., $\rho_\text{AF} \sqrt{p_{A,i}} {\tt E} \left\{{{\bf g}}_{AR,i}^{\text T} {\bf F} {{\bf g}}_{AR,i}\right\} x_{A,i}$, the received signal at ${\text T}_{A,i}$ can be re-expressed as
\begin{align}
  {\hat z}_{A,i}^\text{AF} &= z_{A,i}^\text{AF} - \rho_\text{AF} \sqrt{p_{A,i}} {\tt E} \left\{ {\bf g}_{AR,i}^{\text T} {\bf F} {\bf g}_{AR,i} \right\} x_{A,i}\\
          &= \underbrace{\rho_\text{AF} \sqrt{p_{B,i}} {\tt E} \left\{ {\bf g}_{AR,i}^{\text T} {\bf F} {\bf g}_{RB,i} \right\} x_{B,i}}_{\text{desired signal}}
          + \underbrace{\rho_\text{AF} \sqrt{p_{B,i}} \left({\bf g}_{AR,i}^{\text T} {\bf F} {\bf g}_{RB,i} - {\tt E} \left\{ {\bf g}_{AR,i}^{\text T} {\bf F} {\bf g}_{RB,i} \right\}   \right) x_{B,i}}_{\text{estimation error}} \\
          &+ \underbrace{\rho_\text{AF} \sqrt{p_{A,i}} \left( {\bf g}_{AR,i}^{\text T} {\bf F} {\bf g}_{AR,i} - {\tt E} \left\{ {\bf g}_{AR,i}^{\text T} {\bf F} {\bf g}_{AR,i} \right\} \right) x_{A,i}}_{\text{residual self-interference}} \\
          &+ \underbrace{\rho_\text{AF} \sum\limits_{j\neq i} \left( \sqrt{p_{A,i}} {\bf g}_{AR,i}^{\text T} {\bf F} {\bf g}_{AR,j} x_{A,j} + \sqrt{p_{B,i}} {\bf g}_{AR,i}^{\text T} {\bf F} {\bf g}_{RB,j} x_{B,j} \right)}_{\text{inter-user interference}}
          + \underbrace{\rho_\text{AF} {\bf g}_{AR,i}^{\text T} {\bf F} {\bf n}_R + n_{A,i}}_{\text{compound noise}}.
\end{align}

Note that, ${\text T}_{A,i}$ tries to utilize the statistical information ${\tt E} \left\{{{\bf g}}_{AR,i}^{\text T} {\bf F} {{\bf g}}_{AR,i}\right\}$ to cancel partial interference. Focusing on this, we have
\begin{align}
 {\tt E} \left\{ {{\bf g}}_{AR,i}^{T} {\bf F} {{\bf g}}_{AR,i}\right\} &=  {\tt E} \left\{ \sum\limits_{n=1}^{N} \left( {{\bf g}}_{AR,i}^{T} {\hat{\bf g}}_{RB,n}^* {\hat{\bf g}}_{AR,n}^H {{\bf g}}_{AR,i} + {{\bf g}}_{AR,i}^{T} {\hat{\bf g}}_{AR,n}^* {\hat{\bf g}}_{RB,n}^H {{\bf g}}_{AR,i} \right) \right\},\\
  &= {\tt E} \left\{ {\hat{\bf g}}_{AR,i}^{T} {\hat{\bf g}}_{RB,i}^* {\hat{\bf g}}_{AR,i}^H {\hat{\bf g}}_{AR,i} + {\hat{\bf g}}_{AR,i}^{T} {\hat{\bf g}}_{AR,i}^* {\hat{\bf g}}_{RB,i}^H {\hat{\bf g}}_{AR,i} \right\} = 0,
\end{align}
which indicates that no effective self-interference cancellation can be achieved by ${\text T}_{A,i}$ if only statistical CSI is available. In sharp contrast, if ${\text T}_{A,i}$ knows perfect or estimated CSI, it is capable of completely canceling the self-interference \cite{S.Jin}, or $\rho_\text{AF} \sqrt{p_{A,i}} {\hat{\bf g}}_{AR,i}^{\text T} {\bf F} {\hat{\bf g}}_{AR,i} x_{A,i}$ \cite{F.Gao,C.Wang}, respectively. Nevertheless, the overhead associated with CSI acquisition at users may overweigh the performance gains brought by effective self-interference cancellation, particularly in massive MIMO systems.

Using a standard approach as in \cite{H.Q.Ngo1,J.Hoydis}, an ergodic achievable spectral efficiency of ${\text T}_{A,i}$ is
\begin{align}\label{eq:rate:R}
R_{A,i}^\text{AF} = \frac{1}{2} \log_2\left(  1 + \frac{A_i^\text{AF}}{B_i^\text{AF} + C_i^\text{AF} + D_i^\text{AF} + E_i^\text{AF}} \right),
\end{align}
where
\begin{align}
  A_i^\text{AF} &= p_{B,i} |{\tt E} \left\{ {\bf g}_{AR,i}^{\text T} {\bf F} {\bf g}_{RB,i}\right\}|^2,\\
  B_i^\text{AF} &= p_{B,i}{\text{Var}} \left({\bf g}_{AR,i}^{\text T} {\bf F} {\bf g}_{RB,i}\right),\\
  C_i^\text{AF} &= p_{A,i}{\text{Var}} \left({\bf g}_{AR,i}^{\text T} {\bf F} {\bf g}_{AR,i}\right),\\
    D_i^\text{AF} &= \sum\limits_{j\neq i} \left( {\tt E} \left\{ p_{A,i} |{\bf g}_{AR,i}^{\text T} {\bf F} {\bf g}_{AR,j}|^2 + p_{B,i} |{\bf g}_{AR,i}^{\text T} {\bf F} {\bf g}_{RB,j}|^2 \right\} \right),\\
  E_i^\text{AF} &= {\tt E} \left\{ ||{\bf g}_{AR,i}^{\text T} {\bf F}||^2 \right\} + \frac{1}{\rho_\text{AF}^2}.
\end{align}

Thus, the ergodic sum spectral efficiency of the multipair two-way AF relaying system is given by
\begin{align}\label{eq:rate:R:sum}
  R^\text{AF} = \frac{\tau_c-\tau_p}{\tau_c} \sum\limits_{i = 1}^{N}\left( R_{A,i}^\text{AF} + R_{B,i}^\text{AF} \right),
\end{align}
 where $R_{B,i}^\text{AF}$ is the spectral efficiency of ${\text T}_{B,i}$, which can be derived in a similar fashion.

 In the following, we present an exact analysis of the spectral efficiency based on \eqref{eq:rate:R}.
 \begin{theorem}\label{theor:0}
 With the AF protocol, the ergodic spectral efficiency $R_{A,i}^\text{AF}$ for an arbitrary number of relay antennas is given by \eqref{eq:rate:R} with
  \begin{align}\label{eq:theor:0:A}
  &{ A}_i^\text{AF} = p_{B,i} M^2(M + 1)^2 \sigma_{AR,i}^4 \sigma_{RB,i}^4,\\ \label{eq:theor:0:B}
  &{ B}_i^\text{AF} = p_{B,i} 2 M \left(M + 1 \right) \beta_{AR,i} \beta_{RB,i} \sum\limits_{n=1}^{N} \sigma_{AR,n}^2 \sigma_{RB,n}^2 \\ \notag
  &+ p_{B,i} M \left(M+1 \right)\sigma_{AR,i}^2 \sigma_{RB,i}^2 \left(\beta_{AR,i}\sigma_{RB,i}^2 + \beta_{RB,i}\sigma_{AR,i}^2 \right) + p_{B,i} M^2 \sigma_{AR,i}^2 \sigma_{RB,i}^2 {\tilde \sigma}_{AR,i}^2 {\tilde \sigma}_{RB,i}^2 \\ \notag
  &+ p_{B,i} 2 M \left(M + 1\right)^2 \sigma_{AR,i}^4 \sigma_{RB,i}^4 + p_{B,i}2 M \left(M + 1\right) \sigma_{AR,i}^4 {\tilde \sigma}_{AR,i}^2 \sigma_{RB,i}^2 + p_{B,i} 2 M \left(M + 1\right) \sigma_{AR,i}^2 {\tilde \sigma}_{AR,i}^2 \sigma_{RB,i}^4 \\ \notag
  &+ p_{B,i} 2M\sigma_{AR,i}^2 {\tilde \sigma}_{AR,i}^2 \sigma_{RB,i}^2 {\tilde \sigma}_{RB,i}^2 + p_{B,i} M^2\left(M + 1\right) \sigma_{AR,i}^4 {\tilde \sigma}_{AR,i}^2 \sigma_{RB,i}^2 \\ \notag
  &+ p_{B,i} M^2\left(M + 1\right) \sigma_{AR,i}^2 {\tilde \sigma}_{AR,i}^2 \sigma_{RB,i}^4 + p_{B,i} M^2\sigma_{AR,i}^2 {\tilde \sigma}_{AR,i}^2 \sigma_{RB,i}^2 {\tilde \sigma}_{RB,i}^2,\\ \label{eq:theor:0:C}
  &C_i^\text{AF} =  4 p_{A,i} M (M + 1)\beta_{AR,i}^2 \sum\limits_{n=1}^{N} \sigma_{AR,n}^2 \sigma_{RB,n}^2 \\ \notag
  &+ 4 p_{A,i} \sigma_{AR,i}^2 \sigma_{RB,i}^2 M \left(M + 1 \right) \left( \left(M + 2 \right) \sigma_{AR,i}^4 + \left(M + 5\right) \sigma_{AR,i}^2 {\tilde\sigma}_{AR,i}^2 + {\tilde\sigma}_{AR,i}^4 \right),\\ \label{eq:theor:0:D}
 & { D}_i^\text{AF} = \sum\limits_{j \neq i}
  2 M \left(M + 1\right) \beta_{AR,i} \left( p_{A,i} \beta_{AR,j} + p_{B,i} \beta_{RB,j}\right) \sum\limits_{n \neq i,j} \sigma_{AR,n}^2 \sigma_{RB,n}^2 \\ \notag
  &+ \sum\limits_{j \neq i}
  M \sigma_{AR,i}^2 \sigma_{RB,i}^2 \left(  p_{A,i} \beta_{AR,j} +  p_{B,i} \beta_{RB,j}\right) \left( \left(M + 1\right) \left(M +3 \right) \sigma_{AR,i}^2 + 2 \left(M + 1\right) {\tilde\sigma}_{AR,i}^2 \right) \\ \notag
  &+ \sum\limits_{j \neq i}
  M \beta_{AR,i} \sigma_{AR,j}^2 \sigma_{RB,j}^2 \left( \left(M + 1\right) \left(M +3 \right) \left( p_{A,i} \sigma_{AR,j}^2 +  p_{B,i} \sigma_{RB,j}^2 \right) + 2 \left(M + 1\right) \left( p_{A,i} {\tilde\sigma}_{AR,j}^2 +  p_{B,i} {\tilde\sigma}_{RB,j}^2 \right) \right),\\ \label{eq:theor:0:E}
  & E_i^\text{AF} = 2 M \left(M + 1\right) \beta_{AR,i} \sum\limits_{n \neq i} \sigma_{AR,n}^2 \sigma_{RB,n}^2 \\ \notag
  &+ M  \sigma_{AR,i}^2 \sigma_{RB,i}^2 \left( \left(M + 1\right) \left(M +3 \right) \sigma_{AR,i}^2 + 2 \left(M + 1\right) {\tilde\sigma}_{AR,i}^2 \right)\\ \notag
  &+ \frac{1}{p_r}\sum\limits_{i =1}^N M \sigma_{AR,i}^2 \sigma_{RB,i}^2 \left( \left(M + 1\right) \left(M +3 \right) \left(\sigma_{AR,i}^2 p_{A,i} + \sigma_{RB,i}^2 p_{B,i} \right) + 2 \left(M + 1\right) \left( {\tilde\sigma}_{AR,i}^2 p_{A,i} + {\tilde\sigma}_{RB,i}^2 p_{B,i} \right) \right) \\ \notag
  &+ \frac{1}{p_r}\sum\limits_{i =1}^N 2 M \left(M + 1\right) \left(\beta_{AR,i} p_{A,i} + \beta_{RB,i} p_{B,i} \right) \sum\limits_{n \neq i} \sigma_{AR,n}^2 \sigma_{RB,n}^2 + \frac{1}{p_r} 2M \left(M +1\right) \sum\limits_{n=1}^{N} \sigma_{AR,n}^2. \sigma_{RB,n}^2.
\end{align}
 \end{theorem}
\proof See Appendix \ref{app:theor:0}.\endproof

 Theorem \ref{theor:0} presents an exact spectral efficiency in closed-form, which is applicable to arbitrary system configurations. However, the expression is too complicated to provide useful insights. Using the fact that the relay is equipped with a massive antenna array, we now obtain a simple and accurate approximation for the spectral efficiency.

\begin{theorem}\label{theor:1}
  With the AF protocol, as the number of relay antennas grows to infinity, then we have $R_{A,i}^\text{AF} -  {\tilde R}_{A,i}^\text{AF} \overset{M \rightarrow \infty}{\longrightarrow} 0$, where ${\tilde R}_{A,i}^\text{AF}$ is given by
  \begin{align}\label{eq:theor:1:R}
  {\tilde R}_{A,i}^\text{AF} \triangleq \frac{1}{2}\log_2\left( 1 + \frac{p_{B,i}M}{{\tilde B}_i^\text{AF} + {\tilde C}_i^\text{AF} + {\tilde D}_i^\text{AF} + {\tilde E}_i^\text{AF}} \right),
  \end{align}
where
\begin{align}\label{eq:theor:1:A}
  &{\tilde B}_i^\text{AF} \triangleq p_{B,i} \left(\frac{\beta_{RB,i}}{\sigma_{RB,i}^2} + \frac{\beta_{AR,i}}{\sigma_{AR,i}^2}\right),\\
  &{\tilde C}_i^\text{AF} \triangleq \frac{4p_{A,i}\beta_{AR,i}}{\sigma_{RB,i}^2},\\
 & {\tilde D}_i^\text{AF} \triangleq \sum\limits_{j \neq i} p_{A,j} \left(\frac{\beta_{AR,j}}{\sigma_{RB,i}^2} + \frac{\sigma_{AR,j}^4 \sigma_{RB,j}^2 \beta_{AR,i} }{\sigma_{AR,i}^4 \sigma_{RB,i}^4} \right) + \sum\limits_{j \neq i} p_{B,j} \left(\frac{\beta_{RB,j}}{\sigma_{RB,i}^2} + \frac{\sigma_{AR,j}^2 \sigma_{RB,j}^4 \beta_{AR,i} }{\sigma_{AR,i}^4 \sigma_{RB,i}^4} \right),\\
 & {\tilde E}_i^\text{AF} \triangleq \frac{1}{\sigma_{RB,i}^2} + \frac{1}{p_r \sigma_{AR,i}^4 \sigma_{RB,i}^4} \sum\limits_{n = 1}^{N} \sigma_{AR,n}^2 \sigma_{RB,n}^2 \left( p_{A,n} \sigma_{AR,n}^2 + p_{B,n} \sigma_{RB,n}^2 \right).
\end{align}
\end{theorem}
\proof
When $M$ is infinitely large, the lower order terms in \eqref{eq:theor:0:A}, \eqref{eq:theor:0:B}, \eqref{eq:theor:0:C}, \eqref{eq:theor:0:D}, and \eqref{eq:theor:0:E} are trivial. Thus, by removing them and keeping only the highest order terms, we complete the proof.
\endproof

Theorem \ref{theor:1} presents a large-scale approximation of the \emph{i}-th user's spectral efficiency. Despite being obtained under the massive array assumption, the approximation turns out to be very accurate even for finite number of relay antennas. In addition, it is easy to observe the impact of various factors on the spectral efficiency. For instance, ${\tilde B}_i^\text{AF}$ represents the influence of the estimation error, ${\tilde C}_i$ denotes the residual self-interference, ${\tilde D}_i^\text{AF}$ stands for the inter-user interference caused by other user pairs, and ${\tilde E}_i^\text{AF}$ is composed of the SNR at the relay and the end ${\text T}_{A,i}$.

From Theorem \ref{theor:1}, we observe that ${\tilde R}_{A,i}^\text{AF}$ is an increasing function with respect to $M$, while a decreasing function with respect to ${\tilde B}_i^\text{AF}$, ${\tilde C}_i^\text{AF}$, ${\tilde D}_i^\text{AF}$, and ${\tilde E}_i^\text{AF}$. Also, focusing on the term ${\tilde D}_i^\text{AF}$, it can be seen that the individual user spectral efficiency ${\tilde R}_{A,i}^\text{AF}$ decreases with the number of user pairs $N$; this is anticipated since a higher number of users increases the amount of inter-user interference. Now, we focus on studying the impact of the transmit power of \emph{i}-th user pair $p_{A,i}$ and $p_{B,i}$, the transmit power of the relay $p_r$, and the transmit power of each pilot symbol $p_p$ on the system performance. As can be seen, when $p_{A,i} \rightarrow \infty$ and $p_{B,i} \rightarrow \infty$, the spectral efficiency is limited by $p_r$ and $p_p$; in contrast, it is limited by $p_{A,i}$, $p_{B,i}$, and $p_p$ when $p_r \rightarrow \infty$. Moreover, when $p_{A,i} \rightarrow \infty$, $p_{B,i} \rightarrow \infty$, and $p_r \rightarrow \infty$, ${\tilde E}_i^\text{AF} \rightarrow 0$, which indicates that the noise at the relay and ${\text T}_{A,i}$ can be neglected when the transmit powers of each user and the relay are large enough. 
\subsection{DF protocol}
With the DF protocol, in the first phase, a linear processing matrix ${\bf W}^T$ is applied to the received signals prior to signal detection, hence, the post-processing signals at the relay are given by
\begin{align}
{\bf r}^\text{DF} = \left[ {\begin{array}{*{20}{c}}
\sum\limits_{i = 1}^N \left( \sqrt{p_{A,i}} {\bf \hat G}_{AR}^H {\bf g}_{AR,i} {x}_{A,i} + \sqrt{p_{B,i}} {\bf \hat G}_{AR}^H {\bf g}_{RB,i} { x}_{B,i} \right) + {\bf \hat G}_{AR}^H {\bf n}_R\\
\sum\limits_{i = 1}^N \left( \sqrt{p_{A,i}} {\bf \hat G}_{RB}^H {\bf g}_{AR,i} {x}_{A,i} + \sqrt{p_{B,i}} {\bf \hat G}_{RB}^H {\bf g}_{RB,i} {x}_{B,i} \right) + {\bf \hat G}_{RB}^H {\bf n}_R
\end{array}} \right],
\end{align}
where the top $N$ elements of ${\bf r}^\text{DF}$ stand for the signals from ${\text T}_{A,i}$ ($i = 1,\ldots,N$), while the bottom $N$ elements of ${\bf r}^\text{DF}$ represent the signals from ${\text T}_{B,i}$ ($i = 1,\ldots,N$). Without loss of generality, we focus only on the \emph{i}-th pair of users, i.e., ${\text T}_{A,i}$ and ${\text T}_{B,i}$. This is a superposition of the \emph{i}-th and \emph{(N+i)}-th elements of $\bf r^\text{DF}$, and can be expressed as
\begin{align}
  {\tilde r}_i^\text{DF} &= r_i^\text{DF} + r_{N + i}^\text{DF},\\
  &= \sum\limits_{j = 1}^{N} \left( \sqrt{p_{A,j}} \left( {\bf \hat g}_{AR,i}^H{\bf g}_{AR,j} + {\bf \hat g}_{RB,i}^H{\bf g}_{AR,j}\right)x_{A,j}
  + \sqrt{p_{B,j}} \left( {\bf \hat g}_{AR,i}^H{\bf g}_{RB,j} + {\bf \hat g}_{RB,i}^H{\bf g}_{RB,j}\right)x_{B,j} \right) \\
   &+ \left({\bf\hat g}_{AR,i}^H + {\bf \hat g}_{RB,i}^H \right){\bf n}_R,\\
   &= \underbrace{\sqrt{p_{A,i}} \left( {\bf \hat g}_{AR,i}^H{\bf \hat g}_{AR,i} + {\bf \hat g}_{RB,i}^H{\bf \hat g}_{AR,i}\right)x_{A,i} + \sqrt{p_{B,i}}  \left( {\bf \hat g}_{AR,i}^H{\bf \hat g}_{RB,i} + {\bf \hat g}_{RB,i}^H{\bf \hat g}_{RB,i}\right)x_{B,i}}_{\text{desired signal}}\\
   &+ \underbrace{\sqrt{p_{A,i}} \left( {\bf \hat g}_{AR,i}^H{\bf e}_{AR,i} + {\bf \hat g}_{RB,i}^H{\bf e}_{AR,i}\right)x_{A,i} + \sqrt{p_{B,i}}  \left( {\bf \hat g}_{AR,i}^H{\bf e}_{RB,i} + {\bf \hat g}_{RB,i}^H{\bf e}_{RB,i}\right)x_{B,i}}_{\text{estimation error}}\\
   &+ \underbrace{ \sum\limits_{j \neq i} \left( \sqrt{p_{A,j}} \left( {\bf \hat g}_{AR,i}^H{\bf g}_{AR,j} + {\bf \hat g}_{RB,i}^H{\bf g}_{AR,j}\right)x_{A,j}
  + \sqrt{p_{B,j}} \left( {\bf \hat g}_{AR,i}^H{\bf g}_{RB,j} + {\bf \hat g}_{RB,i}^H{\bf g}_{RB,j}\right)x_{B,j}\right)}_{\text{inter-user interference}}\\
  &+ \underbrace{\left({\bf\hat g}_{AR,i}^H + {\bf \hat g}_{RB,i}^H \right){\bf n}_R}_{\text{compound noise}}.
\end{align}
Since the relay has estimated CSI, it treats the channel estimates as the true channels to decode the signals. To this end, using a standard bound based on the worst-case uncorrelated additive noise \cite{J.Hoydis} yields the ergodic spectral efficiency of the \emph{i}-th user pair in the first phase
\begin{align}\label{eq:R1:DF}
  R_{1,i}^\text{DF} = \frac{1}{2} {\tt E} \left\{ \log_2 \left( 1 + \frac{A_i^\text{DF} + B_i^\text{DF}}{ {\tt E} \left\{ \left( C_i^\text{DF} + D_i^\text{DF} + E_i^\text{DF} \right)| {\hat{\bf G}}_{AR},{\hat{\bf G}}_{RB} \right\} } \right) \right\},
\end{align}
where the inner and outer expectations are taken over the estimation errors and channel estimates, respectively, and
\begin{align}
  A_i^\text{DF} &= p_{A,i} \left(|{\bf \hat g}_{AR,i}^H{\bf \hat g}_{AR,i}|^2 + |{\bf \hat g}_{RB,i}^H{\bf \hat g}_{AR,i}|^2 \right),\\
  B_i^\text{DF} &= p_{B,i} \left(|{\bf \hat g}_{AR,i}^H{\bf \hat g}_{RB,i}|^2 + |{\bf \hat g}_{RB,i}^H{\bf \hat g}_{RB,i}|^2 \right),\\
  C_i^\text{DF} &= p_{A,i} \left( |{\bf \hat g}_{AR,i}^H{\bf e}_{AR,i}|^2 + |{\bf \hat g}_{RB,i}^H{\bf e}_{AR,i}|^2 \right) + p_{B,i} \left( |{\bf \hat g}_{AR,i}^H{\bf e}_{RB,i}|^2 + |{\bf \hat g}_{RB,i}^H{\bf e}_{RB,i}|^2 \right),\\
  D_i^\text{DF} &= \sum\limits_{j \neq i} \left( p_{A,j} \left( |{\bf \hat g}_{AR,i}^H{\bf g}_{AR,j}|^2 + |{\bf \hat g}_{RB,i}^H{\bf g}_{AR,j}|^2\right)
  + p_{B,j} \left( |{\bf \hat g}_{AR,i}^H{\bf g}_{RB,j}|^2 + |{\bf \hat g}_{RB,i}^H{\bf g}_{RB,j}|^2\right)\right),\\
  E_i^\text{DF} &= ||{\bf\hat g}_{AR,i}||^2 + ||{\bf \hat g}_{RB,i}||^2.
\end{align}

In addition, the ergodic spectral efficiency of the ${\text T}_{X,i} \rightarrow {\text T}_R$ ($X \in \left\{A,B\right\}$) link can be obtained as
\begin{align}\label{eq:R:AR:DF}
R_{XR,i}^\text{DF} = \frac{1}{2} {\tt E} \left\{ \log_2 \left( 1 + \frac{X_i^\text{DF}}{ {\tt E} \left\{ \left( C_i^\text{DF} + D_i^\text{DF} + E_i^\text{DF} \right)| {\hat{\bf G}}_{AR},{\hat{\bf G}}_{RB} \right\} } \right) \right\}.
\end{align}

In the second phase, the relay broadcasts to all users using the MR principle; hence the received signal at ${\text T}_{X,i}$ is given by
\begin{align}
  {\bf z}_{X,i}^\text{DF} = \rho_\text{DF} \sum\limits_{j = 1}^{N} \left( {\bf g}_{XR,i}^T{\bf \hat g}_{RB,j}^* {\bf x}_{A,j} + {\bf g}_{XR,i}^T{\bf \hat g}_{AR,j}^* {\bf x}_{B,j}\right) + {\bf n}_{X,i}.
\end{align}

Similar to the AF protocol, partial self-interference cancellation according to the statistical knowledge of the channel gains can be performed at ${\text T}_{A,i}$ and ${\text T}_{B,i}$ after receiving the superimposed signal from ${\text T}_R$. Thus the post-processing signals at ${\text T}_{A,i}$ and ${\text T}_{B,i}$ can be re-expressed as,
\begin{align}
  {\hat z}_{A,i}^\text{DF} &= {z}_{A,i}^\text{DF}- \rho_\text{DF} {\tt E}\left\{ {\bf g}_{AR,i}^T{\bf \hat g}_{RB,i}^* \right\} {x}_{A,i}\\
  &= \underbrace{\rho_\text{DF} {\tt E}\left\{ {\bf g}_{AR,i}^T{\bf \hat g}_{AR,i}^* \right\}{x}_{B,i}}_{\text{desired signal}}
  + \underbrace{\rho_\text{DF}  \left( {\bf g}_{AR,i}^T{\bf \hat g}_{AR,i}^*  - {\tt E}\left\{{\bf g}_{AR,i}^T{\bf \hat g}_{AR,i}^* \right\} \right) {x}_{B,i} }_{\text{estimation error}}\\
  &+ \underbrace{\rho_\text{DF} \left({\bf g}_{AR,i}^T{\bf \hat g}_{RB,i}^* - {\tt E}\left\{{\bf g}_{AR,i}^T{\bf \hat g}_{RB,i}^* \right\} \right){x}_{A,i}}_{\text{residual self-interference}} \\
  &+ \underbrace{\rho_\text{DF} \sum\limits_{j \neq i} \left( {\bf g}_{AR,i}^T{\bf \hat g}_{RB,j}^* {x}_{A,j} + {\bf g}_{AR,i}^T{\bf \hat g}_{AR,j}^* {x}_{B,j}\right)}_{\text{inter-user interference}} + \underbrace{{n}_{A,i}}_{\text{noise}}.\\
  {\hat z}_{B,i}^\text{DF} &= {z}_{B,i}^\text{DF}- \rho_\text{DF} {\tt E}\left\{ {\bf g}_{RB,i}^T{\bf \hat g}_{AR,i}^* \right\} {x}_{B,i}\\
  &= \underbrace{\rho_\text{DF} {\tt E}\left\{ {\bf g}_{RB,i}^T{\bf \hat g}_{RB,i}^* \right\}{x}_{A,i}}_{\text{desired signal}}
  + \underbrace{\rho_\text{DF}  \left( {\bf g}_{RB,i}^T{\bf \hat g}_{RB,i}^*  - {\tt E}\left\{{\bf g}_{RB,i}^T{\bf \hat g}_{RB,i}^* \right\} \right) {x}_{A,i} }_{\text{estimation error}}\\
  &+ \underbrace{\rho_\text{DF} \left({\bf g}_{RB,i}^T{\bf \hat g}_{AR,i}^* - {\tt E}\left\{{\bf g}_{RB,i}^T{\bf \hat g}_{AR,i}^* \right\} \right){x}_{B,i}}_{\text{residual self-interference}} \\
  &+ \underbrace{\rho_\text{DF} \sum\limits_{j \neq i} \left( {\bf g}_{RB,i}^T{\bf \hat g}_{RB,j}^* {x}_{A,j} + {\bf g}_{RB,i}^T{\bf \hat g}_{AR,j}^* {x}_{B,j}\right)}_{\text{inter-user interference}} + \underbrace{{n}_{B,i}}_{\text{noise}}.
\end{align}

Therefore, the ergodic spectral efficiency of the ${\text T}_R \rightarrow {\text T}_{X,i}$ link is expressed as
\begin{align}
  R_{RX,i}^\text{DF} = \frac{1}{2} \log_2 \left(1 + {\text{SINR}_{RX,i}^{\text{DF}}} \right),
\end{align}
where
\begin{align}
  {\text{SINR}_{RX,i}^{\text{DF}}} = \frac{|{\tt E}\left\{ {\bf g}_{XR,i}^T{\bf \hat g}_{XR,i}^* \right\}|^2}{{\text {Var}}\left({\bf g}_{XR,i}^T{\bf \hat g}_{XR,i}^* \right) + {\text {Var}}\left({\bf g}_{AR,i}^T{\bf \hat g}_{RB,i}^* \right) + \sum\limits_{j \neq i} \left( {\tt E}\left\{|{\bf g}_{XR,i}^T{\bf \hat g}_{RB,j}^*|^2\right\} + {\tt E}\left\{|{\bf g}_{XR,i}^T{\bf \hat g}_{AR,j}^*|^2\right\}\right) + \frac{1}{\rho_{\text{DF}}^2}}.
\end{align}

Now, according to \cite{B.Rankov,J.Gao1,J.Gao2,J.Gao3,I.Hammerstrom,A.Alsharoa}, the ergodic spectral efficiency of the \emph{i}-th user pair can be expressed as
\begin{align}
  R_i^\text{DF} = {\text{min}}\left( R_{1,i}^\text{DF}, R_{2,i}^\text{DF}\right),
\end{align}
where
\begin{align}
  R_{2,i}^\text{DF} = {\text{min}} \left( R_{AR,i}^\text{DF}, R_{RB,i}^\text{DF}\right) + {\text{min}} \left( R_{BR,i}^\text{DF}, R_{RA,i}^\text{DF}\right)
\end{align}

Thus, the ergodic sum spectral efficiency of the multipair two-way DF relaying system is given by
\begin{align}\label{eq:rate:DF:R}
  R^\text{DF} = \frac{\tau_c-\tau_p}{\tau_c}  \sum\limits_{i = 1}^{N} R_i^\text{DF}.
\end{align}

When ${\text T}_R$ employs a very large antenna array, i.e., $M \rightarrow \infty$, the large-scale approximation of the spectral efficiency of the \emph{i}-th user pair is presented in the following theorem.
\begin{theorem}\label{theor:1:DF}
With the DF protocol, as the number of relay antennas grows to infinity, then we have $R_i^\text{DF} -  {\tilde R}_i^\text{DF} \overset{M \rightarrow \infty}{\longrightarrow} 0$
, where ${\tilde R}_i^\text{DF}$ is given by
\begin{align}\label{eq:R:DF}
{\tilde R}_i^\text{DF} \triangleq {\text{min}}\left( {\tilde R}_{1,i}^\text{DF}, {\tilde R}_{2,i}^\text{DF}\right),
\end{align}
with
\begin{align}
  &{\tilde R}_{1,i}^\text{DF} \triangleq \frac{1}{2}\log_2 \left(1 + \frac{p_{A,i} \left(M \sigma_{AR,i}^4 + \sigma_{AR,i}^2 \sigma_{RB,i}^2 \right) + p_{B,i} \left( M\sigma_{RB,i}^4 + \sigma_{AR,i}^2 \sigma_{RB,i}^2 \right)}{\left(\sigma_{AR,i}^2 + \sigma_{RB,i}^2\right)\left(p_{A,i}{\tilde\sigma}_{AR,i}^2 + p_{B,i} {\tilde\sigma}_{RB,i}^2 + \sum\limits_{j \neq i}\left(p_{A,j} \beta_{AR,j} + p_{B,j}\beta_{RB,j}\right) + 1\right)}\right),\\
  &{\tilde R}_{2,i}^\text{DF} \triangleq {\text{min}} \left( {\tilde R}_{AR,i}^\text{DF}, {\tilde R}_{RB,i}^\text{DF}\right) + {\text{min}} \left( {\tilde R}_{BR,i}^\text{DF}, {\tilde R}_{RA,i}^\text{DF}\right),
\end{align}
where
\begin{align}
  &{\tilde R}_{AR,i}^\text{DF} \triangleq \frac{1}{2}\log_2 \left(1 + \frac{p_{A,i} \left(M \sigma_{AR,i}^4 + \sigma_{AR,i}^2 \sigma_{RB,i}^2 \right)}{\left(\sigma_{AR,i}^2 + \sigma_{RB,i}^2\right)\left(p_{A,i}{\tilde\sigma}_{AR,i}^2 + p_{B,i} {\tilde\sigma}_{RB,i}^2 + \sum\limits_{j \neq i}\left(p_{A,j} \beta_{AR,j} + p_{B,j}\beta_{RB,j}\right) + 1\right)}\right),\\
  &{\tilde R}_{RA,i}^\text{DF} \triangleq \frac{1}{2}\log_2 \left(1 + \frac{p_r M \sigma_{AR,i}^4}{\left(p_r\beta_{AR,i} + 1\right)\sum\limits_{j = 1}^{N}\left(\sigma_{AR,j}^2 + \sigma_{RB,j}^2 \right)} \right),
\end{align}
and ${\tilde R}_{BR,i}^\text{DF}$ and ${\tilde R}_{RB,i}^\text{DF}$ are obtained by replacing the transmit powers $p_{A,i}$, $p_{B,i}$, and the subscripts ``AR'', ``RB'' with the transmit powers $p_{B,i}$, $p_{A,i}$, and the subscripts ``RB'', ``AR'' in ${\tilde R}_{AR,i}^\text{DF}$ and ${\tilde R}_{RA,i}^\text{DF}$, respectively.
\end{theorem}
\proof See Appendix \ref{app:theor:1:DF}.\endproof
Theorem \ref{theor:1:DF} provides a large-scale approximation of the \emph{i}-th user pair's spectral efficiency. More specifically, ${\tilde R}_{1,i}^\text{DF}$, ${\tilde R}_{AR,i}^\text{DF}$, and ${\tilde R}_{BR,i}^\text{DF}$ are computed by utilizing the law of large numbers, while ${\tilde R}_{RA,i}^\text{DF}$ and ${\tilde R}_{RB,i}^\text{DF}$ are the exact expressions for ${R}_{RA,i}^\text{DF}$ and ${R}_{RB,i}^\text{DF}$. From this approximation, we can see that ${\tilde R}_i^\text{DF}$ is an increasing function with respect to $p_{A,i}$, $p_{B,i}$, and $p_r$. However, when $p_{A,i} \rightarrow \infty$, $p_{B,i} \rightarrow \infty$, and/or $p_r \rightarrow \infty$, ${\tilde R}_i^\text{DF}$ converges to a non-zero limit, due to strong inter-user interference. Moreover, we observe that ${\tilde R}_i^\text{DF}$ increases with the number of relay antennas $M$, indicating the strong advantage of employing massive antenna arrays at the relay, while decreases with the number of user pairs $N$, which is expected since larger number of users increases the amount of inter-user interference. Finally, when $p_r \rightarrow 0$ the bottleneck of spectral efficiency occurs in the second phase, while the opposite holds when $p_{A,i} \rightarrow 0$ and $p_{B,i} \rightarrow 0$, where we have
\begin{align}
  {\tilde R}_i^\text{DF} - \frac{1}{2}\log_2 \left(1 + \frac{p_{A,i} \sigma_{AR,i}^2 \left( M\sigma_{AR,i}^2 + \sigma_{RB,i}^2 \right) + p_{B,i}\sigma_{RB,i}^2 \left( \sigma_{AR,i}^2 + M \sigma_{RB,i}^2 \right)}{\sigma_{AR,i}^2 + \sigma_{RB,i}^2 }\right) \rightarrow 0. \notag
\end{align}

It is of interest to compare the achievable sum rates of AF and DF protocols in the low $p_{A,i}$ and $p_{B,i}$ regime for massive MIMO systems which have the potential to save an order of magnitude in transmit power. Then, we have the following corollary.
\begin{corollary}\label{coro:AF:vs:DF}
In the low SNR regime, i.e.,  $p_{A,i}\rightarrow 0$ and $p_{B,i}\rightarrow 0$, we have ${\tilde R}_{A,i}^\text{AF} + {\tilde R}_{B,i}^\text{AF} > {\tilde R}_i^\text{DF}$.
\end{corollary}
\proof
When $p_{A,i} \rightarrow 0$ and $p_{B,i} \rightarrow 0$, we have
\begin{align}
 {\tilde R}_{A,i}^\text{AF} - \frac{1}{2}\log_2\left( 1 + p_{B,i} M \sigma_{RB,i}^2\right) \rightarrow 0,\\
  {\tilde R}_{B,i}^\text{AF} - \frac{1}{2}\log_2\left( 1 + p_{A,i} M \sigma_{AR,i}^2\right) \rightarrow 0.
\end{align}
To this end, it can be shown that
\begin{align}
 & 1 + \frac{p_{A,i} \sigma_{AR,i}^2 \left( M\sigma_{AR,i}^2 + \sigma_{RB,i}^2 \right) + p_{B,i}\sigma_{RB,i}^2 \left( \sigma_{AR,i}^2 + M \sigma_{RB,i}^2 \right)}{\sigma_{AR,i}^2 + \sigma_{RB,i}^2 },\\
    &< 1 + \frac{p_{A,i} \sigma_{AR,i}^2 M \left(\sigma_{AR,i}^2 + \sigma_{RB,i}^2 \right) + p_{B,i}\sigma_{RB,i}^2 M \left( \sigma_{AR,i}^2 + \sigma_{RB,i}^2 \right)}{\sigma_{AR,i}^2 + \sigma_{RB,i}^2 },\\
    & = 1 + p_{A,i} M \sigma_{AR,i}^2 +  p_{B,i} M \sigma_{RB,i}^2,\\
    &< \left( 1 + p_{A,i} M \sigma_{AR,i}^2\right) \left( 1 + p_{B,i} M \sigma_{RB,i}^2\right),
\end{align}
which completes the proof.
\endproof
Corollary \ref{coro:AF:vs:DF} indicates that the AF protocol outperforms the DF protocol in the low SNR regime.
\subsection{Numerical Results}
We now present numerical results to validate the above analytical results. For all illustrative examples, the following set of parameters are used in simulation. The length of the coherence interval is $\tau_c = 196$ (symbols), chosen by the LTE standard. The length of the pilot sequences is $\tau_p = 2N$ which is the minimum requirement. For simplicity, we set the large-scale fading coefficient $\beta_{AR} = \beta_{RB} = 1$, and assume that each user has the same transmit power, i.e., $p_{A,i} = p_{B,i} = p_u$.
\subsubsection{Validation of analytical expressions}
We assume that $p_p = p_u$, and that the total transmit power of the $N$ user pairs is equal to the transmit power of the relay, i.e., $p_r = 2Np_u$.

\begin{figure}[!ht]
    \centering
    \includegraphics[scale=0.6]{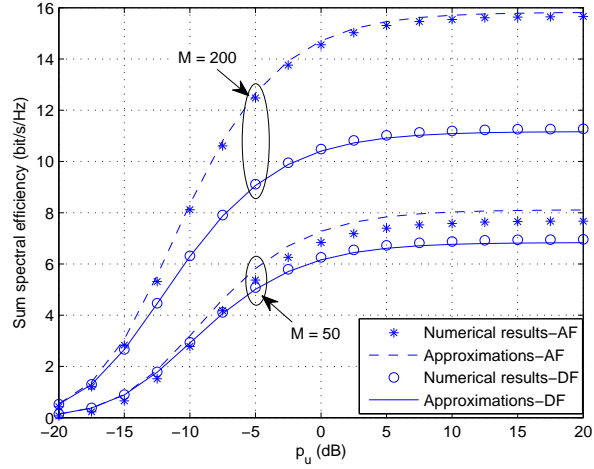}
    \caption{Sum spectral efficiency versus $p_u$ for $N = 5$, $p_p = p_u$ and $p_r = 2Np_u$.}\label{fig:rate_snr}
  \end{figure}

Fig. \ref{fig:rate_snr} shows the sum spectral efficiency versus the transmit power of each user $p_u$ for different number of relay antennas. Note that the ``Approximations'' curves are obtained by using \eqref{eq:theor:1:R} and \eqref{eq:R:DF}, and the ``Numerical results'' curves are generated by Monte-Carlo simulation according to \eqref{eq:rate:R:sum} and \eqref{eq:rate:DF:R} by averaging over $10^4$ independent channel realizations, for the AF and DF protocols, respectively. As can be readily observed, the large-scale approximations are very accurate, especially for large antenna arrays. Moreover, we can see that increasing the number of relay antennas significantly yields higher spectral efficiency, as expected.
\subsubsection{Comparison of the AF and DF protocols}
We now compare the sum spectral efficiency of the AF and DF protocols for different system configurations, i.e., different transmit powers $p_u$, $p_r$, and $p_p$, and different number of relay antennas $M$ and user pairs $N$.

  \begin{figure}[!ht]
    \centering
    \includegraphics[scale=0.6]{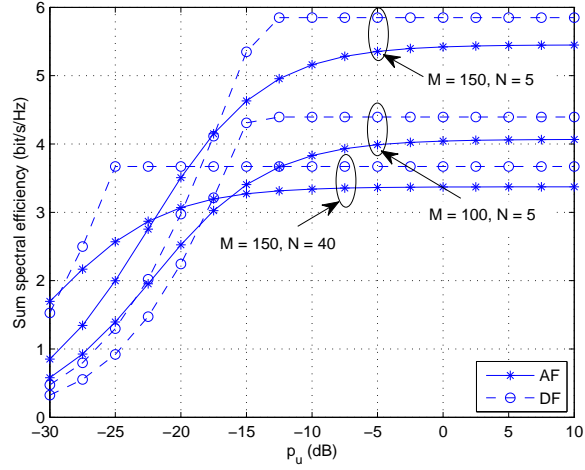}
    \caption{Spectral efficiency versus $p_u$ for $p_r = -10$ dB and $p_p = 10$ dB.}\label{fig:rate_pu}
  \end{figure}
Fig. \ref{fig:rate_pu} shows the sum spectral efficiency versus the transmit power of each user $p_u$ for different $M$ and $N$ with $p_r = -10$ dB and $p_p = 10$ dB. We can observe that for small $p_u$, the AF protocol outperforms the DF protocol, which is consistent with the result in Corollary \ref{coro:AF:vs:DF}. The reason is that when $p_u$ is small, the spectral efficiency of the DF protocol is limited by the performance in the first phase. On the other hand, when $p_u$ is large, the noise amplification phenomenon of the AF protocol will significantly affect the spectral efficiency of the relay to the destination link; this makes DF outperform the AF protocol by eliminating the noise and preventing interference accumulation at the end users.

\begin{figure}[!ht]
    \centering
    \includegraphics[scale=0.6]{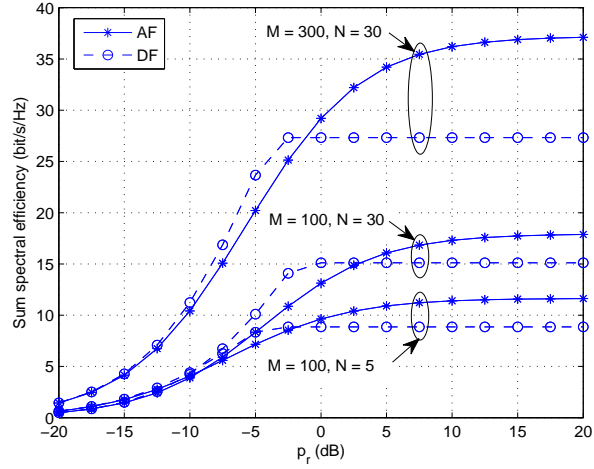}
    \caption{Sum spectral efficiency versus the transmit power of the relay $p_r$ for $p_u = 10$ dB and $p_p = 10$ dB.}\label{fig:rate_pr}
\end{figure}
Fig. \ref{fig:rate_pr} provides the sum spectral efficiency versus the transmit power of the relay $p_r$ for different $M$ and $N$ with $p_u = 10$ dB and $p_p = 10$ dB. As we can observe, the DF protocol is superior to the AF protocol in the low $p_r$ regime but becomes inferior in the high $p_r$ regime. This is due to the fact that low $p_r$ makes the AF protocol suffer severe noise amplification effect and thus leads to spectral efficiency reduction. In addition, focusing on the particular operating point $p_r = 0$ dB, we see that the DF protocol achieves higher sum spectral efficiency when $M=100$ and $N = 30$, while the AF protocol becomes better when $M = 300$ and $N = 30$ or $M = 100$ and $N = 30$. The reason is that the residual interference due to inaccurate channel estimation is a key performance limiting factor for the AF protocol; in other words, the larger the $N$, the stronger the interference. As such, less user pairs are preferable for the AF protocol. However, it turns out that increasing the number of relay antennas is an effective way to mitigate such a detrimental effect.

  \begin{figure}[!ht]
    \centering
    \includegraphics[scale=0.6]{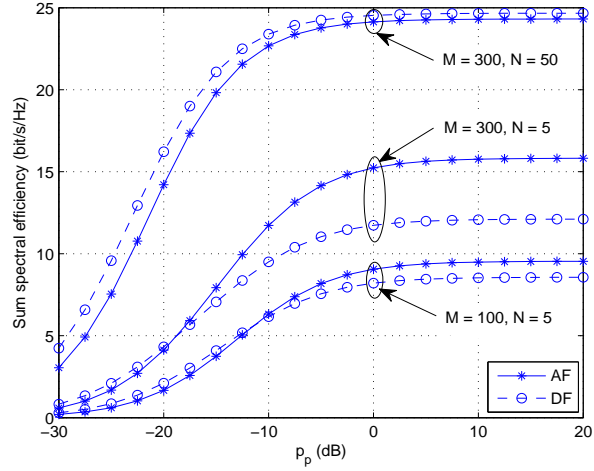}
    \caption{Sum spectral efficiency versus the transmit power of each pilot symbol $p_p$ for $p_u = 0$ dB and $p_r = 0$ dB.}\label{fig:rate_pp}
  \end{figure}

  Fig. \ref{fig:rate_pp} presents the sum spectral efficiency versus the transmit power of each pilot symbol $p_p$ for different $M$ and $N$ with $p_u = 0$ dB and $p_r = 0$ dB. Similarly, we observe that for fixed $N = 5$ and $M$, the DF protocol outperforms the AF protocol in the low $p_p$ regime while the converse holds in the high $p_p$ regime. In addition, focusing on the curves associated with $M = 300$ and $N = 50$, we see that the spectral efficiency of the DF protocol is higher than that of the AF protocol, indicating that a large $N$ is preferred for the DF protocol.


%

\begin{figure}[h!]
\centering
\subfigure[$p_r=0$ dB]{
\label{fig:fixedpr}
\includegraphics[scale=0.55]{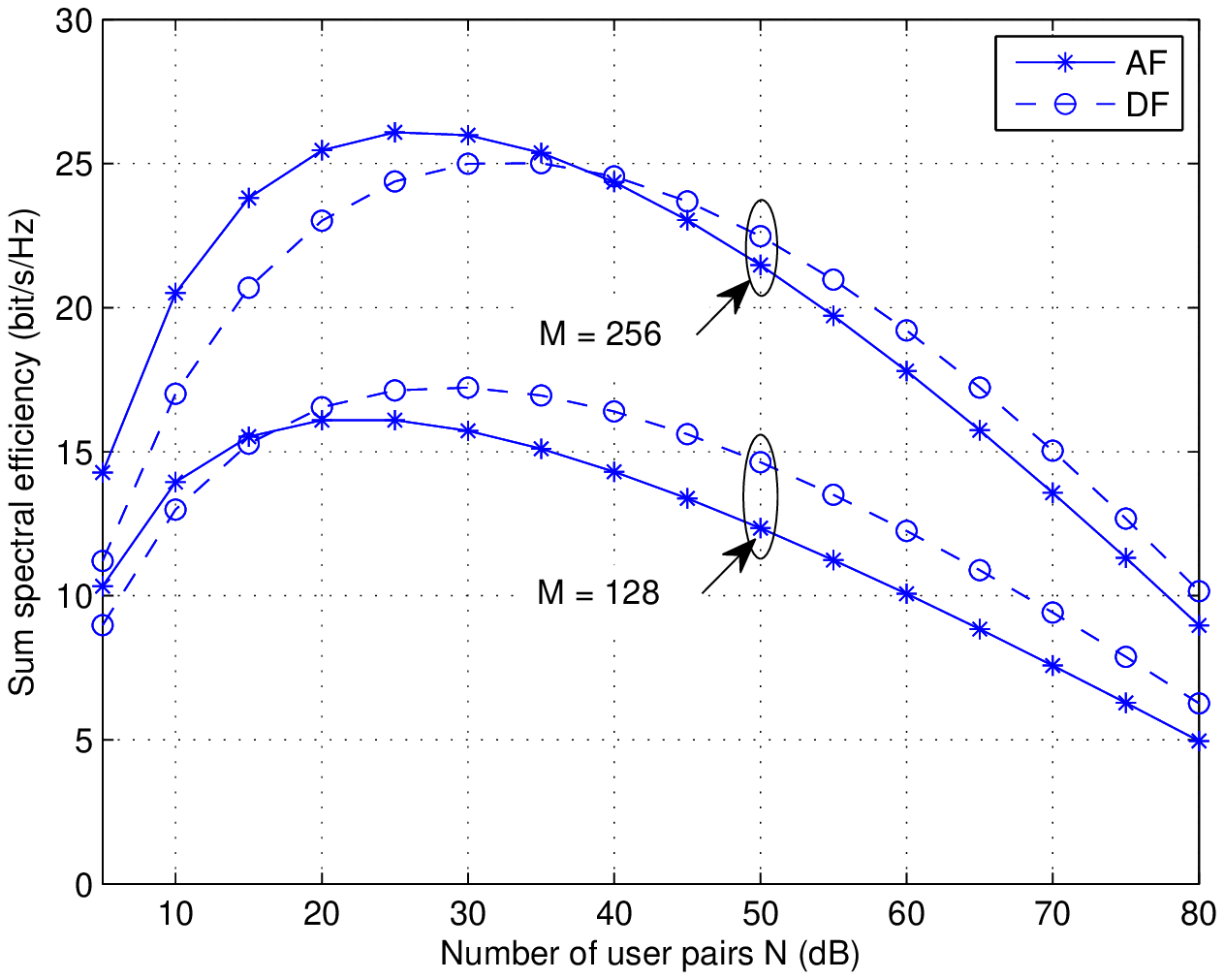}}
\subfigure[$p_r=2Np_u$]{
\label{fig:variablepr}
\includegraphics[scale=0.55]{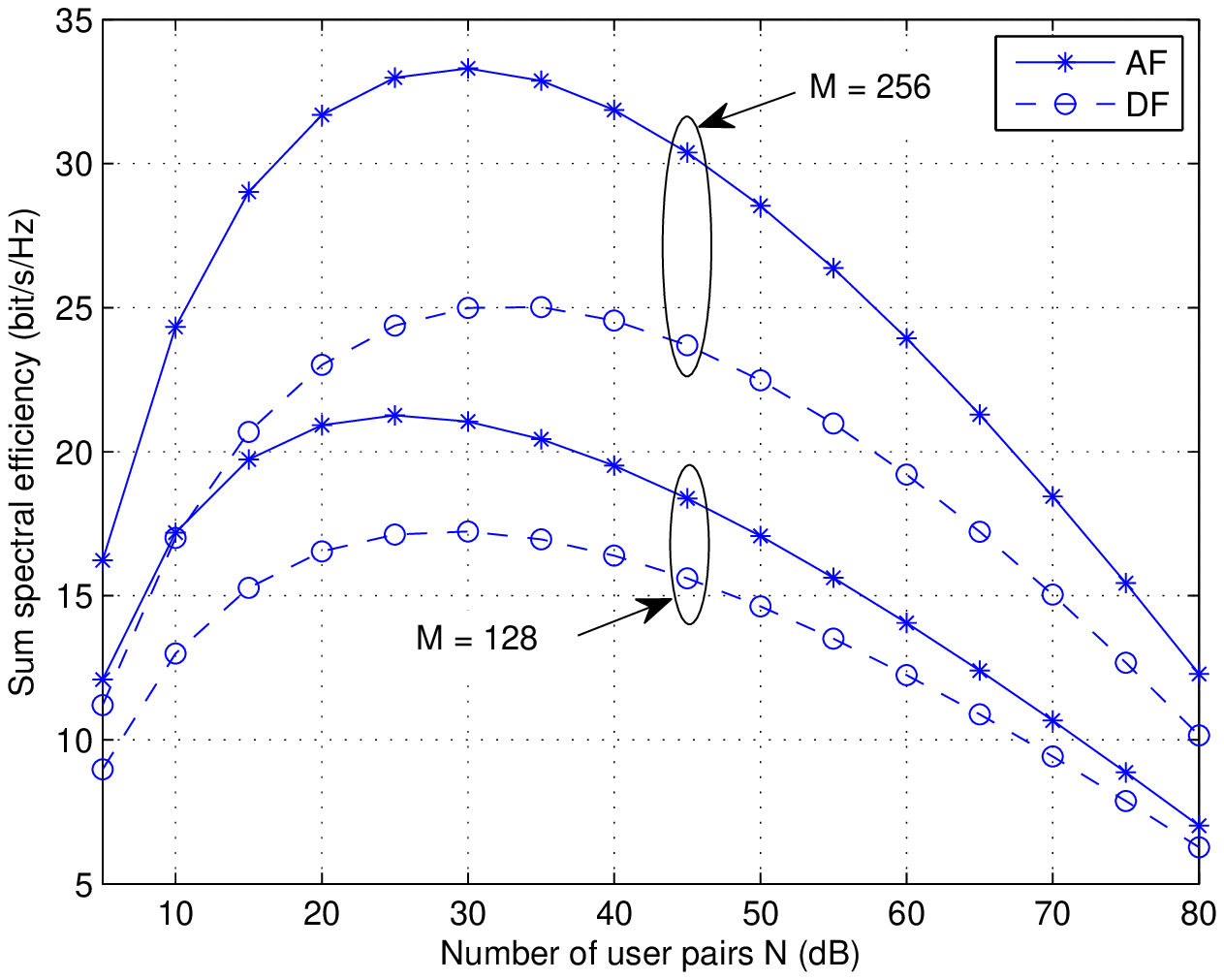}}
\caption{Sum spectral efficiency versus the number of user pairs $N$ for $p_p = 0$ dB, $p_u = 0$ dB.}
\label{Fig.label1}
\end{figure}

Fig. \ref{Fig.label1} illustrates the impact of number of user pairs $N$ on the sum spectral efficiency when $p_p = 0$ dB and $p_u = 0$ dB. As expected, for each system configuration, there exists an optimal number of user pairs $N$ maximizing the spectral efficiency of both the AF and DF protocols. With fixed $p_r$, as shown in Fig. \ref{fig:fixedpr}, the DF protocol achieves higher spectral efficiency than the AF protocol when $N$ is large. However, this is not the case if $p_r$ scales linearly with $N$, as shown in Fig. \ref{fig:variablepr}, where the AF protocol always outperforms the DF protocol. In addition, the performance gap widens when the number of antennas $M$ increases.

\section{Power Scaling Laws}\label{section:4}
In this section, we pursue a detailed investigation of the power scaling laws of both the AF and DF protocols; that is, how the powers can be reduced with $M$ while retaining non-zero spectral efficiency. Since we are interested in the general user's power scaling law rather than a particular user's behavior, we assume that all the users have the same transmit power, i.e., $p_{A,i} = p_{B,i} = p_u$. Then, we characterize the interplay between the relay's transmit power $p_r$, the user's transmit power $p_u$, and the transmit power of each pilot symbol $p_p$, as the number of relay antennas $M$ grows to infinity. More precisely, we consider three different scenarios:
\begin{itemize}
  \item Scenario A: Fixed $p_u$ and $p_r$, while $p_p = \frac{E_p}{M^\gamma}$ with $\gamma > 0$, and $E_p$ being a constant. Such a scenario represents the potential of power saving in the training phase.
  \item Scenario B: Fixed $p_p$, while $p_u = \frac{E_u}{M^\alpha}$, $p_r = \frac{E_r}{M^\beta}$, with $\alpha \geq 0$ and $\beta \geq 0$, and $E_u$, $E_r$ are constants. Hence, the channel estimation accuracy remains unchanged in Scenario B, and the objective is to study the potential power savings in the data transmission phase, as well as, the interplay between the user and relay transmit powers.
  \item Scenario C: This is the most general case where $p_u = \frac{E_u}{M^\alpha}$, $p_r = \frac{E_r}{M^\beta}$, and $p_p = \frac{E_p}{M^\gamma}$, with $\alpha \geq 0$, $\beta \geq 0$, and $\gamma > 0$, $E_u$, $E_r$, and $E_p$ are constants.
\end{itemize}
%
\subsection{AF protocol}
\subsubsection{Scenario A}
The AF protocol gives the following result.
\begin{theorem}\label{theor:2}
  With the AF protocol, for fixed $p_u$, $p_r$ and $E_p$, when $p_p = \frac{E_p}{M^\gamma}$ with $\gamma > 0$, as $M \rightarrow \infty$, we have
    \begin{align}\label{eq:theor:2}
    { R}_{A,i}^\text{AF} - \frac{1}{2}\log_2\left( 1 + \frac{\tau_p E_p M^{1-\gamma}}{{\hat B}_i^\text{AF} + {\hat C}_i^\text{AF} + {\hat D}_i^\text{AF} + {\hat E}_i^\text{AF}} \right) \overset{M \rightarrow \infty}{\longrightarrow} 0,
  \end{align}
where
\begin{align}
  &{\hat B}_i^\text{AF} \triangleq \frac{1}{\beta_{RB,i}} + \frac{1}{\beta_{AR,i}},\\
  &{\hat C}_i^\text{AF} \triangleq \frac{4\beta_{AR,i}}{\beta_{RB,i}^2},\\
 & {\hat D}_i^\text{AF} \triangleq \sum\limits_{j \neq i} \left( \frac{\beta_{AR,j} + \beta_{RB,j}}{\beta_{RB,i}^2}  + \frac{\beta_{AR,j}^4 \beta_{RB,j}^2 + \beta_{AR,j}^2 \beta_{RB,j}^4}{\beta_{AR,i}^3\beta_{RB,i}^4} \right),\\
 & {\hat E}_i^\text{AF} \triangleq \frac{1}{p_u \beta_{RB,i}^2} + \frac{1}{p_r \beta_{AR,i}^4 \beta_{RB,i}^4} \sum\limits_{n = 1}^{N}  \beta_{AR,n}^2 \beta_{RB,n}^2 \left( \beta_{AR,n}^2 + \beta_{RB,n}^2 \right).
\end{align}
\end{theorem}
\proof See Appendix \ref{app:theor:2}. \endproof

Theorem \ref{theor:2} implies that the large-scale approximation of the spectral efficiency ${R}_{A,i}^\text{AF}$ depends on the choice of $\gamma$. When $\gamma > 1$, ${R}_{A,i}^\text{AF}$ reduces to zero due to the poor channel estimation accuracy caused by over-reducing the pilot transmit power. In contrast, when $0<\gamma < 1$, ${ R}_{A,i}^\text{AF}$ grows without bound, which indicates that the transmit power of each pilot symbol can be scaled down further. Finally, when $\gamma = 1$, ${ R}_{A,i}^\text{AF}$ converges to a non-zero limit, which suggests that with large antenna arrays, the transmit power of each pilot symbol can be scaled down at most by $1/M$ to maintain a given quality-of-service (QoS).



\subsubsection{Scenario B}
A corresponding scaling law for Scenario B is obtained as follows.
\begin{theorem}\label{theor:3}
  With the AF protocol, for fixed $p_p$, $E_u$, and $E_r$, when $p_u = \frac{E_u}{M^\alpha}$, $p_r = \frac{E_r}{M^\beta}$, with $\alpha \geq 0$, $\beta \geq 0$, as $M \rightarrow \infty$, we have
  \begin{align}\label{eq:theor:3}
    { R}_{A,i}^\text{AF} - \frac{1}{2}\log_2\left(1 + \frac{1}{\underbrace{\frac{M^{\alpha -1}}{E_u \sigma_{RB,i}^2}}_{\text{Part I}} + \underbrace{\frac{M^{\beta - 1}}{E_r \sigma_{AR,i}^4 \sigma_{RB,i}^4} \sum\limits_{n = 1}^{N} \sigma_{AR,n}^2 \sigma_{RB,n}^2 \left( \sigma_{AR,n}^2 + \sigma_{RB,n}^2 \right)}_{\text{Part II}}} \right) \overset{M \rightarrow \infty}{\longrightarrow} 0.
  \end{align}
\end{theorem}
\proof
Substituting $p_u = \frac{E_u}{M^\alpha}$ and $p_r = \frac{E_r}{M^\beta}$ into \eqref{eq:theor:1:R}, it is easy to show that $\frac{{\tilde B}_i^\text{AF}}{{\tilde E}_i^\text{AF}} \rightarrow 0$, $\frac{{\tilde C}_i^\text{AF}}{{\tilde E}_i^\text{AF}} \rightarrow 0$, and $\frac{{\tilde D}_i^\text{AF}}{{\tilde E}_i^\text{AF}} \rightarrow 0$, as $M \rightarrow \infty$. Hence, keeping the most significant term ${\tilde E}_i^\text{AF}$, omitting the non-significant terms, namely, ${\tilde B}_i^\text{AF}$, ${\tilde C}_i^\text{AF}$, and ${\tilde D}_i^\text{AF}$, and utilizing the fact that $R^\text{AF}_{A,i} - {\tilde R}^\text{AF}_{A,i} \overset{M \rightarrow \infty}{\longrightarrow} 0$ yields the desired result.
\endproof

Theorem \ref{theor:3} reveals that in Scenario B, the estimation error, the residual self-interference, and the inter-user interference vanish completely, and only the compound noise remains, as $M \rightarrow \infty$. The reason is that the compound noise becomes significant as $M \rightarrow \infty$, compared to the estimation error, residual self-interference, and inter-user interference. Moreover, it is observed that the compound noise consists of two parts, namely Part I and Part II as shown in \eqref{eq:theor:3}, which represent the noise at the relay and the noise at the user ${\text T}_{A,i}$, respectively. This observation can be interpreted as, when both the transmit powers of each user and the relay are scaled down inversely proportional to $M$, the effect of noise becomes increasingly significant. In addition, we can also see that when the channel estimation accuracy is fixed, the large-scale approximation of the spectral efficiency ${R}_{A,i}^\text{AF}$ depends on the value of $\alpha$ and $\beta$. When we cut down the transmit powers of the relay and/or of each user too much, namely, 1) $\alpha > 1$, and $\beta \geq 0$, 2) $\alpha \geq 0$, and $\beta > 1$, 3) $\alpha > 1$, and $\beta > 1$, ${R}_{A,i}^\text{AF}$ converges to zero. On the other hand, when we cut down both the transmit powers of the relay and of each user moderately, namely, $0\leq\alpha<1$ and $0\leq\beta<1$, ${R}_{A,i}^\text{AF}$ grows unboundedly. Only if $\alpha=1$ and/or $\beta=1$, ${ R}_{A,i}^\text{AF}$ converges to a finite limit as detailed in the following corollaries.
\begin{corollary}\label{coro:3}
  With the AF protocol, for fixed $p_p$, $E_u$, and $E_r$, when $\alpha = \beta = 1$, namely, $p_u = \frac{E_u}{M}$, $p_r = \frac{E_r}{M}$, as $M \rightarrow \infty$, the spectral efficiency has the limit
  \begin{align}
    {R}_{A,i}^\text{AF} \ {\rightarrow}\ \frac{1}{2}\log_2\left(1 + \frac{1}{\frac{1}{E_u \sigma_{RB,i}^2} + \frac{1}{E_r \sigma_{AR,i}^4 \sigma_{RB,i}^4} \sum\limits_{n = 1}^{N} \left( \sigma_{AR,n}^2 \sigma_{RB,n}^2 \left( \sigma_{AR,n}^2 + \sigma_{RB,n}^2 \right) \right)} \right).
  \end{align}
\end{corollary}

From Corollary \ref{coro:3}, we observe that when both the transmit powers of the relay and of each user are scaled down with the same speed, i.e., $1/M$, ${R}_{A,i}^\text{AF}$ converges to a non-zero limit. Moreover, this non-zero limit increases with $E_u$ and $E_r$ as expected. Now consider the special case of all the links having the same large-scale fading, e.g., $\beta_{AR,i} = \beta_{RB,i} = 1$, for $i = 1, \ldots, N$, then the sum spectral efficiency of the system reduces to
\begin{align}\label{eq:theorem:beta:1}
   {R}^\text{AF} \ {\rightarrow}\ \frac{\tau_c-\tau_p}{\tau_c} N \log_2\left( 1 + \frac{\sigma_1^2 E_u E_r}{E_r + 2N E_u} \right),
\end{align}
where $\sigma_1^2 = \frac{\tau_p p_p}{\tau_p p_p +1}$. Therefore, the sum spectral efficiency in \eqref{eq:theorem:beta:1} is equal to the one of $N$ parallel single-input single-output channels with transmit power $\frac{\sigma_1^2 E_u E_r}{E_r + 2N E_u}$, without interference and small-scale fading. Note that we only need $\frac{2N\left(E_u + \tau_p E_p\right) + E_r}{M}$ power (the transmit power of each user is $\frac{E_u}{M}$, the transmit power of each pilot sequence is $\frac{\tau_p E_p}{M}$, and the transmit power of the relay is $\frac{E_r}{M}$) in multipair two-way AF relaying systems when using very large number of relay antennas. This represents a remarkable power reduction, thereby showcasing the huge benefits from the perspective of radiated energy efficiency by deploying large antenna arrays.
\begin{corollary}\label{coro:4}
  With the AF protocol, for fixed $p_p$, $E_u$ and $E_r$, when $\alpha = 1$ and $0 \leq \beta < 1$, namely, $p_u = \frac{E_u}{M}$, $p_r = \frac{E_r}{M^\beta}$, as $M \rightarrow \infty$, the spectral efficiency has the limit
  \begin{align}
    {R}_{A,i}^\text{AF} \ {\rightarrow}\ \frac{1}{2}\log_2\left(1 + E_u \sigma_{RB,i}^2 \right).
  \end{align}
\end{corollary}

Corollary \ref{coro:4} presents an interesting phenomenon, that if the transmit power of each user is overly cut down compared to the reduction of the relay transmit power, ${R}_{A,i}^\text{AF}$ converges to a non-zero limit that is determined by the noise at the relay. This observation is intuitive, since when both ${\text T}_{A,i}$ and ${\text T}_{B,i}$ transmit with extremely low power, the effect of noise at the relay becomes the performance limiting factor. Similarly, when $0\leq\alpha < \beta = 1$, we have the following corollary.
\begin{corollary}\label{coro:5}
  With the AF protocol, for fixed $p_p$, $E_u$, and $E_r$, when $0 \leq \alpha < 1$ and $\beta = 1$, namely, $p_u = \frac{E_u}{M^\alpha}$, $p_r = \frac{E_r}{M}$,  as $M \rightarrow \infty$, the spectral efficiency has the limit
  \begin{align}
    {R}_{A,i}^\text{AF} \ {\rightarrow}\ \frac{1}{2}\log_2\left(1 + \frac{E_r \sigma_{AR,i}^4 \sigma_{RB,i}^4}{\sum\limits_{n = 1}^{N} \left( \sigma_{AR,n}^2 \sigma_{RB,n}^2 \left( \sigma_{AR,n}^2 + \sigma_{RB,n}^2 \right) \right)} \right).
  \end{align}
\end{corollary}

Similar to the analysis in Corollary \ref{coro:4}, if the down-scaling of the relay transmit power in the second phase is faster than that of the user transmit power in the first phase, the limit of ${R}_{A,i}^\text{AF}$ will only depend on the noise at the users.
\subsubsection{Scenario C}
Next, we consider the most general case.
\begin{theorem}\label{theor:4}
  With the AF protocol, for fixed $E_u$, $E_r$, and $E_p$, when $p_u = \frac{E_u}{M^\alpha}$, $p_r = \frac{E_r}{M^\beta}$, and $p_p = \frac{E_p}{M^\gamma}$, with $\alpha \geq 0$, $\beta \geq 0$, and $\gamma > 0$, as $M \rightarrow \infty$, we have
 \begin{align}\label{eqn:ccc}
    {R}_{A,i}^\text{AF} - \frac{1}{2}\log_2\left(1 + \frac{1}{\frac{M^{\alpha + \gamma -1}}{\tau_p E_p E_u \beta_{RB,i}^2} + \frac{M^{\beta + \gamma - 1}}{\tau_p E_p E_r \beta_{AR,i}^4 \beta_{RB,i}^4} \sum\limits_{n = 1}^{N} \beta_{AR,n}^2 \beta_{RB,n}^2 \left( \beta_{AR,n}^2 + \beta_{RB,n}^2 \right)} \right) \overset{M \rightarrow \infty}{\longrightarrow} 0.
  \end{align}
\end{theorem}
\proof The desired result can be obtained by following the similar lines as in the proof of Theorem \ref{theor:2}. \endproof

Theorem \ref{theor:4} reveals the coupled relationship between the training power and user (or relay) transmit power. When $\alpha + \gamma > 1$ and/or $\beta + \gamma > 1$, ${R}_{A,i}^\text{AF}$ converges to zero, due to either poor estimation accuracy or low user/relay transmit power. On the other hand, when $0< \alpha + \gamma < 1$ and $0< \beta + \gamma <1$, ${R}_{A,i}^\text{AF}$ grows without bound. Only if $\alpha + \gamma = 1$ and/or $\beta + \gamma =1$, ${R}_{A,i}^\text{AF}$ converges to a non-zero limit. In the following, we take a closer look at these particular cases of interest.
\begin{corollary}\label{coro:6}
 With the AF protocol, for fixed $E_u$, $E_r$, and $E_p$, when $\alpha = \beta > 0$ and $\alpha + \gamma = 1$, namely, $p_u = \frac{E_u}{M^\alpha}$, $p_r = \frac{E_r}{M^\beta}$, and $p_p = \frac{E_p}{M^\gamma}$, with $\gamma > 0$, as $M \rightarrow \infty$, the spectral efficiency has the limit
 \begin{align}
    {R}_{A,i}^\text{AF} \ {\rightarrow}\ \frac{1}{2}\log_2\left(1 + \frac{1}{\frac{1}{\tau_p E_p E_u \beta_{RB,i}^2} + \frac{1}{\tau_p E_p E_r \beta_{AR,i}^4 \beta_{RB,i}^4} \sum\limits_{n = 1}^{N} \left( \beta_{AR,n}^2 \beta_{RB,n}^2 \left( \beta_{AR,n}^2 + \beta_{RB,n}^2 \right) \right)} \right).
  \end{align}
\end{corollary}

Corollary \ref{coro:6} suggests that no matter how $\alpha$, $\beta$, and $\gamma$ change, as long as the overall power reduction rate at the user/relay and pilot symbol remains the same, i.e., $\alpha + \gamma = 1$, the same asymptotic spectral efficiency can be attained. In other words, it is possible to balance between the pilot symbol power to the user/relay transmit power.
\begin{corollary}\label{coro:7}
  With the AF protocol, for fixed $E_u$, $E_r$, and $E_p$, when $\alpha > \beta \geq 0$ and $\alpha + \gamma = 1$, namely, $p_u = \frac{E_u}{M^\alpha}$, $p_r = \frac{E_r}{M^\beta}$, and $p_p = \frac{E_p}{M^\gamma}$, with $\gamma > 0$, as $M \rightarrow \infty$, the spectral efficiency has the limit
 \begin{align}
    {R}_{A,i}^\text{AF} \ {\rightarrow}\ \frac{1}{2}\log_2\left(1 + \tau_p E_p E_u \beta_{RB,i}^2 \right).
  \end{align}
\end{corollary}
\proof
When $\alpha > \beta \geq 0$ and $\alpha + \gamma = 1$, $\beta + \gamma < 1$, hence, the first term in the denominator of (\ref{eqn:ccc}) becomes the dominant item, yielding the desired result. \endproof

From Corollary \ref{coro:7}, we observe the same trade-off between $\alpha$ and $\gamma$ as in Corollary \ref{coro:6}, when $\alpha > \beta \geq 0$. However, the spectral efficiency is only related to the noise at the relay. In addition, we also see that the limit of ${R}_{A,i}^\text{AF}$ is independent of the number of users; thus, we conclude that the sum spectral efficiency of the system is an increasing function with respect to $N$. Specifically, it is a linear function of $N$ when assuming all the users have the same large-scale fading, e.g., setting $\beta_{AR,i} = \beta_{RB,i} = 1$. So in this case, a large number of users will boost the sum spectral efficiency.

\begin{corollary}\label{coro:8}
 With the AF protocol, for fixed $E_u$, $E_r$, and $E_p$, when $0 \leq \alpha < \beta$ and $\beta + \gamma = 1$, namely, $p_u = \frac{E_u}{M^\alpha}$, $p_r = \frac{E_r}{M^\beta}$, and $p_p = \frac{E_p}{M^\gamma}$, with $\gamma > 0$, as $M \rightarrow \infty$, the spectral efficiency has the limit
 \begin{align}
    {R}_{A,i}^\text{AF} \ {\rightarrow}\ \frac{1}{2}\log_2\left(1 + \frac{\tau_p E_p E_r \beta_{AR,i}^4 \beta_{RB,i}^4}{\sum\limits_{n = 1}^{N} \beta_{AR,n}^2 \beta_{RB,n}^2 \left( \beta_{AR,n}^2 + \beta_{RB,n}^2 \right) } \right).
  \end{align}
\end{corollary}

From Corollary \ref{coro:8}, we see that when we cut down the transmit power of the relay more compared with the transmit power of each user, i.e., $0\leq \alpha < \beta$, to obtain a constant limit spectral efficiency, the trade-off between the transmit powers of the relay and of each pilot symbol should be satisfied, namely, $\beta + \gamma = 1$. This informative trade-off provides valuable insights, since we can adjust the transmit powers of the relay and of each pilot symbol flexibly based on different demands, to meet the same limit. In addition, Corollary \ref{coro:8} also shows that ${R}_{A,i}^\text{AF}$ is an increasing function with respect to $E_p$ and $E_r$, while an decreasing function of $N$. In other words, when the number of user pairs $N$ increases, the relay and/or each pilot symbol should increase their power in order to maintain the same performance. This is due to the fact that large transmit power of the relay and/or more accurate channel estimation can compensate the individual rate loss caused by stronger inter-user interference.
\subsection{DF protocol}
\subsubsection{Scenario A}
Similarly, we present the following power scaling law for Scenario A with the DF protocol.
\begin{theorem}\label{theor:2:DF}
  With the DF protocol, for fixed $p_u$, $p_r$ and $E_p$, when $p_p = \frac{E_p}{M^\gamma}$ with $\gamma > 0$, as $M \rightarrow \infty$, we have
\begin{align}\label{eq:theor:2:DF}
  {R}_i^\text{DF} - {\text{min}}\left( {\bar R}_{1,i}^\text{DF}, {\bar R}_{2,i}^\text{DF}\right) \overset{M \rightarrow \infty}{\longrightarrow} 0,
\end{align}
where
\begin{align}
  &{\bar R}_{1,i}^\text{DF} \triangleq \frac{1}{2}\log_2 \left(1 + \frac{p_u \frac{\tau_p E_p}{M^{\gamma-1}} \left(\beta_{AR,i}^4 + \beta_{RB,i}^4 \right)}{\left(\beta_{AR,i}^2 + \beta_{RB,i}^2\right)\left(p_u \sum\limits_{j = 1}^{N}\left(\beta_{AR,j} + \beta_{RB,j}\right) + 1\right)}\right),\\
  &{\bar R}_{2,i}^\text{DF} \triangleq {\text{min}} \left( {\bar R}_{AR,i}^\text{DF}, {\bar R}_{RB,i}^\text{DF}\right) + {\text{min}} \left( {\bar R}_{BR,i}^\text{DF}, {\bar R}_{RA,i}^\text{DF}\right),
\end{align}
with
\begin{align}
  &{\bar R}_{AR,i}^\text{DF} \triangleq \frac{1}{2}\log_2 \left(1 + \frac{p_u \frac{\tau_p E_p}{M^{\gamma - 1}} \beta_{AR,i}^4 }{\left(\beta_{AR,i}^2 + \beta_{RB,i}^2 \right)\left(p_u\sum\limits_{j = 1}^{N}\left(\beta_{AR,j} + \beta_{RB,j}\right) + 1\right)} \right),\\
  &{\bar R}_{RA,i}^\text{DF} \triangleq \frac{1}{2}\log_2 \left(1 + \frac{p_r \frac{\tau_p E_p}{M^{\gamma - 1}} \beta_{AR,i}^4}{\left(p_r\beta_{AR,i} + 1\right)\sum\limits_{j = 1}^{N}\left(\beta_{AR,j}^2 + \beta_{RB,j}^2 \right) } \right),
\end{align}
and ${\bar R}_{BR,i}^\text{DF}$ and ${\bar R}_{RB,i}^\text{DF}$ are obtained by replacing the subscripts ``AR'', ``RB'' in ${\bar R}_{AR,i}^\text{DF}$ and ${\bar R}_{RA,i}^\text{DF}$ with the subscripts ``RB'', ``AR'', respectively.
\end{theorem}

Similar to the AF protocol, the large-scale approximation of the spectral efficiency ${R}_i^\text{DF}$ in Scenario A also depends on the choice of $\gamma$. When we cut down $p_p$ too much, i.e., $\gamma > 1$, ${R}_i^\text{DF}$ converges to zero. On the other hand, when $0<\gamma<1$, ${R}_i^\text{DF}$ grows unboundedly. Finally, when $\gamma = 1$, ${R}_{A,i}^\text{AF}$ converges to a non-zero limit. 



%
\subsubsection{Scenario B}
Next, we turn our attention to Scenario B and present the following result.
\begin{theorem}\label{theor:3:DF}
With the DF protocol, for fixed $p_p$, $E_u$, and $E_r$, when $p_u = \frac{E_u}{M^\alpha}$, $p_r = \frac{E_r}{M^\beta}$, with $\alpha \geq 0$, $\beta \geq 0$, as $M \rightarrow \infty$, we have
\begin{align}
  {R}_i^\text{DF} - {\text{min}}\left( {\bar R}_{1,i}^\text{DF}, {\bar R}_{2,i}^\text{DF}\right) \overset{M \rightarrow \infty}{\longrightarrow} 0,
\end{align}
where
\begin{align}
  &{\bar R}_{1,i}^\text{DF} = \frac{1}{2}\log_2 \left(1 + \frac{E_u}{M^{\alpha - 1}} \frac{\sigma_{AR,i}^4 + \sigma_{RB,i}^4}{\sigma_{AR,i}^2 + \sigma_{RB,i}^2}\right),\\
  &{\bar R}_{2,i}^\text{DF} = {\text{min}} \left( {\bar R}_{AR,i}^\text{DF}, {\bar R}_{RB,i}^\text{DF}\right) + {\text{min}} \left( {\bar R}_{BR,i}^\text{DF}, {\bar R}_{RA,i}^\text{DF}\right),
\end{align}
with
\begin{align}
  &{\bar R}_{AR,i}^\text{DF} = \frac{1}{2}\log_2 \left(1 + \frac{E_u}{M^{\alpha - 1}} \frac{\sigma_{AR,i}^4}{\sigma_{AR,i}^2 + \sigma_{RB,i}^2}\right),\\
  &{\bar R}_{RA,i}^\text{DF} = \frac{1}{2}\log_2 \left(1 + \frac{E_r}{M^{\beta - 1}} \frac{\sigma_{AR,i}^4}{\sum\limits_{j = 1}^{N}\left(\sigma_{AR,j}^2 + \sigma_{RB,j}^2\right)}\right),
\end{align}
and ${\bar R}_{BR,i}^\text{DF}$ and ${\bar R}_{RB,i}^\text{DF}$ are obtained by replacing the subscripts ``AR'', ``RB'' in ${\bar R}_{AR,i}^\text{DF}$ and ${\bar R}_{RA,i}^\text{DF}$ with the subscripts ``RB'', ``AR'', respectively.
\end{theorem}

Similar to the AF case, Theorem \ref{theor:3:DF} indicates that when both the transmit power of each user $p_u$ and the transmit power of the relay $p_r$ are scaled down inversely proportional to $M$ ($M \rightarrow \infty$), the effect of estimation error, residual self-interference, and inter-user interference vanishes, and the only remaining impairment comes from the noise at users and the relay. Moreover, when each user's transmit power is sufficiently large, i.e., $E_u \rightarrow \infty$, the large-scale approximation of ${R}_i^\text{DF}$ is determined only by ${\bar R}_{RA,i}^\text{DF}$ and ${\bar R}_{RB,i}^\text{DF}$, suggesting that the bottleneck of spectral efficiency appears in the second phase. In contrast, when the relay's transmit power becomes large, i.e., $E_r \rightarrow \infty$, the large-scale approximation of ${R}_i^\text{DF}$ is determined only by ${\bar R}_{1,i}^\text{DF}$, ${\bar R}_{AR,i}^\text{DF}$, ${\bar R}_{BR,i}^\text{DF}$, indicating that the bottleneck of spectral efficiency occurs in the first phase.

Also, as in the AF case, when we cut down the transmit powers of the relay and/or of each user too much, namely, 1) $\alpha > 1$, and $\beta \geq 0$, 2) $\alpha \geq 0$, and $\beta > 1$, 3) $\alpha > 1$, and $\beta > 1$, ${R}_i^\text{DF}$ converges to zero. On the contrary, when we cut down both the transmit powers of the relay and of each user moderately, i.e., $0\leq\alpha<1$ and $0\leq\beta<1$, ${R}_i^\text{DF}$ grows unboundedly. So the most important task is how to select the parameters $\alpha$ and $\beta$ to make ${R}_i^\text{DF}$ converge to a non-zero finite limit. We discuss this in the following corollaries.

\begin{corollary}\label{coro:3:DF}
  With the DF protocol, for fixed $p_p$, $E_u$, and $E_r$, when $\alpha = \beta = 1$, namely, $p_u = \frac{E_u}{M}$, $p_r = \frac{E_r}{M}$,  as $M \rightarrow \infty$, the spectral efficiency of the \emph{i}-th user pair has the limit
  \begin{align}
  {R}_i^\text{DF} \rightarrow {\text{min}}\left( {\bar R}_{1,i}^\text{DF}, {\bar R}_{2,i}^\text{DF}\right),
\end{align}
where
\begin{align}
  &{\bar R}_{1,i}^\text{DF} = \frac{1}{2}\log_2 \left(1 + \frac{E_u\left(\sigma_{AR,i}^4 + \sigma_{RB,i}^4\right)}{\sigma_{AR,i}^2 + \sigma_{RB,i}^2}\right),\\
  &{\bar R}_{2,i}^\text{DF} = {\text{min}} \left( {\bar R}_{AR,i}^\text{DF}, {\bar R}_{RB,i}^\text{DF}\right) + {\text{min}} \left( {\bar R}_{BR,i}^\text{DF}, {\bar R}_{RA,i}^\text{DF}\right),
\end{align}
with
\begin{align}
  &{\bar R}_{AR,i}^\text{DF} = \frac{1}{2}\log_2 \left(1 + \frac{E_u\sigma_{AR,i}^4}{\sigma_{AR,i}^2 + \sigma_{RB,i}^2}\right),\\
  &{\bar R}_{RA,i}^\text{DF} = \frac{1}{2}\log_2 \left(1 + \frac{E_r\sigma_{AR,i}^4}{\sum\limits_{j = 1}^{N}\left(\sigma_{AR,j}^2 + \sigma_{RB,j}^2\right)}\right),
\end{align}
and ${\bar R}_{BR,i}^\text{DF}$ and ${\bar R}_{RB,i}^\text{DF}$ are obtained by replacing the subscripts ``AR'', ``RB'' in ${\bar R}_{AR,i}^\text{DF}$ and ${\bar R}_{RA,i}^\text{DF}$ with the subscripts ``RB'', ``AR'', respectively.
\end{corollary}

Corollary \ref{coro:3:DF} reveals that when both the transmit powers of the relay and of each user are scaled down with the same speed, i.e., $1/M$, ${R}_i^\text{DF}$ converges to a non-zero limit. Moreover, this non-zero limit is an increasing function with respect to $E_u$ and $E_r$, while a decreasing function with respect to the number of user pairs $N$.
\begin{corollary}\label{coro:4:DF}
  With the DF protocol, for fixed $p_p$, $E_u$, and $E_r$, when $\alpha = 1$ and $0 \leq \beta < 1$, namely, $p_u = \frac{E_u}{M}$, $p_r = \frac{E_r}{M^\beta}$, as $M \rightarrow \infty$, the spectral efficiency of the \emph{i}-th user pair has the limit
  \begin{align}
    {R}_i^\text{DF} \rightarrow \frac{1}{2}\log_2 \left(1 + \frac{E_u\left(\sigma_{AR,i}^4 + \sigma_{RB,i}^4\right)}{\sigma_{AR,i}^2 + \sigma_{RB,i}^2}\right).
  \end{align}
\end{corollary}
\proof
When $\alpha = 1$ and $0 \leq \beta < 1$, we can easily show that ${\bar R}_{AR,i}^\text{DF} \ll {\bar R}_{RB,i}^\text{DF}$ and ${\bar R}_{BR,i}^\text{DF} \ll {\bar R}_{RA,i}^\text{DF}$, thus the large-scale approximation of the spectral efficiency ${R}_i^\text{DF}$ becomes
\begin{align}
  {\text{min}}\left(\frac{1}{2}\log_2 \left(1 + \frac{E_u\left(\sigma_{AR,i}^4 + \sigma_{RB,i}^4\right)}{\sigma_{AR,i}^2 + \sigma_{RB,i}^2}\right), \frac{1}{2}\log_2 \left(1 + \frac{E_u\sigma_{AR,i}^4}{\sigma_{AR,i}^2 + \sigma_{RB,i}^2}\right) + \frac{1}{2}\log_2 \left(1 + \frac{E_u\sigma_{RB,i}^4}{\sigma_{AR,i}^2 + \sigma_{RB,i}^2}\right)  \right).
\end{align}

Utilizing the fact $\log\left(1 + \frac{a_1 + a_2}{b}\right) < \log\left(1 + \frac{a_1}{b}\right) + \log\left(1 + \frac{a_2}{b}\right)$ for $a_1,a_2,b>0$, we arrive at the desired result.
\endproof
Corollary \ref{coro:4:DF} suggests that when we cut down the transmit power of each user too much, i.e., $0\leq\beta<\alpha=1$, the large-scale approximation of the spectral efficiency is determined by the performance in the first phase, i.e., ${\bar R}_{1,i}^\text{DF}$, which depends only on $E_u$, and is independent of $E_r$. This result is expected since when the transmit power of each user is much less than the transmit power of the relay, the bottleneck of spectral efficiency occurs in the first phase. On the other hand, when the transmit power of the relay is cut down more compared with that of each user, i.e., $0\leq\alpha<\beta=1$, the bottleneck of spectral efficiency appears in the second phase, thus ${R}_i^\text{DF}$ is determined by ${\bar R}_{RA,i}^\text{DF}$ and ${\bar R}_{RB,i}^\text{DF}$ as shown in the following corollary.

\begin{corollary}\label{coro:5:DF}
 With the DF protocol, for fixed $p_p$, $E_u$, and $E_r$, when $0 \leq \alpha < 1$ and $\beta = 1$, namely, $p_u = \frac{E_u}{M^\alpha}$, $p_r = \frac{E_r}{M}$, as $M \rightarrow \infty$, the spectral efficiency of the \emph{i}-th user pair has the limit
  \begin{align}
    {R}_i^\text{DF} \rightarrow \frac{1}{2}\log_2 \left(1 + \frac{E_r\sigma_{AR,i}^4}{\sum\limits_{j = 1}^{N}\left(\sigma_{AR,j}^2 + \sigma_{RB,j}^2\right)}\right) + \frac{1}{2}\log_2 \left(1 + \frac{E_r\sigma_{RB,i}^4}{\sum\limits_{j = 1}^{N}\left(\sigma_{AR,j}^2 + \sigma_{RB,j}^2\right)}\right).
  \end{align}
\end{corollary}

\subsubsection{Scenario C}
Finally, a corresponding power scaling law for Scenario C is obtained as follows.
\begin{theorem}\label{theor:4:DF}
  With the DF protocol, for fixed $E_u$, $E_r$, and $E_p$, when $p_u = \frac{E_u}{M^\alpha}$, $p_r = \frac{E_r}{M^\beta}$, and $p_p = \frac{E_p}{M^\gamma}$, with $\alpha \geq 0$, $\beta \geq 0$, and $\gamma > 0$, as $M \rightarrow \infty$, we have
\begin{align}\label{eq:Scenario:C:DF}
  {R}_i^\text{DF} - {\text{min}}\left( {\bar R}_{1,i}^\text{DF}, {\bar R}_{2,i}^\text{DF}\right) \overset{M \rightarrow \infty}{\longrightarrow} 0,
\end{align}
where
\begin{align}
  &{\bar R}_{1,i}^\text{DF} = \frac{1}{2}\log_2 \left(1 + \frac{\tau_p E_u E_p}{M^{\alpha + \gamma - 1}} \frac{\beta_{AR,i}^4 + \beta_{RB,i}^4}{\beta_{AR,i}^2 + \beta_{RB,i}^2}\right),\\
  &{\bar R}_{2,i}^\text{DF} = {\text{min}} \left( {\bar R}_{AR,i}^\text{DF}, {\bar R}_{RB,i}^\text{DF}\right) + {\text{min}} \left( {\bar R}_{BR,i}^\text{DF}, {\bar R}_{RA,i}^\text{DF}\right),
\end{align}
with
\begin{align}
  &{\bar R}_{AR,i}^\text{DF} = \frac{1}{2}\log_2 \left(1 + \frac{\tau_p E_u E_p}{M^{\alpha + \gamma - 1}} \frac{\beta_{AR,i}^4}{\beta_{AR,i}^2 + \beta_{RB,i}^2}\right),\\
  &{\bar R}_{RA,i}^\text{DF} = \frac{1}{2}\log_2 \left(1 + \frac{\tau_p E_r E_p}{M^{\beta +\gamma - 1}} \frac{\beta_{AR,i}^4}{\sum\limits_{j = 1}^{N}\left(\beta_{AR,j}^2 + \beta_{RB,j}^2\right)}\right),
\end{align}
and ${\bar R}_{BR,i}^\text{DF}$ and ${\bar R}_{RB,i}^\text{DF}$ are obtained by replacing the subscripts ``AR'', ``RB'' in ${\bar R}_{AR,i}^\text{DF}$ and ${\bar R}_{RA,i}^\text{DF}$ with the subscripts ``RB'', ``AR'', respectively.
\end{theorem}

As expected, the large-scale approximation of the spectral efficiency ${R}_i^\text{DF}$ depends on the relationship between $\alpha$, $\beta$, and $\gamma$. Moreover, the term $\alpha+\gamma$ determines the spectral efficiency in the first phase, while $\beta+\gamma$ determines the spectral efficiency in the second phase, as elaborated in the following corollaries.

\begin{corollary}\label{coro:6:DF}
  With the DF protocol, for fixed $E_u$, $E_r$, and $E_p$, when $\alpha = \beta > 0$ and $\alpha + \gamma = 1$, namely, $p_u = \frac{E_u}{M^\alpha}$, $p_r = \frac{E_r}{M^\beta}$, and $p_p = \frac{E_p}{M^\gamma}$, with $\gamma > 0$, as $M \rightarrow \infty$, the spectral efficiency of the \emph{i}-th user pair has the limit
\begin{align}
  {R}_i^\text{DF} \rightarrow {\text{min}}\left( {\bar R}_{1,i}^\text{DF}, {\bar R}_{2,i}^\text{DF}\right),
\end{align}
where
\begin{align}
  &{\bar R}_{1,i}^\text{DF} = \frac{1}{2}\log_2 \left(1 + \frac{\tau_p E_u E_p\left(\beta_{AR,i}^4 + \beta_{RB,i}^4\right)}{\beta_{AR,i}^2 + \beta_{RB,i}^2}\right),\\
  &{\bar R}_{2,i}^\text{DF} = {\text{min}} \left( {\bar R}_{AR,i}^\text{DF}, {\bar R}_{RB,i}^\text{DF}\right) + {\text{min}} \left( {\bar R}_{BR,i}^\text{DF}, {\bar R}_{RA,i}^\text{DF}\right),
\end{align}
with
\begin{align}
  &{\bar R}_{AR,i}^\text{DF} = \frac{1}{2}\log_2 \left(1 + \frac{\tau_p E_u E_p\beta_{AR,i}^4}{\beta_{AR,i}^2 + \beta_{RB,i}^2}\right),\\
  &{\bar R}_{RA,i}^\text{DF} = \frac{1}{2}\log_2 \left(1 + \frac{\tau_p E_r E_p\beta_{AR,i}^4}{\sum\limits_{j = 1}^{N}\left(\beta_{AR,j}^2 + \beta_{RB,j}^2\right)}\right),
\end{align}
and ${\bar R}_{BR,i}^\text{DF}$ and ${\bar R}_{RB,i}^\text{DF}$ are obtained by replacing the subscripts ``AR'', ``RB'' in ${\bar R}_{AR,i}^\text{DF}$ and ${\bar R}_{RA,i}^\text{DF}$ with the subscripts ``RB'', ``AR'', respectively.
\end{corollary}

Similar to the AF protocol, Corollary \ref{coro:6:DF} also presents an informative trade-off between the transmit powers of each pilot symbol and of each user/the relay. In other words, if we cut down the transmit power of each pilot symbol too much, which causes poor channel estimation accuracy, the transmit power of each user/the relay should be increased to compensate this imperfection and maintain the same asymptotic spectral efficiency.

\begin{corollary}\label{coro:7:DF}
  With the DF protocol, for fixed $E_u$, $E_r$, and $E_p$, when $\alpha > \beta \geq 0$ and $\alpha + \gamma = 1$, namely, $p_u = \frac{E_u}{M^\alpha}$, $p_r = \frac{E_r}{M^\beta}$, and $p_p = \frac{E_p}{M^\gamma}$, with $\gamma > 0$, as $M \rightarrow \infty$, the spectral efficiency of the \emph{i}-th user pair has the limit
 \begin{align}
     {R}_i^\text{DF} \ {\rightarrow}\ \frac{1}{2}\log_2 \left(1 + \frac{\tau_p E_u E_p\left(\beta_{AR,i}^4 + \beta_{RB,i}^4\right)}{\beta_{AR,i}^2 + \beta_{RB,i}^2}\right).
  \end{align}
\end{corollary}

From Corollary \ref{coro:7:DF}, we can see that the limit of ${R}_i^\text{DF}$ is an increasing function with respect to $E_u$ and $E_p$, indicating that we can boost the spectral efficiency by increasing the transmit power of each user and of each pilot symbol. In addition, similar to the AF protocol, the limit of ${R}_i^\text{DF}$ is also independent of $N$, indicating that the sum spectral efficiency of the system is an increasing function with respect to $N$.

\begin{corollary}\label{coro:8:DF}
  With the DF protocol, for fixed $E_u$, $E_r$, and $E_p$, when $0 \leq \alpha < \beta$ and $\beta + \gamma = 1$, namely, $p_u = \frac{E_u}{M^\alpha}$, $p_r = \frac{E_r}{M^\beta}$, and $p_p = \frac{E_p}{M^\gamma}$, with $\gamma > 0$, as $M \rightarrow \infty$, the spectral efficiency of the \emph{i}-th user pair has the limit
 \begin{align}
    {R}_i^\text{DF} \ {\rightarrow}\ \frac{1}{2}\log_2 \left(1 + \frac{\tau_p E_r E_p\beta_{AR,i}^4}{\sum\limits_{j = 1}^{N}\left(\beta_{AR,j}^2 + \beta_{RB,j}^2\right)}\right) + \frac{1}{2}\log_2 \left(1 + \frac{\tau_p E_r E_p\beta_{RB,i}^4}{\sum\limits_{j = 1}^{N}\left(\beta_{AR,j}^2 + \beta_{RB,j}^2\right)}\right).
  \end{align}
\end{corollary}

Corollary \ref{coro:8:DF} provides the trade-off between the transmit powers of the relay and of each pilot symbol, which is the same as for the AF protocol.
\subsection{Numerical Results}
In this subsection, we provide numerical simulation results to verify the power scaling laws presented in the previous subsections, and investigate the potential for power saving when employing large number of antennas at the relay. Note that the curves labelled as ``Approximations'' are obtained according to Theorems \ref{theor:1} and \ref{theor:1:DF} for the AF and DF protocols, respectively.
\begin{figure}[!ht]
    \centering
    \includegraphics[scale=0.6]{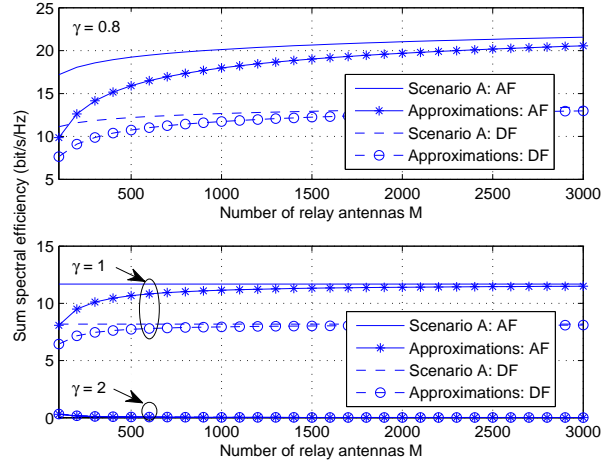}
    \caption{Sum spectral efficiency versus the number of relay antennas $M$ for $N = 5$, $p_u = 10$ dB, $p_r = 20$ dB, and $p_p = E_p/M^\gamma$ with $E_p = 10$ dB.}\label{fig:scenario_A}
  \end{figure}
\subsubsection{Scenario A}
Fig. \ref{fig:scenario_A} verifies the analytical results for Scenario A. The curves labelled as ``Scenario A: AF'', and ``Scenario A: DF'', are plotted according to Theorems \ref{theor:2} and \ref{theor:2:DF} for the AF and DF protocols, respectively. It can be readily observed that in the asymptotic large $M$ regime, the asymptotic curves converge to the exact curves, demonstrating the accuracy of the asymptotic analysis. In addition, when $\gamma > 1$, i.e., $\gamma = 2$, the spectral efficiency of both the AF and DF schemes gradually approaches zero. In contrast, when $0< \gamma < 1$, i.e., $\gamma = 0.8$, the spectral efficiency of both schemes grows unbounded. Finally, when $\gamma = 1$, the spectral efficiency converges to a non-zero limit for both schemes.
\subsubsection{Scenario B}
\begin{figure}[h!]
\centering
\subfigure[Case I and II]{
\label{fig:scenario_B_Case1_Case2}
\includegraphics[scale=0.55]{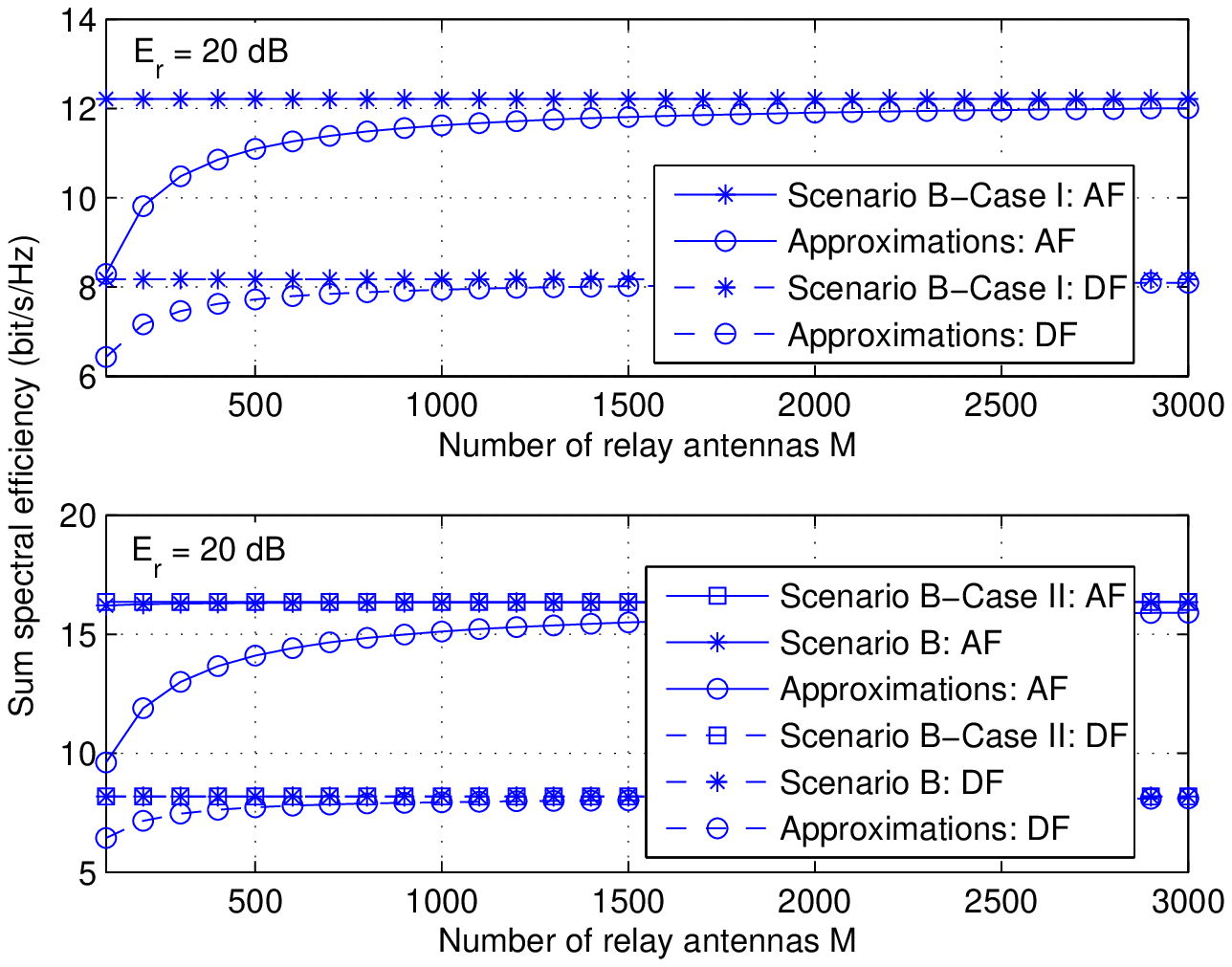}}
\subfigure[Case III]{
\label{fig:scenario_B_Case3_Er}
\includegraphics[scale=0.55]{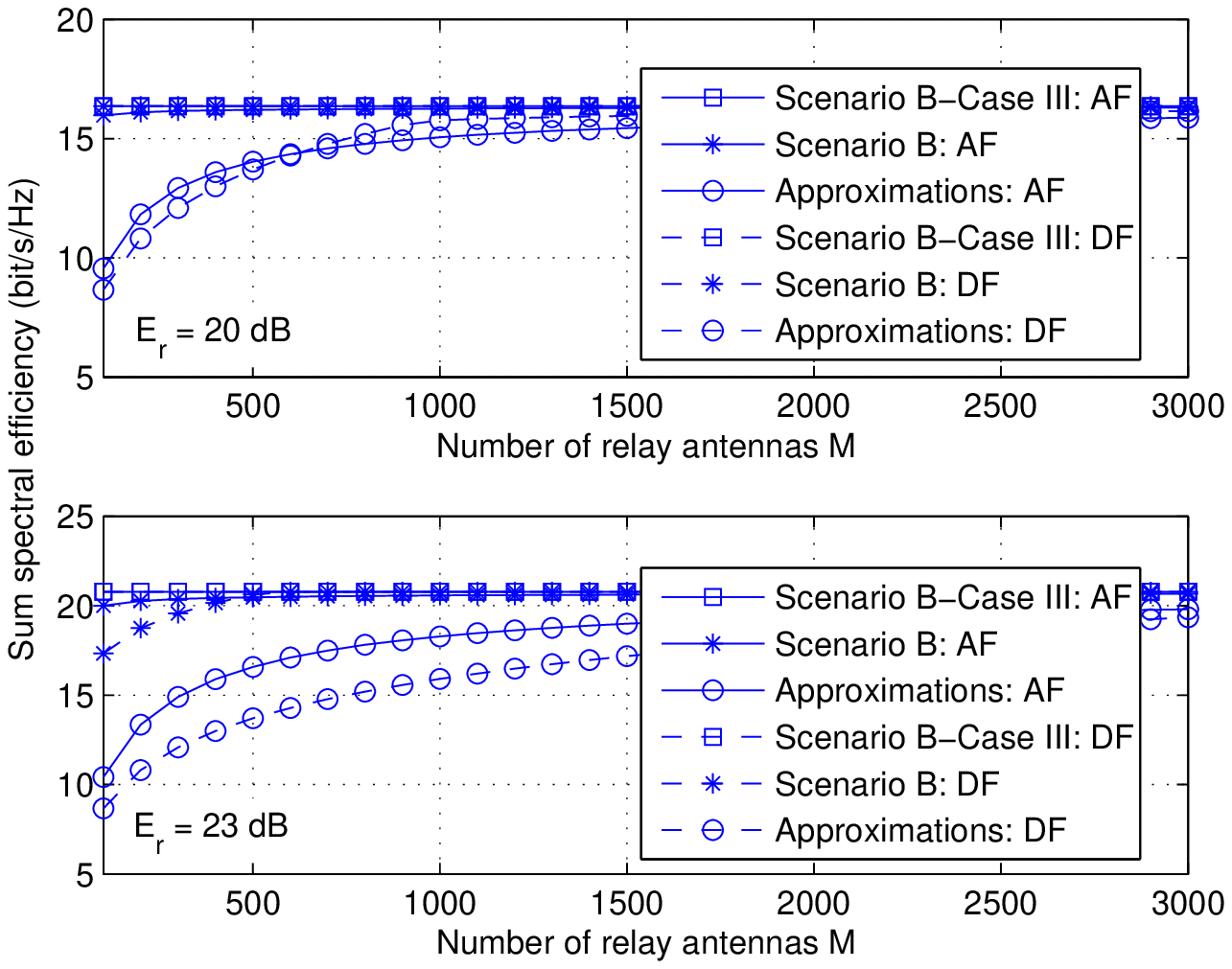}}
\caption{Sum spectral efficiency versus the number of relay antennas $M$ for $p_p = 10$ dB, $p_u = E_u/M^\alpha$ with $E_u = 10$ dB, and $p_r = E_r/M^\beta$.}
\label{Fig.lable}
\end{figure}

Fig. \ref{Fig.lable} investigates how the scaling of the transmit power of each user $p_u = \frac{E_u}{M^\alpha}$ and the transmit power of the relay $p_r = \frac{E_r}{M^\beta}$ affects the achievable spectral efficiency. In particular, three different cases are studied according to the values of $\alpha$ and $\beta$, as presented in TABLE I. Note that the curves labelled as ``Scenario B: AF'' and ``Scenario B: DF'', are generated by using Theorems \ref{theor:3} and \ref{theor:3:DF}, while the curves labelled as ``Scenario B-Case X: AF'' and ``Scenario B- Case X: DF'' with ${\text X} \in \left\{{\text I}, {\text {II}}, {\text {III}} \right\}$ are plotted according to Corollaries \ref{coro:3}--\ref{coro:5} and Corollaries \ref{coro:3:DF}--\ref{coro:5:DF}, for the AF and DF protocols, respectively.
\begin{table}[htbp]\label{table:scenario:B}
\caption{Power scaling cases in Scenario B}
\centering{}%
\begin{tabular}[c]{|c|c|}
\hline
Case & The values of $\alpha$ and $\beta$ \tabularnewline
\hline
I  & $\alpha = \beta = 1$\tabularnewline
\hline
II  & $\alpha = 1, \beta = 0.2$\tabularnewline
\hline
III  & $\alpha = 0.4, \beta = 1$\tabularnewline
\hline
\end{tabular}
\end{table}

In agreement with Corollaries \ref{coro:3}--\ref{coro:5} and Corollaries \ref{coro:3:DF}--\ref{coro:5:DF}, the sum spectral efficiency of both the AF and DF schemes saturates in the asymptotical large $M$ regime for all the three cases. As readily observed, among the three cases, the sum spectral efficiency of Case I, i.e., $\alpha = \beta = 1$, is the lowest, due to simultaneously cutting the transmit powers of each user and of the relay. Moreover, Fig. \ref{fig:scenario_B_Case3_Er} suggests that increasing the relay power improves the sum spectral efficiency. In addition, with sufficiently large number of relay antennas, i.e., $M > 800$, the DF protocol outperforms the AF protocol when $E_r = 20$ dB, while becomes inferior when $E_r = 23$ dB, which is in line with the results presented in Fig. \ref{fig:rate_pr}.

%

\begin{figure}[h!]
\centering
\subfigure[Zero limit spectral efficiency]{
\label{fig:scenario_B_Coverge_0}
\includegraphics[scale=0.55]{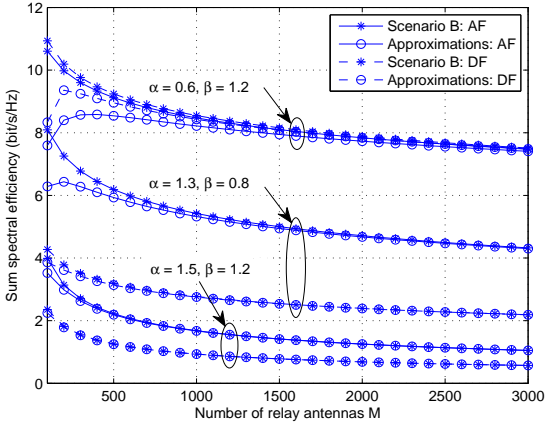}}
\subfigure[Unbounded spectral efficiency]{
\label{fig:scenario_B_Unbounded}
\includegraphics[scale=0.55]{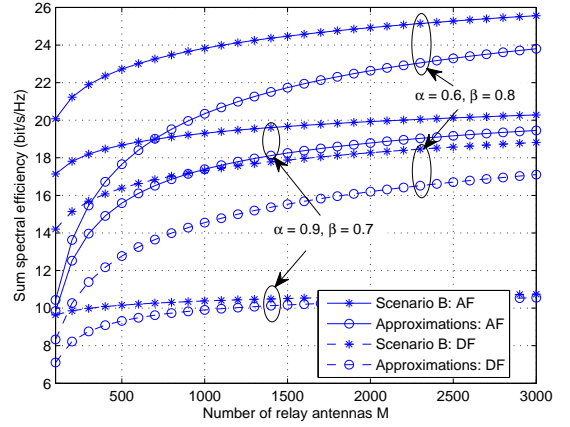}}
\caption{Sum spectral efficiency versus the number of relay antennas $M$ for $p_p = 10$ dB, $p_u = E_u/M^\alpha$ with $E_u = 10$ dB, and $p_r = E_r/M^\beta$ with $E_r = 20$ dB.}
\label{Fig.lable1}
\end{figure}

Fig. \ref{Fig.lable1} illustrates the other two extreme scenarios where the transmit power down-scaling is either too aggressive or too moderate. For the former scenario, three different cases are studied, i.e., $\alpha > 1, \beta \geq 0$, $\alpha \geq 0, \beta > 1$, and $\alpha > 1, \beta > 1$. As expected, when the number of relay antennas increases, the sum spectral efficiency gradually reduces to zero. However, the speed of reduction varies significantly depending on the scaling parameters. The larger the scaling parameters, the faster the decay rate. On the other hand, Fig. \ref{fig:scenario_B_Unbounded} shows that when we cut down the transmit powers of each user and of the relay moderately, the sum spectral efficiency of both the AF and DF schemes grows unboundedly.

%
\subsubsection{Scenario C}

\begin{figure}[h!]
\centering
\subfigure[Zero limit spectral efficiency]{
\label{fig:Scenario_C_Converge0_new}
\includegraphics[scale=0.55]{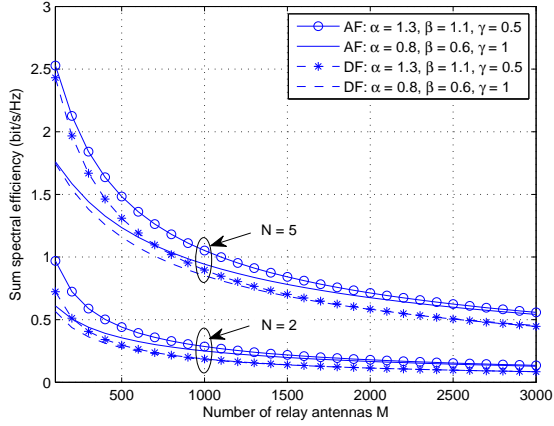}}
\subfigure[Unbounded spectral efficiency]{
\label{fig:Scenario_C_unbounded_new}
\includegraphics[scale=0.55]{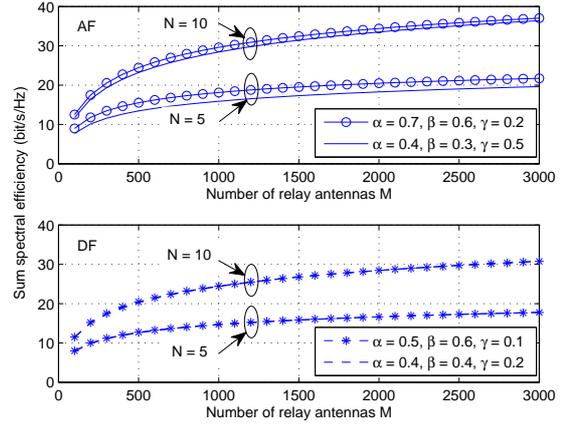}}
\caption{Sum spectral efficiency versus the number of relay antennas $M$ for $p_u = E_u/M^\alpha$ with $E_u = 10$ dB, $p_r = E_r/M^\beta$ with $E_r = 15$ dB, and $p_p = E_p/M^\gamma$ with $E_p = 0$ dB.}
\label{Fig.label3}
\end{figure}

Fig. \ref{Fig.label3} demonstrates the fundamental tradeoff between the user/relay power and the pilot symbol power. For illustration purposes, two extreme scenarios where the transmit power down-scaling is either too aggressive or too moderate are considered. For the former scenario, two sets of curves are drawn according to $\alpha = 1.3$, $\beta = 1.1$, $\gamma = 0.5$ and $\alpha = 0.8$, $\beta = 0.6$, $\gamma = 1$, which satisfy $\alpha + \gamma = 1.8$ and $\beta + \gamma = 1.6$. When the number of relay antennas grows large, the sum spectral efficiency of all system configurations smoothly converges to zero, as predicted. Moreover, the gaps between the two sets of curves reduce with $M$ and eventually vanish. This indicates that as long as $\alpha + \gamma$ and $\beta + \gamma$ are the same, the asymptotic sum spectral efficiency remains unchanged. Now, let us focus on the two curves associated with the AF protocol with $N=5$. Interestingly, we see that the curve associated with $\gamma=0.5$ yields better sum spectral efficiency in the finite antenna regime, despite the fact that the user or relay power is over-reduced compared to the $\gamma=1$ case, which suggests that it is of crucial importance to improve the channel estimation accuracy. The same behavior appears for the DF protocol as shown in Fig. \ref{fig:Scenario_C_Converge0_new}, and the unbounded spectral efficiency scenario as shown in Fig. \ref{fig:Scenario_C_unbounded_new}.


\section{Power Allocation}\label{section:5}
Power control is an effective means to enhance the sum spectral efficiency of the system. In this section, we formulate a power allocation problem maximizing the sum spectral efficiency of both the AF and DF protocols subject to a total power constraint, i.e., $\sum\limits_{i = 1}^N\left( p_{A,i} + p_{B,i} \right) + p_r \leq P$. For notational simplicity, we define ${\cal N} \triangleq \left\{ 1,\ldots,N \right\}$, ${\bf p}_A \triangleq \left[p_{A,1},\ldots,p_{A,N}\right]^T$, and ${\bf p}_B \triangleq \left[p_{B,1},\ldots,p_{B,N}\right]^T$.
\subsection{AF protocol}
For mathematical tractability, instead of using the exact sum spectral efficiency expression in Theorem \ref{theor:0}, we work with the large-scale approximation from Theorem \ref{theor:1} which has shown be tight for even moderate $M$. Thus, the power allocation optimization problem is formulated as
\begin{align}\label{problem1:AF}
\mathop{\text{maximize}}\limits_{{\bf p}_A, {\bf p}_B, p_r} \quad &\frac{\tau_c - \tau_p}{\tau_c} \sum\limits_{i = 1}^{N}  \left( {\tilde R}_{A,i}^\text{AF} + {\tilde R}_{B,i}^\text{AF} \right) \\
  {\text{subject to}} \quad  &\sum\limits_{i = 1}^N\left( p_{A,i} + p_{B,i} \right) + p_r \leq P,\\
  &{\bf p}_A \geq {\bf 0}, {\bf p}_B \geq {\bf 0},p_r \geq 0,
\end{align}
where ${\tilde R}_{B,i}^\text{AF}$ is the large-scale approximation for the spectral efficiency of ${\text T}_{B,i}$, which can be derived in a similar fashion.

Since $\log(\cdot)$ is an increasing function, \eqref{problem1:AF} can be equivalently reformulated as ${\cal{P}}^\text{AF}_1$:
\begin{align}\label{problem2:AF}
 \mathop{\text{minimize}}\limits_{{{\bf p}_A, {\bf p}_B, p_r}\atop{ {\bf\gamma}_{A}^\text{AF},{\bf\gamma}_{B}^\text{AF}}} \quad & \prod_{i = 1}^{N}  \left( 1 + \gamma_{A,i}^\text{AF} \right)^{-1} \left(1 + \gamma_{B,i}^\text{AF} \right)^{-1} \\
 {\text{subject to}}\quad  & \gamma_{A,i}^\text{AF} \leq \frac{p_{B,i}}{\sum\limits_{j = 1}^N\left(a_{i,j}^\text{AF} p_{A,j} + b_{i,j}^\text{AF} p_{B,j}\right) + p_r^{-1}\sum\limits_{j = 1}^N\left(c_{i,j}^\text{AF} p_{A,j} + d_{i,j}^\text{AF} p_{B,j}\right) + e_i},  i \in {\cal N}\\
 & \gamma_{B,i}^\text{AF} \leq \frac{p_{A,i}}{\sum\limits_{j = 1}^N\left({\tilde a}_{i,j}^\text{AF} p_{A,j} + {\tilde b}_{i,j}^\text{AF} p_{B,j}\right) + p_r^{-1}\sum\limits_{j = 1}^N\left({\tilde c}_{i,j}^\text{AF} p_{A,j} + {\tilde d}_{i,j}^\text{AF} p_{B,j}\right) + {\tilde e}_i}, i \in {\cal N}\\
  &\sum\limits_{i = 1}^N\left( p_{A,i} + p_{B,i} \right) + p_r \leq P,\\
  &{\bf p}_A \geq {\bf 0}, {\bf p}_B \geq {\bf 0},p_r \geq 0,
\end{align}
where ${\bf\gamma}_{A}^\text{AF} \triangleq \left[\gamma_{A,1}^{\text AF},\ldots, \gamma_{A,N}^{\text AF}\right]^T$, ${\bf\gamma}_{B}^\text{AF} \triangleq \left[\gamma_{B,1}^{\text AF},\ldots, \gamma_{B,N}^{\text AF}\right]^T$
$ a_{i,j}^\text{AF} = \frac{1}{M}
\begin{cases}
  \frac{4\beta_{AR,i}}{\sigma_{RB,i}^2}, & j = i,\\
  \frac{\beta_{AR,j}}{\sigma_{RB,i}^2} + \frac{\sigma_{AR,j}^4\sigma_{RB,j}^2\beta_{AR,i}}{\sigma_{AR,i}^4\sigma_{RB,i}^4}, & j \neq i,
\end{cases}$,
$ b_{i,j}^\text{AF} = \frac{1}{M}
\begin{cases}
 \frac{\beta_{RB,i}}{\sigma_{RB,i}^2} + \frac{\beta_{AR,i}}{\sigma_{AR,i}^2} , & j = i,\\
  \frac{\beta_{RB,j}}{\sigma_{RB,i}^2} + \frac{\sigma_{AR,j}^2\sigma_{RB,j}^4\beta_{AR,i}}{\sigma_{AR,i}^4\sigma_{RB,i}^4}, & j \neq i,
\end{cases}$,
$c_{i,j}^\text{AF} = \frac{\sigma_{AR,j}^4\sigma_{RB,j}^2}{M\sigma_{AR,i}^4\sigma_{RB,i}^4}$, $d_{i,j}^\text{AF} = \frac{\sigma_{AR,j}^2\sigma_{RB,j}^4}{M\sigma_{AR,i}^4\sigma_{RB,i}^4}$, $e_{i}^\text{AF} = \frac{1}{M\sigma_{RB,i}^2}$,
and ${\tilde a}_{i,j}^\text{AF}$, ${\tilde b}_{i,j}^\text{AF}$, ${\tilde c}_{i,j}^\text{AF}$, ${\tilde d}_{i,j}^\text{AF}$, ${\tilde e}_{i,j}^\text{AF}$ are obtained by replacing the subscripts ``AR'', ``RB'' with ``RB'', ``AR'' in ${b}_{i,j}^\text{AF}$, ${a}_{i,j}^\text{AF}$, ${d}_{i,j}^\text{AF}$, ${c}_{i,j}^\text{AF}$, ${e}_{i,j}^\text{AF}$, respectively.

The above problem ${\cal{P}}^\text{AF}_1$ is identified as a complementary geometric programming problem, which is nonconvex \cite{M.Avriel}. Also, $\gamma_{AR,i}^\text{AF}$ and $\gamma_{RB,i}^\text{AF}$ are considered as the signal-to-interference-plus-noise ratio (SINR) of ${\tilde R}_{A,i}^\text{AF}$ and ${\tilde R}_{B,i}^\text{AF}$, respectively. In addition, we have replaced the equality ``$=$'' with ``$\leq$'' in the first two constraints of problem ${\cal{P}}^\text{AF}_1$; however, this does not change or relax the original problem \eqref{problem1:AF}, since the objective function is decreasing with respect to $\gamma_{AR,i}^\text{AF}$ and $\gamma_{RB,i}^\text{AF}$. Therefore, we can guarantee that these two constraints must be active at any optimal solution of ${\cal{P}}^\text{AF}_1$.

It can be observed that all the inequality constraints can be transformed into posynomial functions, hence, if the objective function is also a posynomial function, the problem ${\cal{P}}^\text{AF}_1$ becomes a standard GP problem, and can be solved very efficiently with standard optimization tools such as CVX \cite{M.Grant} or ggplab \cite{A.Mutapcic}. However, the objective function is not a posynomial function. Fortunately, capitalizing on the technique proposed in \cite[Lemma 1]{P.C.Weeraddana}, we can find a close local optimum of the original problem ${\cal{P}}^\text{AF}_1$ by solving a sequence of GPs. The key idea is to use a monomial function $\omega_{X,i}^\text{AF}\left(\gamma_{X,i}^\text{AF}\right)^{\mu_{X,i}^\text{AF}}$ to approximate $1 + \gamma_{X,i}^\text{AF}$ near an arbitrary point ${\hat\gamma}_{X,i}^\text{AF} > 0$, where $X \in \left\{A,B\right\}$, ${\mu_{X,i}^\text{AF}} = \frac{{\hat\gamma}_{X,i}^\text{AF}}{1 + {\hat\gamma}_{X,i}^\text{AF}}$ and $\omega_{X,i}^\text{AF} = \left({\hat\gamma}_{X,i}^\text{AF}\right)^{-{\mu_{X,i}^\text{AF}}} \left(1 + {\hat\gamma}_{X,i}^\text{AF}\right)$. At each iteration, the GP is obtained by replacing the posynomial objective function with its best local monomial approximation near the solution obtained at the previous iteration. Now, we outline the successive approximation algorithm to solve the original problem ${\cal{P}}^\text{AF}_1$ in the following.
\begin{algorithm}[ht!]
  \caption{Successive approximation algorithm for ${\cal{P}}^\text{AF}_1$}\label{algor:AF}

    1) {\it Initialization}. Define a tolerance $\epsilon$ and parameter $\theta$. Set $k = 1$, the initial values of ${\hat\gamma}_{A,i}^\text{AF}$ and ${\hat\gamma}_{B,i}^\text{AF}$ are chosen according to the SINR in Theorem \ref{theor:1} when letting $p_{A,i} = p_{B,i} = \frac{P}{4N}$, and $p_r = \frac{P}{2}$.

    2) {\it Iteration $k$}. Compute ${\mu_{A,i}^\text{AF}} = \frac{{\hat\gamma}_{A,i}^\text{AF}}{1 + {\hat\gamma}_{A,i}^\text{AF}}$ and ${\mu_{B,i}^\text{AF}} = \frac{{\hat\gamma}_{B,i}^\text{AF}}{1 + {\hat\gamma}_{B,i}^\text{AF}}$. Then, solve the following GP problem ${\cal{P}}^\text{AF}_2$:
    \begin{align}\label{problem3:AF} \notag
  \mathop{\text{minimize}}\limits_{{{\bf p}_A, {\bf p}_B, p_r}\atop{ {\bf\gamma}_{A}^\text{AF},{\bf\gamma}_{B}^\text{AF}}} \quad & \prod_{i = 1}^{N}  \left(\gamma_{A,i}^\text{AF} \right)^{-{\mu_{A,i}^\text{AF}}} \left(\gamma_{B,i}^\text{AF} \right)^{-{\mu_{B,i}^\text{AF}}} \\ \notag
 {\text{subject to}} \quad &\theta^{-1} {\hat\gamma}_{A,i}^\text{AF} \leq {\gamma}_{A,i}^\text{AF} \leq \theta {\hat\gamma}_{A,i}^\text{AF}, i \in {\cal N}\\ \notag
 & \theta^{-1} {\hat\gamma}_{B,i}^\text{AF} \leq {\gamma}_{B,i}^\text{AF} \leq \theta {\hat\gamma}_{B,i}^\text{AF}, i \in {\cal N}\\ \notag
 & \gamma_{A,i}^\text{AF} {p_{B,i}^{-1}} \left({\sum\limits_{j = 1}^N\left(a_{i,j}^\text{AF} p_{A,j} + b_{i,j}^\text{AF} p_{B,j}\right) + p_r^{-1}\sum\limits_{j = 1}^N\left(c_{i,j}^\text{AF} p_{A,j} + d_{i,j}^\text{AF} p_{B,j}\right) + e_i}\right) \leq 1,  i \in {\cal N}\\ \notag
 & \gamma_{B,i}^\text{AF} {p_{A,i}^{-1}}\left({\sum\limits_{j = 1}^N\left({\tilde a}_{i,j}^\text{AF} p_{A,j} + {\tilde b}_{i,j}^\text{AF} p_{B,j}\right) + p_r^{-1}\sum\limits_{j = 1}^N\left({\tilde c}_{i,j}^\text{AF} p_{A,j} + {\tilde d}_{i,j}^\text{AF} p_{B,j}\right) + {\tilde e}_i}\right) \leq 1, i \in {\cal N}\\ \notag
  &\sum\limits_{i = 1}^N\left( p_{A,i} + p_{B,i} \right) + p_r \leq P,\\
    &{\bf p}_A \geq {\bf 0}, {\bf p}_B \geq {\bf 0},p_r \geq 0.
\end{align}

Denote the optimal solutions by ${\gamma}_{A,i}^{(k),\text{AF}}$ and ${\gamma}_{B,i}^{(k),\text{AF}}$, $i \in {\cal N}$.

    3) {\it Stopping criterion}. If $\max_i |{\gamma}_{A,i}^{(k),\text{AF}} - {\hat\gamma}_{A,i}^\text{AF}| < \epsilon$ and/or $\max_i |{\gamma}_{B,i}^{(k),\text{AF}} - {\hat\gamma}_{B,i}^\text{AF}| < \epsilon$, stop; otherwise, go to step 4).

    4) {\it Update initial values}. Set ${\hat\gamma}_{A,i}^\text{AF} = {\gamma}_{A,i}^{(k),\text{AF}}$ and ${\hat\gamma}_{B,i}^\text{AF} = {\gamma}_{B,i}^{(k),\text{AF}}$, and $k = k + 1$. Go to step 2).
\end{algorithm}

Please note, we have neglected $\omega_{A,i}^\text{AF}$ and $\omega_{B,i}^\text{AF}$ in the objective function of ${\cal{P}}^\text{AF}_2$, since they are constants and do not affect the problem solution. Also, some trust region constraints, i.e., the first two constraints, are added, which limit how much the variables are allowed to differ from the current guess ${\hat\gamma}_{A,i}^\text{AF}$ and ${\hat\gamma}_{B,i}^\text{AF}$. The convergence of the above algorithm to a Karush-Kuhn-Tucker solution of the original nonconvex problem ${\cal P}_1^{\text{AF}}$ is guaranteed, and the detailed proof can be found in \cite{B.R.Marks}. The parameter $\theta > 1$ controls the desired accuracy. More precisely, when $\theta$ is close to 1 it provides good accuracy for the monomial approximation but with slower convergence speed, and vice versa if $\theta$ is large. As discussed in \cite{P.C.Weeraddana,S.Boyd1,S.Boyd2}, $\theta = 1.1$ offers a good tradeoff between the accuracy and convergence speed.

To gain further insights, we now consider the special case when all the users transmit with the same power, i.e., $p_{A,i} = p_{B,i} = p_u$, then, the optimization problem \eqref{problem1:AF} reduces to
\begin{align}
  {\cal{P}}^\text{AF}_3:  \mathop\text{maximize}\limits_{p_u,p_r}  \quad & \frac{\tau_c - \tau_p}{\tau_c} \sum\limits_{i = 1}^{N}  \left( {\tilde R}_{A,i}^\text{AF} + {\tilde R}_{B,i}^\text{AF} \right) \\
  {\text{subject to}} \quad &2N p_u + p_r \leq P,\\
  &p_u\geq 0,p_r\geq0.
\end{align}
\begin{theorem}\label{theor:opt:AF:original}
${\cal{P}}^\text{AF}_3$ is a convex optimization problem.
\end{theorem}
\proof See Appendix \ref{app:theor:opt:AF:original}. \endproof

Since the optimization problem ${\cal{P}}^\text{AF}_3$ is convex, the optimal solutions $p_u^\text{AF,opt} \in \left(0,\frac{P}{2N}\right] $ and $p_r^\text{AF,opt} \in \left(0,P\right]$ maximizing the sum spectral efficiency can be solved efficiently by adopting some standard techniques, such as the bisection method with respect to $P$. However, it is difficult to directly obtain closed-form expressions of $p_u^\text{AF,opt}$ and $p_r^\text{AF,opt}$, since the objective function relies on the statistical characteristics of all the channel vectors. In order to simplify the analysis and provide some further insights, we assume that all the users have the same large-scale fading, e.g., $\beta_{AR,i} = \beta_{RB,i} = 1$, thereby resulting in $\sigma_{A,i}^2 = \sigma_{B,i}^2 = \sigma^2$, ${\tilde \sigma}_{A,i}^2 = {\tilde \sigma}_{B,i}^2 = {\tilde \sigma}^2$, ${\tilde R}_{A,i}^\text{AF} = {\tilde R}_{B,i}^\text{AF}$, and then the optimization problem ${\cal{P}}^\text{AF}_3$ reduces to
\begin{align}\label{eq:opt:AF:no_large}
  {\cal{P}}^\text{AF}_4: \mathop\text{maximize}\limits_{p_u,p_r} \quad  & \frac{2N\left(\tau_c - \tau_p\right)}{\tau_c} {\tilde R}_{A,i}^\text{AF} \\
  {\text{subject to}} \quad & 2N p_u + p_r \leq P,\\
  &p_u\geq 0,p_r\geq0.
\end{align}

The optimal solution of the optimization problem ${\cal{P}}^\text{AF}_4$ can be analytically characterized in the following theorem.
\begin{theorem}\label{theor:opt:AF}
With the AF protocol, the optimization problem ${\cal{P}}^\text{AF}_4$ is solved by
\begin{equation}
\begin{cases}
  p_u^{\text{AF,opt}} &= \frac{P}{4N},\\
  p_r^{\text{AF,opt}} &= \frac{P}{2}.
\end{cases}
\end{equation}
\end{theorem}
\proof See Appendix \ref{app:theor:opt:AF}. \endproof

Theorem \ref{theor:opt:AF} suggests that for a given power budget $2N p_u + p_r \leq P$, half of the total power should be allocated to the relay regardless of the number of users, and the remaining half should be equally allocated to the $2N$ users. Such a symmetric power allocation strategy is rather intuitive due to the symmetric system setup. In addition, it can be directly inferred that the optimal power $p_u^{\text{AF,opt}}$ decreases monotonically by increasing the number of user pairs $N$, which serves as a useful guideline for practical system design.
\subsection{DF protocol}
For the DF protocol, focusing on the large-scale approximation in Theorem \ref{theor:1:DF}, the power allocation optimization problem is formulated as
\begin{align}\label{problem1:DF}
\mathop{\text{maximize}}\limits_{{\bf p}_A, {\bf p}_B, p_r} \quad &\frac{\tau_c - \tau_p}{\tau_c} \sum\limits_{i = 1}^{N} {\tilde R}_{i}^\text{DF} \\
  {\text{subject to}} \quad  &\sum\limits_{i = 1}^N\left( p_{A,i} + p_{B,i} \right) + p_r \leq P,\\
  &{\bf p}_A \geq {\bf 0}, {\bf p}_B \geq {\bf 0},p_r \geq 0.
\end{align}

Then, following similar procedures as the AF protocol, \eqref{problem1:DF} can be equivalently reformulated as ${\cal{P}}^\text{DF}_1$:
\begin{align}\label{problem2:DF}
\mathop{\text{minimize}}\limits_{{{\bf p}_A, {\bf p}_B, p_r}\atop{{\bf\gamma}^\text{DF},{\bf\gamma}_{A}^\text{DF},{\bf\gamma}_{B}^\text{DF}}} \quad & \prod_{i = 1}^{N}  \left( 1 + \gamma_{i}^\text{DF} \right)^{-1} \\
 {\text{subject to}}\quad  & \gamma_{i}^\text{DF} \leq \frac{a_i^\text{DF} p_{A,i} + b_i^\text{DF} p_{B,i}}{\sum\limits_{j =1}^N\left(c_{i,j}^\text{DF} p_{A,i} + d_{i,j}^\text{DF} p_{B,i}\right) + 1}, i \in {\cal N} \\
 & \gamma_i^\text{DF} \leq \gamma_{A,i}^\text{DF} + \gamma_{B,i}^\text{DF} + \gamma_{A,i}^\text{DF}\gamma_{B,i}^\text{DF}, i \in {\cal N} \\
 & \gamma_{A,i}^\text{DF} \leq \frac{a_i^\text{DF} p_{A,i}}{\sum\limits_{j =1}^N\left(c_{i,j}^\text{DF} p_{A,i} + d_{i,j}^\text{DF} p_{B,i}\right) + 1}, i \in {\cal N} \\
 & \gamma_{A,i}^\text{DF} \leq \frac{p_r}{e_i^\text{DF} p_r + f_i^\text{DF}}, i \in {\cal N} \\
  & \gamma_{B,i}^\text{DF} \leq \frac{b_i^\text{DF} p_{B,i}}{\sum\limits_{j =1}^N\left(c_{i,j}^\text{DF} p_{A,i} + d_{i,j}^\text{DF} p_{B,i}\right) + 1}, i \in {\cal N} \\
 & \gamma_{B,i}^\text{DF} \leq \frac{p_r}{{\tilde e}_i^\text{DF} p_r + {\tilde f}_i^\text{DF}}, i \in {\cal N} \\
  &\sum\limits_{i = 1}^N\left( p_{A,i} + p_{B,i} \right) + p_r \leq P,\\
  &{\bf p}_A \geq {\bf 0}, {\bf p}_B \geq {\bf 0},p_r \geq 0.
\end{align}
where ${\bf\gamma}^\text{DF} \triangleq \left[\gamma_{1}^{\text DF},\ldots, \gamma_{N}^{\text DF}\right]^T$, ${\bf\gamma}_{A}^\text{DF} \triangleq \left[\gamma_{A,1}^{\text DF},\ldots, \gamma_{A,N}^{\text DF}\right]^T$, ${\bf\gamma}_{B}^\text{DF} \triangleq \left[\gamma_{B,1}^{\text DF},\ldots, \gamma_{B,N}^{\text DF}\right]^T$, $a_i^\text{DF} = \frac{M\sigma_{AR,i}^4 + \sigma_{AR,i}^2\sigma_{RB,i}^2}{\sigma_{AR,i}^2 + \sigma_{RB,i}^2}$, $b_i^\text{DF} = \frac{M\sigma_{RB,i}^4 + \sigma_{AR,i}^2\sigma_{RB,i}^2}{\sigma_{AR,i}^2 + \sigma_{RB,i}^2}$,
$ c_{i,j}^\text{AF} =
\begin{cases}
  {\tilde \sigma}_{AR,i}^2 , & j = i,\\
  \beta_{AR,j}, & j \neq i,
\end{cases}$, \\
$e_i^\text{DF} = \frac{\beta_{AR,i}}{M\sigma_{AR,i}^4}\sum\limits_{j = 1}^N\left( \sigma_{AR,j}^2 + \sigma_{RB,j}^2\right)$, $f_i^\text{DF} = \frac{1}{M\sigma_{AR,i}^4}\sum\limits_{j = 1}^N\left( \sigma_{AR,j}^2 + \sigma_{RB,j}^2\right)$,
and ${d}_{i,j}^\text{DF}$, ${\tilde e}_{i}^\text{DF}$, ${\tilde f}_{i}^\text{DF}$ are obtained by replacing the subscripts ``AR'', ``RB'' with ``RB'', ``AR'' in ${c}_{i,j}^\text{DF}$, ${e}_{i}^\text{DF}$, ${f}_{i}^\text{DF}$, respectively.

Since the objective function $1 + \gamma_{i}^\text{DF}$ can be approximated by a monomial function $\omega_{X,i}^\text{DF}\left(\gamma_{i}^\text{DF}\right)^{\mu_{i}^\text{DF}}$, where ${\mu_{i}^\text{DF}} = \frac{{\hat\gamma}_{i}^\text{DF}}{1 + {\hat\gamma}_{i}^\text{DF}}$ and $\omega_{i}^\text{DF} = \left({\hat\gamma}_{i}^\text{DF}\right)^{-{\mu_{i}^\text{DF}}} \left(1 + {\hat\gamma}_{i}^\text{DF}\right)$, the main challenge to convert ${\cal{P}}^\text{DF}_1$ into a GP problem is to transform the first two inequality constraints into the form of posynomials. According to \cite{S.Boyd1,C.He}, the geometric mean is no larger than the arithmetic mean for any set of positive numbers; thus we have
\begin{align}
  a_i^\text{DF} p_{A,i} + b_i^\text{DF} p_{B,i} \geq \left(\frac{a_i^\text{DF} p_{A,i}}{\nu_{A,i}^\text{DF}}\right)^{\nu_{A,i}^\text{DF}} \left(\frac{b_i^\text{DF} p_{B,i}}{\nu_{B,i}^\text{DF}}\right)^{\nu_{B,i}^\text{DF}},
\end{align}
where ${\nu_{A,i}^\text{DF}} = \frac{a_i^\text{DF} {\hat p}_{A,i}}{a_i^\text{DF} {\hat p}_{A,i} + b_i^\text{DF} {\hat p}_{B,i}}$, ${\nu_{B,i}^\text{DF}} = \frac{b_i^\text{DF} {\hat p}_{B,i}}{a_i^\text{DF} {\hat p}_{A,i} + b_i^\text{DF} {\hat p}_{B,i}}$, and ${\hat p}_{A,i}$, ${\hat p}_{B,i}$ are the initialization values.

As a result, the first inequality constraint in ${\cal{P}}^\text{DF}_1$ can be approximated by \cite{C.He}
\begin{align}
  \gamma_{i}^\text{DF} \leq \frac{\left(\frac{a_i^\text{DF} p_{A,i}}{\nu_{A,i}^\text{DF}}\right)^{\nu_{A,i}^\text{DF}} \left(\frac{b_i^\text{DF} p_{B,i}}{\nu_{B,i}^\text{DF}}\right)^{\nu_{B,i}^\text{DF}}}{\sum\limits_{j =1}^N\left(c_{i,j}^\text{DF} p_{A,i} + d_{i,j}^\text{DF} p_{B,i}\right) + 1}, i \in {\cal N}.
\end{align}

Now, we focus on the approximation of the second inequality constraint. Following the idea proposed in \cite[Lemma 1]{P.C.Weeraddana}, we use a monomial function $g(x,y) = \eta x^{\lambda_1}y^{\lambda_2}$ to approximate $f(x,y) = x + y + xy$ near an arbitrary point ${\hat x},{\hat y} > 0$. To make the approximation accurate, we need to ensure that
\begin{equation}
\begin{cases}
  {\hat x} + {\hat y} + {\hat x} {\hat y} = \eta {\hat x}^{\lambda_1}{\hat y}^{\lambda_2},\\
  1 + {\hat y} = \eta \lambda_1 {\hat x}^{\lambda_1 - 1}{\hat y}^{\lambda_2},\\
  1 + {\hat x} = \eta \lambda_2 {\hat x}^{\lambda_1}{\hat y}^{\lambda_2 - 1}.\\
\end{cases}
\end{equation}
As such, the parameters $\eta$, $\lambda_1$, and $\lambda_2$ can be computed as
\begin{equation}
\begin{cases}
  \lambda_1 = \frac{{\hat x}\left(1 + {\hat y} \right)}{{\hat x} + {\hat y} + {\hat x} {\hat y}},\\
\lambda_2 = \frac{{\hat y}\left(1 + {\hat x} \right)}{{\hat x} + {\hat y} + {\hat x} {\hat y}},\\
\eta = \left({\hat x} + {\hat y} + {\hat x} {\hat y}\right){\hat x}^{- \lambda_1} {\hat y}^{- \lambda_2}.
\end{cases}
\end{equation}
To this end, the second inequality constraint in ${\cal{P}}^\text{DF}_1$ can be approximated by
\begin{align}
  \gamma_i^\text{DF} \leq \eta_i^\text{DF} \left(\gamma_{A,i}^\text{DF}\right)^{\lambda_{A,i}^\text{DF}} \left(\gamma_{B,i}^\text{DF}\right)^{\lambda_{B,i}^\text{DF}}, i \in {\cal N},
\end{align}
where $\eta_{i}^\text{DF} = \left({\hat \gamma_{A,i}^\text{DF}} + {\hat \gamma_{B,i}^\text{DF}} + {\hat \gamma_{A,i}^\text{DF}} {\hat \gamma_{B,i}^\text{DF}}\right) \left( {\hat \gamma_{A,i}^\text{DF}}\right)^{- {\lambda_{A,i}^\text{DF}}} \left( {\hat \gamma_{B,i}^\text{DF}}\right)^{- {\lambda_{B,i}^\text{DF}}}$, ${\lambda_{A,i}^\text{DF}} = \frac{{\hat \gamma_{A,i}^\text{DF}}\left(1 + {\hat \gamma_{B,i}^\text{DF}} \right)}{{\hat \gamma_{A,i}^\text{DF}} + {\hat \gamma_{B,i}^\text{DF}} + {\hat \gamma_{A,i}^\text{DF}} {\hat \gamma_{B,i}^\text{DF}}}$,\\${\lambda_{B,i}^\text{DF}} = \frac{{\hat \gamma_{B,i}^\text{DF}}\left(1 + {\hat \gamma_{A,i}^\text{DF}} \right)}{{\hat \gamma_{A,i}^\text{DF}} + {\hat \gamma_{B,i}^\text{DF}} + {\hat \gamma_{A,i}^\text{DF}} {\hat \gamma_{B,i}^\text{DF}}}$, and ${\hat \gamma_{A,i}^\text{DF}}$, ${\hat \gamma_{B,i}^\text{DF}}$ are the initialization values.

We now outline the successive approximation algorithm to solve the original problem ${\cal{P}}^\text{DF}_1$ as follows.

\begin{algorithm}[ht!]
  \caption{Successive approximation algorithm for ${\cal{P}}^\text{DF}_1$}\label{algor:DF}

    1) {\it Initialization}. Define a tolerance $\epsilon$ and parameter $\theta$. Set $k = 1$, the initial values of ${\hat\gamma}_{i}^\text{DF}$, ${\hat\gamma}_{A,i}^\text{DF}$ and ${\hat\gamma}_{B,i}^\text{DF}$ are chosen according to the SINR in Theorem \ref{theor:1:DF}. Also, we set ${\hat p}_{A,i} = {\hat p}_{B,i} = \frac{P}{4N}$,

    2) {\it Iteration $k$}. Compute ${\mu_{i}^\text{DF}} = \frac{{\hat\gamma}_{i}^\text{DF}}{1 + {\hat\gamma}_{i}^\text{DF}}$, ${\nu_{A,i}^\text{DF}} = \frac{a_i^\text{DF} {\hat p}_{A,i}}{a_i^\text{DF} {\hat p}_{A,i} + b_i^\text{DF} {\hat p}_{B,i}}$, ${\nu_{B,i}^\text{DF}} = \frac{b_i^\text{DF} {\hat p}_{B,i}}{a_i^\text{DF} {\hat p}_{A,i} + b_i^\text{DF} {\hat p}_{B,i}}$,
    $\eta_{i}^\text{DF} = \left({\hat \gamma_{A,i}^\text{DF}} + {\hat \gamma_{B,i}^\text{DF}} + {\hat \gamma_{A,i}^\text{DF}} {\hat \gamma_{B,i}^\text{DF}}\right) \left( {\hat \gamma_{A,i}^\text{DF}}\right)^{- {\lambda_{A,i}^\text{DF}}} \left( {\hat \gamma_{B,i}^\text{DF}}\right)^{- {\lambda_{B,i}^\text{DF}}}$, ${\lambda_{A,i}^\text{DF}} = \frac{{\hat \gamma_{A,i}^\text{DF}}\left(1 + {\hat \gamma_{B,i}^\text{DF}} \right)}{{\hat \gamma_{A,i}^\text{DF}} + {\hat \gamma_{B,i}^\text{DF}} + {\hat \gamma_{A,i}^\text{DF}} {\hat \gamma_{B,i}^\text{DF}}}$, ${\lambda_{B,i}^\text{DF}} = \frac{{\hat \gamma_{B,i}^\text{DF}}\left(1 + {\hat \gamma_{A,i}^\text{DF}} \right)}{{\hat \gamma_{A,i}^\text{DF}} + {\hat \gamma_{B,i}^\text{DF}} + {\hat \gamma_{A,i}^\text{DF}} {\hat \gamma_{B,i}^\text{DF}}}$.
    Then, solve the following GP problem ${\cal{P}}^\text{DF}_2$:
    \begin{align}\label{problem3:DF} \notag
  \mathop{\text{minimize}}\limits_{{{\bf p}_A, {\bf p}_B, p_r}\atop{{\bf\gamma}^\text{DF},{\bf\gamma}_{A}^\text{DF},{\bf\gamma}_{B}^\text{DF}}} \quad & \prod_{i = 1}^{N}  \left(\gamma_{i}^\text{DF} \right)^{- {\mu_{i}^\text{DF}}} \\ \notag
 {\text{subject to}}\quad &\theta^{-1} {\hat p}_{A,i}^\text{DF} \leq {p}_{A,i}^\text{DF} \leq \theta {\hat p}_{A,i}^\text{DF},i \in {\cal N}\\ \notag
 &\theta^{-1} {\hat p}_{B,i}^\text{DF} \leq {p}_{B,i}^\text{DF} \leq \theta {\hat p}_{B,i}^\text{DF}, i \in {\cal N}\\ \notag
  &\theta^{-1} {\hat\gamma}_{i}^\text{DF} \leq {\gamma}_{i}^\text{DF} \leq \theta {\hat\gamma}_{i}^\text{DF}, i \in {\cal N}\\ \notag
 &\theta^{-1} {\hat\gamma}_{A,i}^\text{DF} \leq {\gamma}_{A,i}^\text{DF} \leq \theta {\hat\gamma}_{A,i}^\text{DF}, i \in {\cal N}\\ \notag
 &\theta^{-1} {\hat\gamma}_{B,i}^\text{DF} \leq {\gamma}_{B,i}^\text{DF} \leq \theta {\hat\gamma}_{B,i}^\text{DF}, i \in {\cal N}\\ \notag
 & \gamma_{i}^\text{DF}  {\left(\frac{a_i^\text{DF} p_{A,i}}{\nu_{A,i}^\text{DF}}\right)^{-\nu_{A,i}^\text{DF}} \left(\frac{b_i^\text{DF} p_{B,i}}{\nu_{B,i}^\text{DF}}\right)^{-\nu_{B,i}^\text{DF}}} \left( {\sum\limits_{j =1}^N\left(c_{i,j}^\text{DF} p_{A,i} + d_{i,j}^\text{DF} p_{B,i}\right) + 1} \right) \leq 1, i \in {\cal N},\\ \notag
 &   \gamma_i^\text{DF} \left( \eta_i^\text{DF}\right)^{-1} \left(\gamma_{A,i}^\text{DF}\right)^{-\lambda_{A,i}^\text{DF}} \left(\gamma_{B,i}^\text{DF}\right)^{-\lambda_{B,i}^\text{DF}} \leq 1, i \in {\cal N},\\ \notag
 & \gamma_{A,i}^\text{DF} {\left(a_i^\text{DF}\right)^{-1}p_{A,i}^{-1}}\left({\sum\limits_{j =1}^N\left(c_{i,j}^\text{DF} p_{A,i} + d_{i,j}^\text{DF} p_{B,i}\right) + 1}\right) \leq 1, i \in {\cal N} \\ \notag
 & \gamma_{A,i}^\text{DF}  {p_r^{-1}}\left({e_i^\text{DF} p_r + f_i^\text{DF}}\right) \leq 1, i \in {\cal N} \\ \notag
  & \gamma_{B,i}^\text{DF} {\left(b_i^\text{DF}\right)^{-1} p_{B,i}^{-1}}\left({\sum\limits_{j =1}^N\left(c_{i,j}^\text{DF} p_{A,i} + d_{i,j}^\text{DF} p_{B,i}\right) + 1}\right) \leq 1, i \in {\cal N} \\ \notag
 & \gamma_{B,i}^\text{DF} {p_r^{-1}}\left({{\tilde e}_i^\text{DF} p_r + {\tilde f}_i^\text{DF}}\right) \leq 1, i \in {\cal N} \\ \notag
  &\sum\limits_{i = 1}^N\left( p_{A,i} + p_{B,i} \right) + p_r \leq P,\\
  &{\bf p}_A \geq {\bf 0}, {\bf p}_B \geq {\bf 0},p_r \geq 0.
\end{align}

Denote the optimal solutions by ${p}_{A,i}^{(k),\text{DF}}$, ${p}_{B,i}^{(k),\text{DF}}$, ${\gamma}_{i}^{(k),\text{DF}}$, ${\gamma}_{A,i}^{(k),\text{DF}}$, ${\gamma}_{B,i}^{(k),\text{DF}}$, $i \in {\cal N}$.

    3) {\it Stopping criterion}. If $\max_i |{p}_{A,i}^{(k),\text{DF}} - {\hat p}_{A,i}^\text{DF}| < \epsilon$ and/or $\max_i |{p}_{B,i}^{(k),\text{DF}} - {\hat p}_{B,i}^\text{DF}| < \epsilon$ and/or $\max_i |{\gamma}_{i}^{(k),\text{DF}} - {\hat\gamma}_{i}^\text{DF}| < \epsilon$ and/or $\max_i |{\gamma}_{A,i}^{(k),\text{DF}} - {\hat\gamma}_{A,i}^\text{DF}| < \epsilon$ and/or $\max_i |{\gamma}_{B,i}^{(k),\text{DF}} - {\hat\gamma}_{B,i}^\text{DF}| < \epsilon$, stop; otherwise, go to step 4).

    4) {\it Update initial values}. Set ${\hat p}_{A,i}^\text{DF} = {p}_{A,i}^{(k),\text{DF}}$, ${\hat p}_{B,i}^\text{DF} = {p}_{B,i}^{(k),\text{DF}}$, ${\hat\gamma}_{i}^\text{DF} = {\gamma}_{i}^{(k),\text{DF}}$,  ${\hat\gamma}_{A,i}^\text{DF} = {\gamma}_{A,i}^{(k),\text{DF}}$, ${\hat\gamma}_{B,i}^\text{DF} = {\gamma}_{B,i}^{(k),\text{DF}}$, and $k = k + 1$. Go to step 2).
\end{algorithm}

Similar to algorithms for the AF protocol, we have removed $\omega_{i}^\text{DF}$ in the objective function, since they do not affect the optimization problem. Also, five extra inequalities as trust region constraints are included.

Algorithm \ref{algor:DF} focuses on the case where each user transmits with a different power, and yields a close local optimum of the original problem ${\cal{P}}^\text{DF}_1$ by solving a sequence of GPs. Now, we turn our attention to the scenario where all the users transmit with the same power, i.e., $p_{A,i} = p_{B,i} = p_u$; hence, the problem \eqref{problem1:DF} reduces to the following special case:
\begin{align}
  {\cal{P}}^\text{DF}_3: \mathop{\text{maximize}}\limits_{p_u,p_r} \quad  &\frac{\tau_c - \tau_p}{\tau_c} \sum\limits_{i = 1}^{N}  {\tilde R}_{i}^\text{DF} \\
  {\text{subject to}} \quad & 2N p_u + p_r \leq P,\\
  &p_u \geq0, p_r \geq 0.
\end{align}

\begin{theorem}\label{theor:opt:DF:original}
${\cal{P}}^\text{DF}_3$ is a convex optimization problem.
\end{theorem}
\proof See Appendix \ref{app:theor:opt:DF:original}. \endproof

Similar to the AF protocol, the optimal solutions $p_u^\text{DF,opt} \in \left(0,\frac{P}{2N}\right] $ and $p_r^\text{DF,opt} \in \left(0,P\right]$ maximizing the sum spectral efficiency can be solved efficiently by adopting some standard techniques, such as the bisection method with respect to $P$, due to the convexity of the optimization problem ${\cal{P}}^\text{DF}_3$.

\subsection{Numerical Results}

\begin{figure}[!ht]
    \centering
    \includegraphics[scale=0.6]{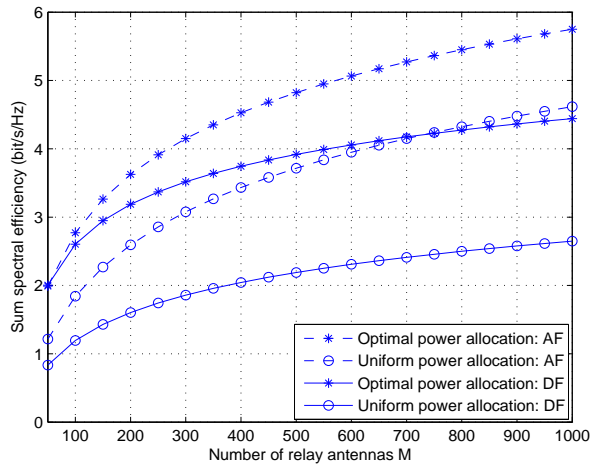}
    \caption{Sum spectral efficiency versus the number of relay antennas $M$ for $P = 10$ dB, $p_p = 10$ dB and $N = 5$.}\label{fig:power_allocation_rate_M}
  \end{figure}

Fig. \ref{fig:power_allocation_rate_M} illustrates the impact of the optimal power allocation scheme on the sum spectral efficiency when all users' large-scale fading are different. The different large-scale fading parameters are arbitrarily generated by $\beta_{AR,i} = z_i \left(r_{AR,i}/r_0\right)^\alpha$ and $\beta_{RB,i} = z_i \left(r_{BR,i}/r_0\right)^\alpha$, where $z_i$ is a log-normal random variable with standard deviation $8$ dB, $r_{AR,i}$ and $r_{RB,i}$ are the locations of ${\text T}_{AR,i}$ and ${\text T}_{RB,i}$ from the relay, $\alpha = 3.8$ is the path loss exponent, and $r_0$ denotes the guard interval which specifies the nearest distance between the users and the relay. The relay is located at the center of a cell with a radius of $1000$ meters and $r_0 = 100$ meters. We choose $P = 10$ dB, $p_p = 10$ dB, $N = 5$, $\beta_{AR} = \left[0.2688, 0.0368, 0.00025, 0.1398, 0.0047 \right]$, and $\beta_{RB} = \left[0.0003, 0.00025, 0.0050, 0.0794, 0.0001 \right]$. The optimal power allocation curves are generated by Algorithm \ref{algor:AF} and Algorithm \ref{algor:DF}. As a benchmark scheme for comparison, we also plot the sum spectral efficiency with uniform power allocation, i.e., the relay transmit power equals to the sum user transmit power. Please note that such a uniform power allocation scheme is optimal for the AF protocol when all the users have the same large-scale fading parameters, as shown in Theorem \ref{theor:opt:AF}. As can be observed, the optimal power allocation policy provides 34.8\% and 89.2\% spectral efficiency enhancement when $M = 300$, for the AF and DF protocols, respectively. This suggests that the improvement of the DF protocol is more prominent.

\begin{figure}[h!]
\centering
\subfigure[AF and DF]{
\label{fig:AF_DF}
\includegraphics[scale=0.55]{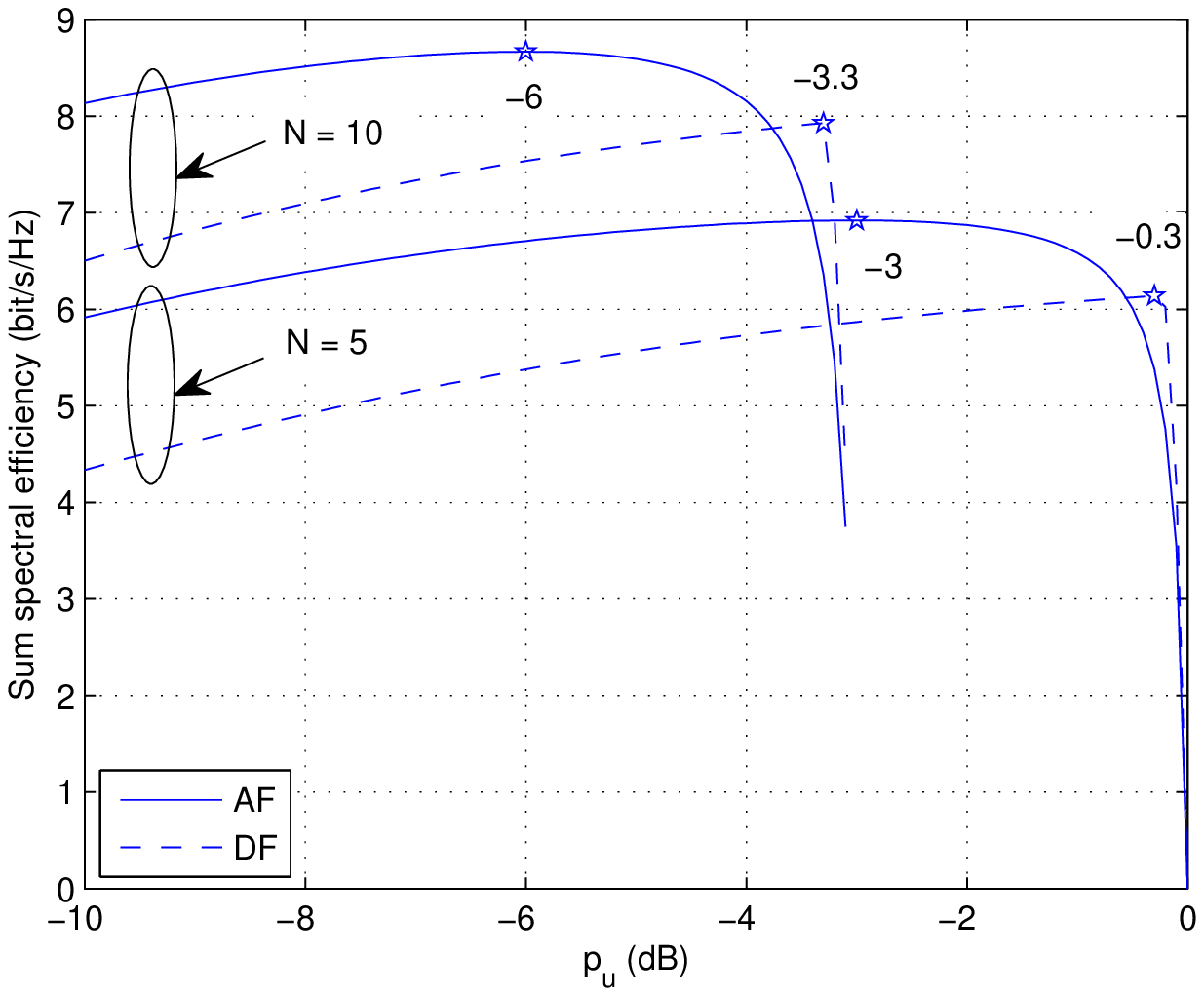}}
\subfigure[DF]{
\label{fig:DF}
\includegraphics[scale=0.55]{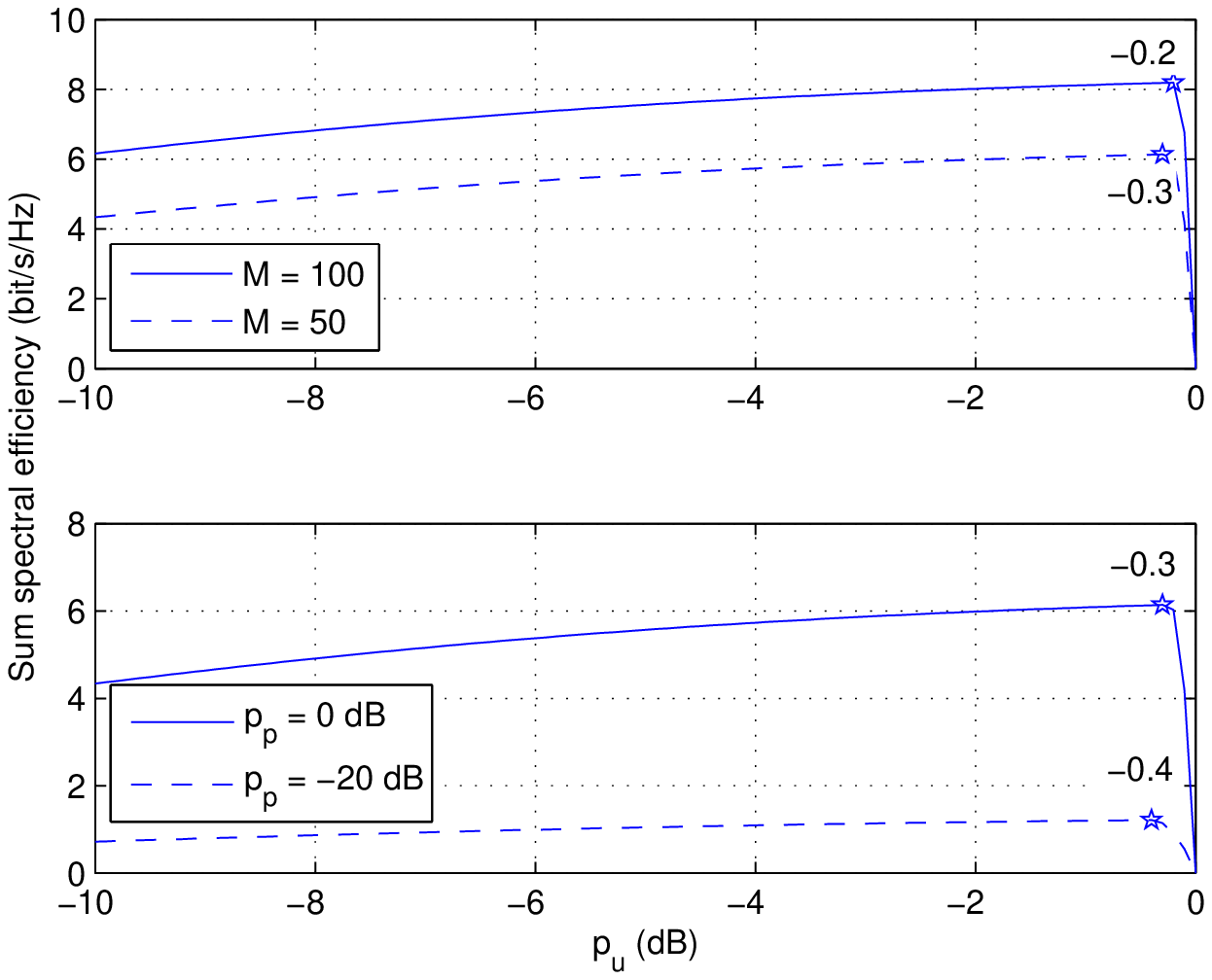}}
\caption{Sum spectral efficiency versus $p_u$ for $P = 10$ dB ($p_p = 0$ dB, $M = 50$ and $N = 5$ unless otherwise specified).}
\label{fig:power_allocation}
\end{figure}

 Fig. \ref{fig:power_allocation} examines the impact of key system parameters such as $M$, $N$, and $p_p$ on the optimal power allocation scheme when every user has the same transmit power, i.e, $p_{A,i} = p_{B,i} = p_u$. As expected, for a given power budget $P = 10$ dB, the optimal user transmit power is a decreasing function with respect to the number of user pairs $N$ for both the AF and DF protocols, as shown in Fig. \ref{fig:AF_DF}. Then, we focus on the DF protocol and depict the outcomes in Fig. \ref{fig:DF}. As can be seen, with fixed number of user pairs $N = 5$, the optimal user transmit power for the DF protocol $p_u^\text{DF,opt}$ with $M = 100$ is larger than that with $M = 50$, suggesting that we should increase the transmit power of each user when the number of relay antennas is large. In addition, when the pilot training power increases, i.e., from $p_p = -20$ dB to $p_p = 0$ dB, the optimal user transmit power $p_u^\text{DF,opt}$ also increases, indicating that when the channel estimation accuracy is improved, we need to use a higher transmit power for each user.
\section{Conclusion}\label{section:6}
  We have investigated the sum spectral efficiency of a multipair two-way half-duplex relaying system employing the MR scheme taking into account of realistic CSI assumption. In particular, an exact expression was derived for the sum spectral efficiency of the AF protocol and closed-form large-scale approximations of the spectral efficiency for both AF and DF protocols were obtained when the number of relay antennas is large. Based on these expressions, a detailed comparison between the AF and DF protocols was conducted, and it was shown that the spectral efficiency of the AF protocol is more sensitive in scenarios with low transmit power of the relay and pilot symbol, small number of relay antennas, and large number of user pairs. Moreover, power scaling laws of the system were characterized, which showed that the transmit powers of user, relay, and pilot symbol can be scaled down inversely proportional to the number of relay antennas. In addition, it was revealed that there exists a fundamental tradeoff between the user/relay transmit power and pilot symbol power, which provides great flexibility for the design of practical systems. Finally, the transmit powers of each user and the relay were optimized to enhance the sum spectral efficiency.

\appendices
\section{Proof of Theorem \ref{theor:0}}\label{app:theor:0}
The end-to-end SINR given in \eqref{eq:rate:R} consists of five terms: 1) desired signal power $A_i^\text{AF}$; 2) estimation error $B_i^\text{AF}$; 3) residual self-interference $C_i^\text{AF}$; 4) inter-user interference $D_i^\text{AF}$; 5) compound noise $E_i^\text{AF}$.

We start by first rewriting $\bf F$ as
\begin{align}\label{eq:F:vector}
  {\bf F} = \sum\limits_{n=1}^{N} \left( {\hat{\bf g}}_{RB,n}^* {\hat{\bf g}}_{AR,n}^H + {\hat{\bf g}}_{AR,n}^* {\hat{\bf g}}_{RB,n}^H \right).
\end{align}

After substituting \eqref{eq:F:vector} into \eqref{eq:rate:R}, we compute the five terms one by one.

1) Compute $A_i^\text{AF}$:
\begin{align}
  {\tt E} \left\{{{\bf g}}_{AR,i}^{T} {\bf F} {{\bf g}}_{RB,i} \right\} &= {\tt E} \left\{ \sum\limits_{n=1}^{N} \left( {{\bf g}}_{AR,i}^{T} {\hat{\bf g}}_{RB,n}^* {\hat{\bf g}}_{AR,n}^H {{\bf g}}_{RB,i} + {{\bf g}}_{AR,i}^{T} {\hat{\bf g}}_{AR,n}^* {\hat{\bf g}}_{RB,n}^H {{\bf g}}_{RB,i} \right) \right\},\\ \notag
  &= {\tt E} \left\{ \left( {{\bf g}}_{AR,i}^{T} {\hat{\bf g}}_{RB,i}^* {\hat{\bf g}}_{AR,i}^H {{\bf g}}_{RB,i} + {{\bf g}}_{AR,i}^{T} {\hat{\bf g}}_{AR,i}^* {\hat{\bf g}}_{RB,i}^H {{\bf g}}_{RB,i} \right) \right\},\\ \notag
  &= {\tt E} \left\{|{\hat{\bf g}}_{AR,i}^H {\hat{\bf g}}_{RB,i}|^2 + ||{\hat{\bf g}}_{AR,i}||^2||{\hat{\bf g}}_{RB,i}||^2 \right\},\\ \notag
  &= M\sigma_{AR,i}^2\sigma_{RB,i}^2 + M^2 \sigma_{AR,i}^2\sigma_{RB,i}^2.
\end{align}

Consequently, we obtain
\begin{align}\label{eq:derm:A}
  A_i^\text{AF} = p_{B,i} M^2(M + 1)^2 \sigma_{AR,i}^4 \sigma_{RB,i}^4.
\end{align}
  2) Compute $B_i^\text{AF}$:
\begin{align}\label{eq:B:AF}
  &{\tt E} \left\{ | {{\bf g}}_{AR,i}^{T} {\bf F} {{\bf g}}_{RB,i} |^2 \right\} \\ \notag
   &= {\tt E} \left\{ \sum\limits_{n=1}^{N} \sum\limits_{l=1}^{N} {{\bf g}}_{AR,i}^{ T} \left({\hat{\bf g}}_{RB,n}^* {\hat{\bf g}}_{AR,n}^H  + {\hat{\bf g}}_{AR,n}^* {\hat{\bf g}}_{RB,n}^H \right) {{\bf g}}_{RB,i} {{\bf g}}_{RB,i}^{H} \left( {\hat{\bf g}}_{AR,l} {\hat{\bf g}}_{RB,l}^T + {\hat{\bf g}}_{RB,l} {\hat{\bf g}}_{AR,l}^T \right) {{\bf g}}_{AR,i}^* \right\}, \notag
\end{align}
which can be decomposed into three different cases:

a) for $n \neq l \neq i$, we have ${\tt E} \left\{ | {{\bf g}}_{AR,i}^{T} {\bf F} {{\bf g}}_{RB,i} |^2 \right\} = 0$.

b) for $n = l \neq i$, we have
\begin{align}
  &{\tt E} \left\{ | {{\bf g}}_{AR,i}^{T} {\bf F} {{\bf g}}_{RB,i} |^2 \right\} \\ \notag
  &= {\tt E} \left\{ \sum\limits_{n=1}^{N} {{\bf g}}_{AR,i}^{ T} \left({\hat{\bf g}}_{RB,n}^* {\hat{\bf g}}_{AR,n}^H  + {\hat{\bf g}}_{AR,n}^* {\hat{\bf g}}_{RB,n}^H \right) {{\bf g}}_{RB,i} {{\bf g}}_{RB,i}^{H} \left( {\hat{\bf g}}_{AR,n} {\hat{\bf g}}_{RB,n}^T + {\hat{\bf g}}_{RB,n} {\hat{\bf g}}_{AR,n}^T \right) {{\bf g}}_{AR,i}^* \right\}.
\end{align}

Since ${{\bf g}}_{AR,i}$, ${{\bf g}}_{RB,i}$, ${\hat{\bf g}}_{AR,i}$, and ${\hat{\bf g}}_{RB,i}$ are independent of each other, we can take expectation over the four vectors separately. Utilizing the fact that ${\tt E} \left\{{\bf g}_{RB,i} {{\bf g}}_{RB,i}^H \right\} = \beta_{RB,i} {\bf I}_M$, ${\tt E} \left\{{\bf g}_{AR,i}^T {{\bf g}}_{AR,i}^* \right\} = M \beta_{AR,i}$, and
\begin{align}
  &{\tt E} \left\{\left({\hat{\bf g}}_{RB,n}^* {\hat{\bf g}}_{AR,n}^H  + {\hat{\bf g}}_{AR,n}^* {\hat{\bf g}}_{RB,n}^H \right) \left( {\hat{\bf g}}_{AR,n} {\hat{\bf g}}_{RB,n}^T + {\hat{\bf g}}_{RB,n} {\hat{\bf g}}_{AR,n}^T \right) \right\} \\ \notag
   &= {\tt E} \left\{ {\hat{\bf g}}_{RB,n}^* {\hat{\bf g}}_{AR,n}^H {\hat{\bf g}}_{AR,n} {\hat{\bf g}}_{RB,n}^T \right\} + {\tt E} \left\{ {\hat{\bf g}}_{RB,n}^* {\hat{\bf g}}_{AR,n}^H {\hat{\bf g}}_{RB,n} {\hat{\bf g}}_{AR,n}^T \right\} + {\tt E} \left\{ {\hat{\bf g}}_{AR,n}^* {\hat{\bf g}}_{RB,n}^H {\hat{\bf g}}_{AR,n} {\hat{\bf g}}_{RB,n}^T \right\} \\ \notag
    & + {\tt E} \left\{ {\hat{\bf g}}_{AR,n}^* {\hat{\bf g}}_{RB,n}^H {\hat{\bf g}}_{RB,n} {\hat{\bf g}}_{AR,n}^T \right\},\\ \notag
    &= M \sigma_{AR,n}^2 \sigma_{RB,n}^2 {\bf I}_M + \sigma_{AR,n}^2 \sigma_{RB,n}^2 {\bf I}_M + \sigma_{AR,n}^2 \sigma_{RB,n}^2 {\bf I}_M + M \sigma_{AR,n}^2 \sigma_{RB,n}^2 {\bf I}_M,\\ \notag
    & =  2 \left(M + 1 \right) \sigma_{AR,n}^2 \sigma_{RB,n}^2 {\bf I}_M,
\end{align}
we have
\begin{align}
  {\tt E} \left\{ | {{\bf g}}_{AR,i}^{T} {\bf F} {{\bf g}}_{RB,i} |^2 \right\} = 2 M \left(M + 1 \right) \beta_{AR,i} \beta_{RB,i} \sum\limits_{n=1}^{N} \sigma_{AR,n}^2 \sigma_{RB,n}^2.
\end{align}

c) for $n = l = i$, we have
\begin{align}\label{eq:B:AF::3:1}
   &{\tt E} \left\{ | {{\bf g}}_{AR,i}^{T} {\bf F} {{\bf g}}_{RB,i} |^2 \right\} \\ \notag
  &= {\tt E} \left\{ {{\bf g}}_{AR,i}^{ T} \left({\hat{\bf g}}_{RB,i}^* {\hat{\bf g}}_{AR,i}^H  + {\hat{\bf g}}_{AR,i}^* {\hat{\bf g}}_{RB,i}^H \right) {{\bf g}}_{RB,i} {{\bf g}}_{RB,i}^{H} \left( {\hat{\bf g}}_{AR,i} {\hat{\bf g}}_{RB,i}^T + {\hat{\bf g}}_{RB,i} {\hat{\bf g}}_{AR,i}^T \right) {{\bf g}}_{AR,i}^* \right\}, \\ \notag
  &= {\tt E} \left\{ {{\bf g}}_{AR,i}^{ T} {\hat{\bf g}}_{RB,i}^* {\hat{\bf g}}_{AR,i}^H  {{\bf g}}_{RB,i} {{\bf g}}_{RB,i}^{H} {\hat{\bf g}}_{AR,i} {\hat{\bf g}}_{RB,i}^T {{\bf g}}_{AR,i}^* \right\} + {\tt E} \left\{ {{\bf g}}_{AR,i}^{ T} {\hat{\bf g}}_{RB,i}^* {\hat{\bf g}}_{AR,i}^H  {{\bf g}}_{RB,i} {{\bf g}}_{RB,i}^{H} {\hat{\bf g}}_{RB,i} {\hat{\bf g}}_{AR,i}^T {{\bf g}}_{AR,i}^* \right\} \\ \notag
  &+ {\tt E} \left\{ {{\bf g}}_{AR,i}^{ T} {\hat{\bf g}}_{AR,i}^* {\hat{\bf g}}_{RB,i}^H {{\bf g}}_{RB,i} {{\bf g}}_{RB,i}^{H}  {\hat{\bf g}}_{AR,i} {\hat{\bf g}}_{RB,i}^T {{\bf g}}_{AR,i}^* \right\} + {\tt E} \left\{ {{\bf g}}_{AR,i}^{ T} {\hat{\bf g}}_{AR,i}^* {\hat{\bf g}}_{RB,i}^H  {{\bf g}}_{RB,i} {{\bf g}}_{RB,i}^{H} {\hat{\bf g}}_{RB,i} {\hat{\bf g}}_{AR,i}^T {{\bf g}}_{AR,i}^* \right\}.
\end{align}

The first term in \eqref{eq:B:AF::3:1} becomes
\begin{align}\label{eq:B:AF:3:1:1}
  &{\tt E} \left\{ {{\bf g}}_{AR,i}^{ T} {\hat{\bf g}}_{RB,i}^* {\hat{\bf g}}_{AR,i}^H {{\bf g}}_{RB,i} {{\bf g}}_{RB,i}^{H} {\hat{\bf g}}_{AR,i} {\hat{\bf g}}_{RB,i}^T {{\bf g}}_{AR,i}^* \right\}\\ \notag
  &= {\tt E} \left\{ {\hat{\bf g}}_{AR,i}^{ T} {\hat{\bf g}}_{RB,i}^* {\hat{\bf g}}_{AR,i}^H {\hat{\bf g}}_{RB,i} {\hat{\bf g}}_{RB,i}^{H} {\hat{\bf g}}_{AR,i} {\hat{\bf g}}_{RB,i}^T {\hat{\bf g}}_{AR,i}^* \right\} + {\tt E} \left\{ {{\bf e}}_{AR,i}^{ T} {\hat{\bf g}}_{RB,i}^* {\hat{\bf g}}_{AR,i}^H {\hat{\bf g}}_{RB,i} {\hat{\bf g}}_{RB,i}^{H} {\hat{\bf g}}_{AR,i} {\hat{\bf g}}_{RB,i}^T {{\bf e}}_{AR,i}^* \right\} \\ \notag
  &+ {\tt E} \left\{ {\hat{\bf g}}_{AR,i}^{ T} {\hat{\bf g}}_{RB,i}^* {\hat{\bf g}}_{AR,i}^H {{\bf e}}_{RB,i} {{\bf e}}_{RB,i}^{H} {\hat{\bf g}}_{AR,i} {\hat{\bf g}}_{RB,i}^T {\hat{\bf g}}_{AR,i}^* \right\} + {\tt E} \left\{ {{\bf e}}_{AR,i}^{ T} {\hat{\bf g}}_{RB,i}^* {\hat{\bf g}}_{AR,i}^H {{\bf e}}_{RB,i} {{\bf e}}_{RB,i}^{H} {\hat{\bf g}}_{AR,i} {\hat{\bf g}}_{RB,i}^T {{\bf e}}_{AR,i}^* \right\}, \\ \notag
  &= {\tt E} \left\{ |{\hat{\bf g}}_{AR,i}^{H} {\hat{\bf g}}_{RB,i} |^4 \right\} + {\tilde \sigma}_{AR,i}^2 {\tt E} \left\{ |{\hat{\bf g}}_{AR,i}^{H} {\hat{\bf g}}_{RB,i}|^2 ||{\hat{\bf g}}_{RB,i}||^2 \right\} + {\tilde \sigma}_{RB,i}^2 {\tt E} \left\{ |{\hat{\bf g}}_{AR,i}^{H} {\hat{\bf g}}_{RB,i}|^2 ||{\hat{\bf g}}_{AR,i}||^2 \right\} \\ \notag
  & + {\tilde \sigma}_{RB,i}^2 {\tt E} \left\{ |{\hat{\bf g}}_{RB,i}^{H} {{\bf e}}_{AR,i}|^2 ||{\hat{\bf g}}_{AR,i}||^2 \right\},\\ \notag
  &= {\tt E} \left\{|{\tilde g}_{AR,i}|^4 ||{\hat{\bf g}}_{RB,i}||^4 \right\} + {\tilde \sigma}_{AR,i}^2 {\tt E} \left\{|{\tilde g}_{AR,i}|^2 ||{\hat{\bf g}}_{RB,i}||^4 \right\} + {\tilde \sigma}_{RB,i}^2 {\tt E} \left\{|{\tilde g}_{RB,i}|^2 ||{\hat{\bf g}}_{AR,i}||^4 \right\} \\ \notag
  & +  {\tilde \sigma}_{RB,i}^2 {\tt E} \left\{ |{\hat{\bf g}}_{RB,i}^{H} {{\bf e}}_{AR,i}|^2 \right\} {\tt E} \left\{ ||{\hat{\bf g}}_{AR,i}||^2 \right\},
\end{align}
where ${\tilde g}_{AR,i} \triangleq \frac{{\hat{\bf g}}_{AR,i}^H {\hat{\bf g}}_{RB,i}}{||{\hat{\bf g}}_{RB,i}||}$ and ${\tilde g}_{RB,i} \triangleq \frac{{\hat{\bf g}}_{AR,i}^H {\hat{\bf g}}_{RB,i}}{||{\hat{\bf g}}_{AR,i}||}$. Conditioned on ${\hat{\bf g}}_{RB,i}$, ${\tilde g}_{AR,i}$ is a Gaussian random variable with zero mean and variance $\sigma_{AR,i}^2$ which does not depend on ${\hat{\bf g}}_{RB,i}$ whereas conditioned on ${\hat{\bf g}}_{AR,i}$, ${\tilde g}_{RB,i}$ is a Gaussian random variable with zero mean and variance $\sigma_{RB,i}^2$ which does not depend on ${\hat{\bf g}}_{AR,i}$. Therefore, we have ${\tt E} \left\{|{\tilde g}_{AR,i}|^4 ||{\hat{\bf g}}_{RB,i}||^4 \right\} = {\tt E} \left\{|{\tilde g}_{AR,i}|^4 \right\} {\tt E} \left\{ ||{\hat{\bf g}}_{RB,i}||^4 \right\}$, ${\tt E} \left\{|{\tilde g}_{AR,i}|^2 ||{\hat{\bf g}}_{RB,i}||^4 \right\} = {\tt E} \left\{|{\tilde g}_{AR,i}|^2 \right\} {\tt E} \left\{ ||{\hat{\bf g}}_{RB,i}||^4 \right\}$, and ${\tt E} \left\{|{\tilde g}_{RB,i}|^2 ||{\hat{\bf g}}_{AR,i}||^4 \right\} = {\tt E} \left\{|{\tilde g}_{RB,i}|^2 \right\} {\tt E} \left\{ ||{\hat{\bf g}}_{AR,i}||^4 \right\}$. Then, by using the fact that ${\tt E} \left\{|{\tilde g}_{AR,i}|^4 \right\}  = 2 \sigma_{AR,i}^4$, ${\tt E} \left\{|{\tilde g}_{RB,i}|^2 \right\}  = \sigma_{RB,i}^4$, and ${\tt E} \left\{ ||{\hat{\bf g}}_{RB,i}||^4 \right\} = M (M + 1)\sigma_{RB,i}^4$, \eqref{eq:B:AF:3:1:1} can be calculated as
\begin{align}
&{\tt E} \left\{ {{\bf g}}_{AR,i}^{ T} {\hat{\bf g}}_{RB,i}^* {\hat{\bf g}}_{AR,i}^H {{\bf g}}_{RB,i} {{\bf g}}_{RB,i}^{H} {\hat{\bf g}}_{AR,i} {\hat{\bf g}}_{RB,i}^T {{\bf g}}_{AR,i}^* \right\}\\ \notag
&= 2 \sigma_{AR,i}^4 M \left(M + 1 \right) \sigma_{RB,i}^4 + {\tilde \sigma}_{AR,i}^2 \sigma_{AR,i}^2 M \left(M + 1 \right) \sigma_{RB,i}^4 \\ \notag
&+ {\tilde \sigma}_{RB,i}^2 \sigma_{RB,i}^2 M \left(M + 1 \right) \sigma_{AR,i}^4 + {\tilde \sigma}_{RB,i}^2 {\tilde \sigma}_{AR,i}^2 \sigma_{RB,i}^2 M^2 \sigma_{AR,i}^2,\\ \notag
&= M \left(M+1 \right)\sigma_{AR,i}^2 \sigma_{RB,i}^2 \left(\beta_{AR,i}\sigma_{RB,i}^2 + \beta_{RB,i}\sigma_{AR,i}^2 \right) + M^2 \sigma_{AR,i}^2 \sigma_{RB,i}^2 {\tilde \sigma}_{AR,i}^2 {\tilde \sigma}_{RB,i}^2.
\end{align}

Following the same procedure, the last three terms in \eqref{eq:B:AF::3:1} can be calculated as
\begin{align}
  & {\tt E} \left\{ {{\bf g}}_{AR,i}^{ T} {\hat{\bf g}}_{RB,i}^* {\hat{\bf g}}_{AR,i}^H  {{\bf g}}_{RB,i} {{\bf g}}_{RB,i}^{H} {\hat{\bf g}}_{RB,i} {\hat{\bf g}}_{AR,i}^T {{\bf g}}_{AR,i}^* \right\} \\ \notag
  &= M \left(M + 1 \right)^2 \sigma_{AR,i}^4 \sigma_{RB,i}^4 + M \left(M + 1\right) \sigma_{AR,i}^4 {\tilde \sigma}_{AR,i}^2 \sigma_{RB,i}^2  \\ \notag
  &+ M \left(M + 1 \right) \sigma_{AR,i}^2 {\tilde \sigma}_{AR,i}^2 \sigma_{RB,i}^4 + M\sigma_{AR,i}^2 {\tilde \sigma}_{AR,i}^2 \sigma_{RB,i}^2 {\tilde \sigma}_{RB,i}^2,\\
  & {\tt E} \left\{ {{\bf g}}_{AR,i}^{ T} {\hat{\bf g}}_{AR,i}^* {\hat{\bf g}}_{RB,i}^H {{\bf g}}_{RB,i} {{\bf g}}_{RB,i}^{H}  {\hat{\bf g}}_{AR,i} {\hat{\bf g}}_{RB,i}^T {{\bf g}}_{AR,i}^* \right\} \\ \notag
  &= M \left(M + 1 \right)^2 \sigma_{AR,i}^4 \sigma_{RB,i}^4 + M \left(M + 1 \right) \sigma_{AR,i}^4 {\tilde \sigma}_{AR,i}^2 \sigma_{RB,i}^2  \\ \notag
  &+ M \left(M + 1 \right) \sigma_{AR,i}^2 {\tilde \sigma}_{AR,i}^2 \sigma_{RB,i}^4 + M\sigma_{AR,i}^2 {\tilde \sigma}_{AR,i}^2 \sigma_{RB,i}^2 {\tilde \sigma}_{RB,i}^2, \\
  & {\tt E} \left\{ {{\bf g}}_{AR,i}^{ T} {\hat{\bf g}}_{AR,i}^* {\hat{\bf g}}_{RB,i}^H  {{\bf g}}_{RB,i} {{\bf g}}_{RB,i}^{H} {\hat{\bf g}}_{RB,i} {\hat{\bf g}}_{AR,i}^T {{\bf g}}_{AR,i}^* \right\} \\ \notag
  &= M^2 \left(M + 1 \right)^2 \sigma_{AR,i}^4 \sigma_{RB,i}^4 + M^2\left(M + 1 \right) \sigma_{AR,i}^4 {\tilde \sigma}_{AR,i}^2 \sigma_{RB,i}^2 \\ \notag
  &+ M^2 \left(M + 1\right) \sigma_{AR,i}^2 {\tilde \sigma}_{AR,i}^2 \sigma_{RB,i}^4 + M^2\sigma_{AR,i}^2 {\tilde \sigma}_{AR,i}^2 \sigma_{RB,i}^2 {\tilde \sigma}_{RB,i}^2.
\end{align}

Finally, combining a), b), and c), we obtain
\begin{align}\label{eq:derm:B}
  &B_i^\text{AF} = p_{B,i} {\tt E} \left\{ | {{\bf g}}_{AR,i}^{T} {\bf F} {{\bf g}}_{RB,i} |^2 \right\} - p_{B,i} M^2 \left(M + 1 \right)^2 \sigma_{AR,i}^4 \sigma_{RB,i}^4,\\ \notag
  &= p_{B,i} 2 M \left(M + 1 \right) \beta_{AR,i} \beta_{RB,i} \sum\limits_{n=1}^{N} \sigma_{AR,n}^2 \sigma_{RB,n}^2 \\ \notag
  &+ p_{B,i} M \left(M+1 \right)\sigma_{AR,i}^2 \sigma_{RB,i}^2 \left(\beta_{AR,i}\sigma_{RB,i}^2 + \beta_{RB,i}\sigma_{AR,i}^2 \right) + p_{B,i} M^2 \sigma_{AR,i}^2 \sigma_{RB,i}^2 {\tilde \sigma}_{AR,i}^2 {\tilde \sigma}_{RB,i}^2 \\ \notag
  &+ p_{B,i} 2 M \left(M + 1\right)^2 \sigma_{AR,i}^4 \sigma_{RB,i}^4 + p_{B,i}2 M \left(M + 1\right) \sigma_{AR,i}^4 {\tilde \sigma}_{AR,i}^2 \sigma_{RB,i}^2 + p_{B,i} 2 M \left(M + 1\right) \sigma_{AR,i}^2 {\tilde \sigma}_{AR,i}^2 \sigma_{RB,i}^4 \\ \notag
  &+ p_{B,i} 2M\sigma_{AR,i}^2 {\tilde \sigma}_{AR,i}^2 \sigma_{RB,i}^2 {\tilde \sigma}_{RB,i}^2 + p_{B,i} M^2\left(M + 1\right) \sigma_{AR,i}^4 {\tilde \sigma}_{AR,i}^2 \sigma_{RB,i}^2 \\ \notag
  &+ p_{B,i} M^2\left(M + 1\right) \sigma_{AR,i}^2 {\tilde \sigma}_{AR,i}^2 \sigma_{RB,i}^4 + p_{B,i} M^2\sigma_{AR,i}^2 {\tilde \sigma}_{AR,i}^2 \sigma_{RB,i}^2 {\tilde \sigma}_{RB,i}^2.
\end{align}

3) Compute $C_i^\text{AF}$:

By utilizing the same technique as in the derivation of $B_i^\text{AF}$, we have

a) for $n \neq l \neq i$, we have ${\tt E} \left\{ | {{\bf g}}_{AR,i}^{T} {\bf F} {{\bf g}}_{AR,i} |^2 \right\} = 0$.

b) for $n = l \neq i$, we have
\begin{align}
  &{\tt E} \left\{ | {{\bf g}}_{AR,i}^{T} {\bf F} {{\bf g}}_{AR,i} |^2 \right\} \\ \notag
  &= {\tt E} \left\{ \sum\limits_{n=1}^{N} {{\bf g}}_{AR,i}^{ T} \left({\hat{\bf g}}_{RB,n}^* {\hat{\bf g}}_{AR,n}^H  + {\hat{\bf g}}_{AR,n}^* {\hat{\bf g}}_{RB,n}^H \right) {{\bf g}}_{AR,i} {{\bf g}}_{AR,i}^{H} \left( {\hat{\bf g}}_{AR,n} {\hat{\bf g}}_{RB,n}^T + {\hat{\bf g}}_{RB,n} {\hat{\bf g}}_{AR,n}^T \right) {{\bf g}}_{AR,i}^* \right\}, \\ \notag
  & = 4 M (M + 1)\beta_{AR,i}^2 \sum\limits_{n=1}^{N} \sigma_{AR,n}^2 \sigma_{RB,n}^2.
\end{align}

c) for $n = l = i$, we have
\begin{align}
   &{\tt E} \left\{ | {{\bf g}}_{AR,i}^{T} {\bf F} {{\bf g}}_{AR,i} |^2 \right\} \\ \notag
  &= {\tt E} \left\{ {{\bf g}}_{AR,i}^{ T} \left({\hat{\bf g}}_{RB,i}^* {\hat{\bf g}}_{AR,i}^H  + {\hat{\bf g}}_{AR,i}^* {\hat{\bf g}}_{RB,i}^H \right) {{\bf g}}_{AR,i} {{\bf g}}_{AR,i}^{H} \left( {\hat{\bf g}}_{AR,i} {\hat{\bf g}}_{RB,i}^T + {\hat{\bf g}}_{RB,i} {\hat{\bf g}}_{AR,i}^T \right) {{\bf g}}_{AR,i}^* \right\}, \\ \notag
  &= 4 \sigma_{AR,i}^2 \sigma_{RB,i}^2 M \left(M + 1 \right) \left( \left(M + 2 \right) \sigma_{AR,i}^4 + \left(M + 5\right) \sigma_{AR,i}^2 {\tilde\sigma}_{AR,i}^2 + {\tilde\sigma}_{AR,i}^4 \right).
\end{align}

Then, by using the following fact
\begin{align}
  {\tt E} \left\{ {{\bf g}}_{AR,i}^{T} {\bf F} {{\bf g}}_{AR,i} \right\} = 0,
\end{align}
we have
\begin{align}\label{eq:derm:C}
  C_i^\text{AF} &=  4 p_{A,i} M (M + 1)\beta_{AR,i}^2 \sum\limits_{n=1}^{N} \sigma_{AR,n}^2 \sigma_{RB,n}^2 \\ \notag
  &+ 4 p_{A,i} \sigma_{AR,i}^2 \sigma_{RB,i}^2 M \left(M + 1 \right) \left( \left(M + 2 \right) \sigma_{AR,i}^4 + \left(M + 5\right) \sigma_{AR,i}^2 {\tilde\sigma}_{AR,i}^2 + {\tilde\sigma}_{AR,i}^4 \right) .
\end{align}

4) Compute $D_i^\text{AF}$:
\begin{align}\label{eq:D1}
  &{\tt E} \left\{ | {{\bf g}}_{AR,i}^{T} {\bf F} {{\bf g}}_{AR,j} |^2 \right\} \\ \notag
     &= {\tt E} \left\{ \sum\limits_{n=1}^{N} \sum\limits_{l=1}^{N} {{\bf g}}_{AR,i}^{ T} \left({\hat{\bf g}}_{RB,n}^* {\hat{\bf g}}_{AR,n}^H  + {\hat{\bf g}}_{AR,n}^* {\hat{\bf g}}_{RB,n}^H \right) {{\bf g}}_{AR,j} {{\bf g}}_{AR,j}^{H} \left( {\hat{\bf g}}_{AR,l} {\hat{\bf g}}_{RB,l}^T + {\hat{\bf g}}_{RB,l} {\hat{\bf g}}_{AR,l}^T \right) {{\bf g}}_{AR,i}^* \right\}, \notag
\end{align}
which can be decomposed into six different cases:

a) for $n \neq l \neq i,j \left( j \neq i \right)$, we have ${\tt E} \left\{ | {{\bf g}}_{AR,i}^{T} {\bf F} {{\bf g}}_{AR,j} |^2 \right\} = 0$.

b) for $n = l \neq i,j \left( j \neq i \right)$, we have
\begin{align}\label{eq:D2}
  &{\tt E} \left\{ | {{\bf g}}_{AR,i}^{T} {\bf F} {{\bf g}}_{AR,j} |^2 \right\} \\ \notag
  &= {\tt E} \left\{\sum\limits_{n \neq i,j} {{\bf g}}_{AR,i}^{ T} \left({\hat{\bf g}}_{RB,n}^* {\hat{\bf g}}_{AR,n}^H  + {\hat{\bf g}}_{AR,n}^* {\hat{\bf g}}_{RB,n}^H \right) {{\bf g}}_{AR,j} {{\bf g}}_{AR,j}^{H} \left( {\hat{\bf g}}_{AR,n} {\hat{\bf g}}_{RB,n}^T + {\hat{\bf g}}_{RB,n} {\hat{\bf g}}_{AR,n}^T \right) {{\bf g}}_{AR,i}^* \right\}, \\ \notag
  &= 2 M \left(M + 1\right) \beta_{AR,i} \beta_{AR,j}\sum\limits_{n \neq i,j} \sigma_{AR,n}^2 \sigma_{RB,n}^2.
\end{align}

c) for $n = l = i \left( j \neq i \right)$, we have
\begin{align}
  &{\tt E} \left\{ | {{\bf g}}_{AR,i}^{T} {\bf F} {{\bf g}}_{AR,j} |^2 \right\} \\ \notag
  &= {\tt E} \left\{{{\bf g}}_{AR,i}^{ T} \left({\hat{\bf g}}_{RB,i}^* {\hat{\bf g}}_{AR,i}^H  + {\hat{\bf g}}_{AR,i}^* {\hat{\bf g}}_{RB,i}^H \right) {{\bf g}}_{AR,j} {{\bf g}}_{AR,j}^{H} \left( {\hat{\bf g}}_{AR,i} {\hat{\bf g}}_{RB,i}^T + {\hat{\bf g}}_{RB,i} {\hat{\bf g}}_{AR,i}^T \right) {{\bf g}}_{AR,i}^* \right\}, \\ \notag
  &= M \beta_{AR,j} \sigma_{AR,i}^2 \sigma_{RB,i}^2 \left( \left(M + 1\right) \left(M +3 \right) \sigma_{AR,i}^2 + 2 \left(M + 1\right) {\tilde\sigma}_{AR,i}^2 \right).
\end{align}

d) for $n = l = j \left( j \neq i \right)$, we have
\begin{align}
  &{\tt E} \left\{ | {{\bf g}}_{AR,i}^{T} {\bf F} {{\bf g}}_{AR,j} |^2 \right\} \\ \notag
  &= {\tt E} \left\{{{\bf g}}_{AR,i}^{ T} \left({\hat{\bf g}}_{RB,j}^* {\hat{\bf g}}_{AR,j}^H  + {\hat{\bf g}}_{AR,j}^* {\hat{\bf g}}_{RB,j}^H \right) {{\bf g}}_{AR,j} {{\bf g}}_{AR,j}^{H} \left( {\hat{\bf g}}_{AR,j} {\hat{\bf g}}_{RB,j}^T + {\hat{\bf g}}_{RB,j} {\hat{\bf g}}_{AR,j}^T \right) {{\bf g}}_{AR,i}^* \right\}, \\ \notag
  &= M \beta_{AR,i} \sigma_{AR,j}^2 \sigma_{RB,j}^2 \left( \left(M + 1\right) \left(M +3 \right) \sigma_{AR,j}^2 + 2 \left(M + 1\right) {\tilde\sigma}_{AR,j}^2 \right).
\end{align}

e) for $n = i$ and $l = j \left( j \neq i \right)$, we have ${\tt E} \left\{ | {{\bf g}}_{AR,i}^{T} {\bf F} {{\bf g}}_{AR,j} |^2 \right\} = 0$.

f) for $n = j$ and $l = i \left( j \neq i \right)$, we have ${\tt E} \left\{ | {{\bf g}}_{AR,i}^{T} {\bf F} {{\bf g}}_{AR,j} |^2 \right\} = 0$.

Altogether, ${\tt E} \left\{ |{{\bf g}}_{AR,i}^{T} {\bf F} {{\bf g}}_{AR,j}|^2 \right\}$ is given by
\begin{align}\label{D:AF:term:1}
  &{\tt E} \left\{ {{\bf g}}_{AR,i}^{T} {\bf F} {{\bf g}}_{AR,j} \right\} = 2 M \left(M + 1\right) \beta_{AR,i} \beta_{AR,j}\sum\limits_{n \neq i,j} \sigma_{AR,n}^2 \sigma_{RB,n}^2 \\ \notag
  &+ M \beta_{AR,j} \sigma_{AR,i}^2 \sigma_{RB,i}^2 \left( \left(M + 1\right) \left(M +3 \right) \sigma_{AR,i}^2 + 2 \left(M + 1\right) {\tilde\sigma}_{AR,i}^2 \right) \\ \notag
  &+ M \beta_{AR,i} \sigma_{AR,j}^2 \sigma_{RB,j}^2 \left( \left(M + 1\right) \left(M +3 \right) \sigma_{AR,j}^2 + 2 \left(M + 1\right) {\tilde\sigma}_{AR,j}^2 \right).
\end{align}

Following the same technique as in deriving \eqref{D:AF:term:1}, we can obtain
\begin{align}\label{D:AF:term:2}
  &{\tt E} \left\{| {{\bf g}}_{AR,i}^{T} {\bf F} {{\bf g}}_{RB,j}|^2 \right\} = 2 M \left(M + 1\right) \beta_{AR,i} \beta_{RB,j}\sum\limits_{n \neq i,j} \sigma_{AR,n}^2 \sigma_{RB,n}^2 \\ \notag
  &+ M \beta_{RB,j} \sigma_{AR,i}^2 \sigma_{RB,i}^2 \left( \left(M + 1\right) \left(M +3 \right) \sigma_{AR,i}^2 + 2 \left(M + 1\right) {\tilde\sigma}_{AR,i}^2 \right) \\ \notag
  &+ M \beta_{AR,i} \sigma_{AR,j}^2 \sigma_{RB,j}^2 \left( \left(M + 1\right) \left(M +3 \right) \sigma_{RB,j}^2 + 2 \left(M + 1\right) {\tilde\sigma}_{RB,j}^2 \right).
\end{align}

As a result, combining \eqref{D:AF:term:1} and \eqref{D:AF:term:2} yields
\begin{align}
  &D_i^\text{AF} = \sum\limits_{j \neq i}
  2 M \left(M + 1\right) \beta_{AR,i} \left( p_{A,i} \beta_{AR,j} + p_{B,i} \beta_{RB,j}\right) \sum\limits_{n \neq i,j} \sigma_{AR,n}^2 \sigma_{RB,n}^2 \\ \notag
  &+ \sum\limits_{j \neq i}
  M \sigma_{AR,i}^2 \sigma_{RB,i}^2 \left(  p_{A,i} \beta_{AR,j} +  p_{B,i} \beta_{RB,j}\right) \left( \left(M + 1\right) \left(M +3 \right) \sigma_{AR,i}^2 + 2 \left(M + 1\right) {\tilde\sigma}_{AR,i}^2 \right) \\ \notag
  &+ \sum\limits_{j \neq i}
  M \beta_{AR,i} \sigma_{AR,j}^2 \sigma_{RB,j}^2 \left( \left(M + 1\right) \left(M +3 \right) \left( p_{A,i} \sigma_{AR,j}^2 +  p_{B,i} \sigma_{RB,j}^2 \right) + 2 \left(M + 1\right) \left( p_{A,i} {\tilde\sigma}_{AR,j}^2 +  p_{B,i} {\tilde\sigma}_{RB,j}^2 \right) \right).
\end{align}
5) Compute $E_i^\text{AF}$:

(a) Compute ${\tt E}\left\{||{\bf g}_{AR,i}^{\text T} {\bf F}||^2\right\}$:

Again, using the same technique as in the derivation of \eqref{D:AF:term:1}, we obtain
\begin{align}\label{eq:E:AF:11}
 & {\tt E}\left\{||{\bf g}_{AR,i}^{\text T} {\bf F}||^2\right\} = 2 M \left(M + 1\right) \beta_{AR,i} \sum\limits_{n \neq i} \sigma_{AR,n}^2 \sigma_{RB,n}^2 \\ \notag
  &+ M  \sigma_{AR,i}^2 \sigma_{RB,i}^2 \left( \left(M + 1\right) \left(M +3 \right) \sigma_{AR,i}^2 + 2 \left(M + 1\right) {\tilde\sigma}_{AR,i}^2 \right).
\end{align}

(b) Compute ${\tt E}\left\{ ||{\bf F}{\bf g}_{AR,i}||^2 \right\}$ and ${\tt E}\left\{ ||{\bf F}{\bf g}_{RB,i}||^2 \right\}$:
\begin{align}\label{eq:noise:5:AR}
 & {\tt E}\left\{||{\bf F} {\bf g}_{AR,i}||^2\right\} = 2 M \left(M + 1\right) \beta_{AR,i} \sum\limits_{n \neq i} \sigma_{AR,n}^2 \sigma_{RB,n}^2 \\ \notag
  &+ M  \sigma_{AR,i}^2 \sigma_{RB,i}^2 \left( \left(M + 1\right) \left(M +3 \right) \sigma_{AR,i}^2 + 2 \left(M + 1\right) {\tilde\sigma}_{AR,i}^2 \right).
\end{align}

Similarly, we have
\begin{align}\label{eq:noise:5:RB}
   & {\tt E}\left\{||{\bf F} {\bf g}_{RB,i}||^2\right\} = 2 M \left(M + 1\right) \beta_{RB,i} \sum\limits_{n \neq i} \sigma_{AR,n}^2 \sigma_{RB,n}^2 \\ \notag
  &+ M  \sigma_{AR,i}^2 \sigma_{RB,i}^2 \left( \left(M + 1\right) \left(M +3 \right) \sigma_{RB,i}^2 + 2 \left(M + 1\right) {\tilde\sigma}_{RB,i}^2 \right).
\end{align}

(c) Compute ${\tt E}\left\{ ||{\bf F}||_{\text F}^2  \right\}$:
\begin{align}\label{eq:noise:5:c}
  &{\tt E}\left\{ ||{\bf F}||_{\text F}^2  \right\} = {\tt E} \left\{ {\text{tr}} \left( {\bf A} {\bf B}^T {\bf B}^{*} {\bf A}^H \right\} \right),\\ \notag
  &= {\text{tr}} \left(   {\tt E}\left\{ {\hat{\bf G}}_{AR} {\hat{\bf G}}_{RB} {\hat{\bf G}}_{RB}^* {\hat{\bf G}}_{AR}^H \right\} \right) + {\text{tr}} \left( {\tt E}\left\{ {\hat{\bf G}}_{AR} {\hat{\bf G}}_{RB} {\hat{\bf G}}_{AR}^* {\hat{\bf G}}_{RB}^H \right\} \right) \\ \notag
  &+ {\text{tr}} \left({\tt E}\left\{ {\hat{\bf G}}_{RB} {\hat{\bf G}}_{AR} {\hat{\bf G}}_{RB}^* {\hat{\bf G}}_{AR}^H \right\} \right) + {\text{tr}} \left({\tt E}\left\{ {\hat{\bf G}}_{RB} {\hat{\bf G}}_{AR} {\hat{\bf G}}_{AR}^* {\hat{\bf G}}_{RB}^H \right\} \right),\\ \notag
  &= 2M \left(M +1\right) \sum\limits_{n=1}^{N} \sigma_{AR,n}^2 \sigma_{RB,n}^2.
\end{align}

Combining \eqref{eq:noise:5:AR}, \eqref{eq:noise:5:RB} and \eqref{eq:noise:5:c}, $\rho_\text{AF}^2$ is expressed as
\begin{align}\label{eq:rho:AF}
  \rho_\text{AF}^2 = \frac{p_r}{ \sum\limits_{i =1}^N \left(a_i + b_i\right) + 2M \left(M +1\right) \sum\limits_{n=1}^{N} \sigma_{AR,n}^2 \sigma_{RB,n}^2},
\end{align}
where $a_i = M \sigma_{AR,i}^2 \sigma_{RB,i}^2 \left( \left(M + 1\right) \left(M +3 \right) \left(\sigma_{AR,i}^2 p_{A,i} + \sigma_{RB,i}^2 p_{B,i} \right) + 2 \left(M + 1\right) \left( {\tilde\sigma}_{AR,i}^2 p_{A,i} + {\tilde\sigma}_{RB,i}^2 p_{B,i} \right) \right)$, $b_i = 2 M \left(M + 1\right) \left(\beta_{AR,i} p_{A,i} + \beta_{RB,i} p_{B,i} \right) \sum\limits_{n \neq i} \sigma_{AR,n}^2 \sigma_{RB,n}^2$.

We arrive at the desired result $E_i^\text{AF}$ by combining \eqref{eq:E:AF:11} and \eqref{eq:rho:AF}.
\section{Proof of Theorem \ref{theor:1:DF}}\label{app:theor:1:DF}
Here we only present the detailed derivation for ${\tilde R}_{1,i}^\text{DF}$ and ${\tilde R}_{RA,i}^\text{DF}$, since ${\tilde R}_{AR,i}^\text{DF}$ and ${\tilde R}_{BR,i}^\text{DF}$ can be obtained in a relatively straightforward way, while ${\tilde R}_{RB,i}^\text{DF}$ can be derived in the same fashion as ${\tilde R}_{RA,i}^\text{DF}$.

First, we focus on \eqref{eq:R1:DF}, which consists of five terms: 1) desired signal power of ${\text T}_{B,i}$ $A_i^\text{DF}$; 2) desired signal power of ${\text T}_{A,i}$ $B_i^\text{DF}$; 3) estimation error $C_i^\text{DF}$; 4) inter-user interference $D_i^\text{DF}$; 5) compound noise $E_i^\text{DF}$. For each of these five terms, we will subsequently derive a deterministic equivalent expression.

Before proceeding, we first review some useful results, which are given in the following lemma.
\begin{lemma}\label{lemma:1}
  Let ${\bf x} \thicksim {\cal {CN}}({\bf 0},\sigma_x^2{\bf I}_M)$ and ${\bf y} \thicksim {\cal {CN}}({\bf 0},\sigma_y^2{\bf I}_M)$. Assume that ${\bf x}$ and ${\bf y}$ are mutually independent. Then, we have
  \begin{align}
    &\frac{1}{M}{\bf x}^\dag {\bf x} \overset{a.s.}{\rightarrow} \sigma_x^2, \ M \rightarrow \infty, \\
    &\frac{1}{M}{\bf x}^\dag {\bf y} \overset{a.s.}{\rightarrow} 0, \ M \rightarrow \infty,\\
    &\frac{1}{M^2}|{\bf x}^\dag {\bf y}|^2 - \frac{1}{M}\sigma_x^2 \sigma_y^2 \overset{a.s.}{\rightarrow} 0, \ M \rightarrow \infty.
  \end{align}
\end{lemma}

Now, we compute the five terms one by one.

1) Deterministic equivalent for $A_i^\text{DF}$:
\begin{align}
  A_i^\text{DF} &= M^2 p_{A,i} \left(\frac{1}{M^2}|{\bf \hat g}_{AR,i}^H{\bf \hat g}_{AR,i}|^2 + \frac{1}{M^2}|{\bf \hat g}_{RB,i}^H{\bf \hat g}_{AR,i}|^2 \right).
\end{align}

By invoking Lemma \ref{lemma:1}, we have
\begin{align}\label{eq:A:DF}
   A_i^\text{DF} -  M p_{A,i} \left(M \sigma_{AR,i}^4 + \sigma_{AR,i}^2\sigma_{RB,i}^2\right) \overset{a.s.}{\rightarrow} 0.
\end{align}

2) Deterministic equivalent for $B_i^\text{DF}$:
\begin{align}
  B_i^\text{DF} &= M^2 p_{B,i} \left(\frac{1}{M^2}|{\bf \hat g}_{AR,i}^H{\bf \hat g}_{RB,i}|^2 + \frac{1}{M^2}|{\bf \hat g}_{RB,i}^H{\bf \hat g}_{RB,i}|^2 \right).
\end{align}

Similarly, we obtain
\begin{align}\label{eq:B:DF}
    B_i^\text{DF} - M p_{B,i} \left(M \sigma_{RB,i}^4 + \sigma_{AR,i}^2\sigma_{RB,i}^2\right) \overset{a.s.}{\rightarrow} 0.
\end{align}

3) Deterministic equivalent for $C_i^\text{DF}$:

Since
\begin{align}
  C_i^\text{DF} &= M^2 p_{A,i} \left( \frac{1}{M^2}|{\bf \hat g}_{AR,i}^H{\bf e}_{AR,i}|^2 + \frac{1}{M^2}|{\bf \hat g}_{RB,i}^H{\bf e}_{AR,i}|^2 \right)\\ \notag
   &+ M^2 p_{B,i} \left( \frac{1}{M^2}|{\bf \hat g}_{AR,i}^H{\bf e}_{RB,i}|^2 + \frac{1}{M^2}|{\bf \hat g}_{RB,i}^H{\bf e}_{RB,i}|^2 \right),
\end{align}
we have
\begin{align}\label{eq:C:DF}
    C_i^\text{DF} -  M \left(\sigma_{AR,i}^2 + \sigma_{RB,i}^2 \right) \left( p_{A,i} {\tilde \sigma}_{AR,i}^2 +  p_{B,i} {\tilde \sigma}_{RB,i}^2  \right) \overset{a.s.}{\rightarrow} 0.
\end{align}
4) Deterministic equivalent for $D_i^\text{DF}$:

For $j\neq i$, we have
\begin{align}
  D_i^\text{DF} &= M^2\sum\limits_{j\neq i}  p_{A,j} \left( \frac{1}{M^2}|{\bf \hat g}_{AR,i}^H{\bf g}_{AR,j}|^2 + \frac{1}{M^2} |{\bf \hat g}_{RB,i}^H{\bf g}_{AR,j}|^2\right) \\ \notag
  &+ M^2\sum\limits_{j\neq i} p_{B,j} \left( \frac{1}{M^2} |{\bf \hat g}_{AR,i}^H{\bf g}_{RB,j}|^2 + \frac{1}{M^2} |{\bf \hat g}_{RB,i}^H{\bf g}_{RB,j}|^2 \right) .
\end{align}

To this end, we obtain
\begin{align}\label{eq:D:DF}
  D_i^\text{DF} - M \sum\limits_{j\neq i}\left( p_{A,j} \left(\sigma_{AR,i}^2\beta_{AR,j} + \sigma_{RB,i}^2\beta_{AR,j} \right) + p_{B,j} \left(\sigma_{AR,i}^2\beta_{RB,j} + \sigma_{RB,i}^2\beta_{RB,j} \right) \right) \overset{a.s.}{\rightarrow} 0.
\end{align}

5) Deterministic equivalent for $E_i^\text{DF}$:
\begin{align}
  E_i^\text{DF} &= M^2\left(\frac{1}{M^2}|{\bf\hat g}_{AR,i}|^2 + \frac{1}{M^2}|{\bf \hat g}_{RB,i}|^2\right).
\end{align}

Then, employing Lemma \ref{lemma:1} yields
\begin{align}\label{eq:E:DF}
 E_i^\text{DF} - M\left(\sigma_{AR,i}^2 + \sigma_{RB,i}^2\right) \overset{a.s.}{\rightarrow} 0.
\end{align}

Substituting \eqref{eq:A:DF}, \eqref{eq:B:DF}, \eqref{eq:C:DF}, \eqref{eq:D:DF}, and \eqref{eq:E:DF} into \eqref{eq:R1:DF}, \eqref{eq:R:AR:DF}, and after some algebraic manipulations, we obtain ${\tilde R}_{1,i}^\text{DF}$, ${\tilde R}_{AR,i}^\text{DF}$, ${\tilde R}_{BR,i}^\text{DF}$.

Now, we turn our attention to derive ${\tilde R}_{RA,i}^\text{DF}$.

1) Compute ${\tt E}\left\{ {\bf g}_{AR,i}^T{\bf \hat g}_{AR,i}^* \right\}$:
\begin{align}\label{eq:DF:RA:signal}
  {\tt E}\left\{ {\bf g}_{AR,i}^T{\bf \hat g}_{AR,i}^* \right\} = {\tt E}\left\{ ||{\bf \hat g}_{AR,i}||^2 \right\} + {\tt E}\left\{ {\bf e}_{AR,i}^T{\bf \hat g}_{AR,i}^* \right\} = M \sigma_{AR,i}^2.
\end{align}

2) Compute ${\text {Var}}\left({\bf g}_{AR,i}^T{\bf \hat g}_{AR,i}^* \right)$
\begin{align}\label{eq:DF:RA:esti}
  &{\text {Var}}\left({\bf g}_{AR,i}^T{\bf \hat g}_{AR,i}^* \right) = {\tt E}\left\{|{\bf g}_{AR,i}^T{\bf \hat g}_{AR,i}^*|^2\right\} - M^2\sigma_{AR,i}^4,\\ \notag
  &= {\tt E}\left\{||{\bf \hat g}_{AR,i}||^4\right\} + {\tt E}\left\{|{\bf e}_{AR,i}^T{\bf \hat g}_{AR,i}^*|^2\right\} - M^2\sigma_{AR,i}^4,\\ \notag
  &= M(M+1) \sigma_{AR,i}^4 + M\sigma_{AR,i}^2{\tilde\sigma}_{AR,i}^2 - M^2\sigma_{AR,i}^4,\\ \notag
  &= M\sigma_{AR,i}^2\beta_{AR,i}.
\end{align}

3) Compute ${\text {Var}}\left({\bf g}_{AR,i}^T{\bf \hat g}_{RB,i}^* \right)$:
\begin{align}\label{eq:DF:RA:self}
  &{\text {Var}}\left({\bf g}_{AR,i}^T{\bf \hat g}_{RB,i}^* \right) = {\tt E}\left\{|{\bf g}_{AR,i}^T{\bf \hat g}_{RB,i}^*|^2\right\} - |{\tt E}\left\{{\bf g}_{AR,i}^T{\bf \hat g}_{RB,i}^*\right\}|^2,\\ \notag
  &= M\sigma_{RB,i}^2\beta_{AR,i}.
\end{align}

4) Compute $\sum\limits_{j \neq i} \left( {\tt E}\left\{|{\bf g}_{AR,i}^T{\bf \hat g}_{RB,j}^*|^2\right\} + {\tt E}\left\{|{\bf g}_{AR,i}^T{\bf \hat g}_{AR,j}^*|^2\right\}\right)$:

For $j\neq i$, we obtain
\begin{align}
  {\tt E}\left\{|{\bf g}_{AR,i}^T{\bf \hat g}_{RB,j}^*|^2\right\} = M\sigma_{RB,j}^2\beta_{AR,i},
\end{align}
and
\begin{align}
  {\tt E}\left\{|{\bf g}_{AR,i}^T{\bf \hat g}_{AR,j}^*|^2\right\} = M\sigma_{AR,j}^2\beta_{AR,i}.
\end{align}

Thus, we have
\begin{align}\label{eq:DF:RA:interf}
  \sum\limits_{j \neq i} \left( {\tt E}\left\{|{\bf g}_{AR,i}^T{\bf \hat g}_{RB,j}^*|^2\right\} + {\tt E}\left\{|{\bf g}_{AR,i}^T{\bf \hat g}_{AR,j}^*|^2\right\}\right) = M\beta_{AR,i}\sum\limits_{j \neq i} \left(\sigma_{AR,j}^2 + \sigma_{RB,j}^2 \right).
\end{align}

Combining \eqref{eq:DF:RA:signal}, \eqref{eq:DF:RA:esti}, \eqref{eq:DF:RA:self}, and \eqref{eq:DF:RA:interf} completes the proof.
\section{Proof of Theorem \ref{theor:2}}\label{app:theor:2}
With fixed $E_p$, when $p_p = \frac{E_p}{M^\gamma}$ and $\gamma > 0$, as $M \rightarrow \infty$, we have
\begin{align}\label{eq:sigma:limit}
  \sigma_{AR,i}^2 -  \frac{\tau_p E_p\beta_{AR,i}^2}{M^\gamma} \overset{a.s.}{\rightarrow} 0, {\tilde\sigma}_{AR,i}^2 \overset{a.s.}{\rightarrow} \beta_{AR,i},\\
  \sigma_{RB,i}^2 -  \frac{\tau_p E_p\beta_{RB,i}^2}{M^\gamma} \overset{a.s.}{\rightarrow} 0, {\tilde\sigma}_{RB,i}^2 \overset{a.s.}{\rightarrow} \beta_{RB,i}.
\end{align}

Substituting \eqref{eq:sigma:limit} into \eqref{eq:theor:1:R}, we have
\begin{align}
  &{\tilde B}_i^\text{AF} - \frac{M^\gamma}{\tau_p E_p} \left(\frac{1}{\beta_{RB,i}} + \frac{1}{\beta_{AR,i}} \right) \overset{a.s.}{\rightarrow} 0,\\
  &{\tilde C}_i^\text{AF} - \frac{M^\gamma}{\tau_p E_p} \frac{4\beta_{AR,i}^2}{\beta_{RB,i}^2} \overset{a.s.}{\rightarrow} 0,\\
 & {\tilde D}_i^\text{AF} - \frac{M^\gamma}{\tau_p E_p} \sum_{j \neq i} \left( \frac{\beta_{AR,j} + \beta_{RB,j}}{\beta_{RB,i}^2}  + \frac{\beta_{AR,j}^4 \beta_{RB,j}^2 \beta_{AR,i} + \beta_{AR,j}^2 \beta_{RB,j}^4 \beta_{RB,i}}{\beta_{AR,i}^4\beta_{RB,i}^4} \right) \overset{a.s.}{\rightarrow} 0,\\
 & {\tilde E}_i^\text{AF} - \frac{M^\gamma}{\tau_p E_p} \left( \frac{1}{p_u \beta_{RB,i}^2} + \frac{1}{p_r \beta_{AR,i}^4 \beta_{RB,i}^4} \sum\limits_{n = 1}^{N} \left( \beta_{AR,n}^2 \beta_{RB,n}^2 \left( \beta_{AR,n}^2 + \beta_{RB,n}^2 \right) \right) \right) \overset{a.s.}{\rightarrow} 0.
\end{align}

Therefore, we have now proved that ${\tilde R}^\text{AF}_{A,i} - {\bar R}^\text{AF}_{A,i} \overset{M \rightarrow \infty}{\longrightarrow} 0$. Since we know that $R^\text{AF}_{A,i} - {\tilde R}^\text{AF}_{A,i} \overset{M \rightarrow \infty}{\longrightarrow} 0$, this implies that also $R^\text{AF}_{A,i} - {\bar R}^\text{AF}_{A,i} \overset{M \rightarrow \infty}{\longrightarrow} 0$.

\section{Proof of Theorem \ref{theor:opt:AF:original}}\label{app:theor:opt:AF:original}
For a given $p_u$, the objective function of the optimization problem ${\cal P}^\text{AF}_3$ is an increasing function with respect to $p_p$, while for a given $p_p$, this function is an increasing function with respect to $p_u$; hence, the objective function is maximized when $2N p_u + p_r = P$ \cite{H.Q.Ngo4}.

Now, focusing on ${\tilde R}_{A,i}^\text{AF}$ and substituting $2N p_u + p_r = P$ into ${\tilde R}_{A,i}^\text{AF}$, we have
\begin{align}
  {\tilde R}_{A,i}^\text{AF}= \frac{1}{2}\log_2\left(1 + \frac{1}{a + \frac{b}{p_u} + \frac{c}{d - p_u}}\right),
\end{align}
where $a = \frac{1}{M}\left( \frac{\beta_{RB,i} + 4\beta_{AR,i}}{\sigma_{RB,i}^2} + \frac{\beta_{AR,i}}{\sigma_{AR,i}^2} + \sum\limits_{j \neq i} \left(\frac{\beta_{AR,j}}{\sigma_{RB,i}^2} + \frac{\sigma_{AR,j}^4 \sigma_{RB,j}^2 \beta_{AR,i} }{\sigma_{AR,i}^4 \sigma_{RB,i}^4} \right) + \sum\limits_{j \neq i} \left(\frac{\beta_{RB,j}}{\sigma_{RB,i}^2} + \frac{\sigma_{AR,j}^2 \sigma_{RB,j}^4 \beta_{AR,i} }{\sigma_{AR,i}^4 \sigma_{RB,i}^4} \right) \right)$, $b = \frac{1}{M\sigma_{RB,i}^2}$, $c = \frac{1}{2MN\sigma_{AR,i}^4\sigma_{RB,i}^4} \sum\limits_{n=1}^N \left(\sigma_{AR,n}^2\sigma_{RB,n}^2\left(\sigma_{AR,n}^2 + \sigma_{RB,i}^2\right)\right)$, and $d = \frac{P}{2N}$.

Taking the second derivative with respect to $p_u$ yields
\begin{align}
  &\frac{\partial {\tilde R}_{A,i}^\text{AF}}{\partial p_u} \\ \notag
  &= -\frac{b^2 \left(2 c d^2 + \left(1 + 2 a\right) \left(d - p_u\right)^3\right) \left(d - p_u\right)}{2\ln2\left(b \left(d - p_u\right) + \left(c + a \left(d - p_u\right)\right) p_u\right)^2 \left(b \left(d -
      p_u\right) + \left(c + \left(1 + a\right) \left(d - p_u\right)\right) p_u\right)^2} \\ \notag
      & -\frac{c \left(c + 2 a c + 2 a \left(1 + a\right) \left(d - p_u\right)\right) p_u^4}{2\ln2\left(b \left(d - p_u\right) + \left(c + a \left(d - p_u\right)\right) p_u\right)^2 \left(b \left(d -
      p_u\right) + \left(c + \left(1 + a\right) \left(d - p_u\right)\right) p_u\right)^2} \\ \notag
      & -\frac{ b p_u \left(c^2 d^2 +
   a \left(1 + a\right) \left(d - p_u\right)^4 + \left(1 + 2 a\right) c \left(d - p_u\right) \left(d^2 - d p_u + p_u^2\right)\right)}{\ln2\left(b \left(d - p_u\right) + \left(c + a \left(d - p_u\right)\right) p_u\right)^2 \left(b \left(d -
      p_u\right) + \left(c + \left(1 + a\right) \left(d - p_u\right)\right) p_u\right)^2} < 0. \notag
\end{align}
Thus, ${\tilde R}_{A,i}^\text{AF}$ is a strictly concave function with respect to $p_u$. Since nonnegative weighted sums preserve convexity \cite{S.Byod}, the objective function $\frac{\tau_c - \tau_p}{\tau_c} \sum\limits_{i = 1}^{N}  \left( {\tilde R}_{A,i}^\text{AF} + {\tilde R}_{B,i}^\text{AF} \right)$ is also a strictly concave function with respect to $p_u$. Recall that the constraints of the optimization problem ${\cal P}^\text{AF}_3$ are all affine functions, and hence ${\cal P}^\text{AF}_3$ is a convex optimization problem.
\section{Proof of Theorem \ref{theor:opt:AF}}\label{app:theor:opt:AF}
Substituting $\sigma_{A,i}^2 = \sigma_{B,i}^2 = \sigma^2$, ${\tilde \sigma}_{A,i}^2 = {\tilde \sigma}_{B,i}^2 = {\tilde \sigma}^2$, ${\tilde R}_{A,i}^\text{AF} = {\tilde R}_{B,i}^\text{AF}$ into the objective function in ${\cal P}^\text{AF}_4$, after some simple algebraic manipulations, ${\cal P}^\text{AF}_4$ reduces to
\begin{align}
  &\arg \max_{p_u} \left\{ \log_2\left(1 + M/f(p_u) \right) \right\} \\
  &{\text{s.t.}} \ 0 \leq p_u \leq \frac{P}{2N},
\end{align}
where $f(p_u) = a + \frac{b}{p_u} + \frac{c}{P-2Np_u}$ with $a = \frac{1}{\sigma^2}\left(4 N + 2\right)$, $b = \frac{1}{\sigma^2}$, and $c = \frac{2N}{\sigma^2}$.

It is easy to show that $f''(p_u) = \frac{2b}{p_u^3} + \frac{8N^2c}{\left(P - 2Np_u\right)^3} \geq 0$, hence $f(p_u)$ is a convex function in $0 \leq p_u \leq \frac{P}{2N}$. Therefore, the optimal solution can be obtained by solving $f'(p_u) = 0$.
\section{Proof of Theorem \ref{theor:opt:DF:original}}\label{app:theor:opt:DF:original}
Using the same argument as in the proof of Theorem \ref{theor:opt:DF:original}, it can be proved that the objective function in ${\cal P}^\text{DF}_3$ is maximized when $2Np_u + p_r = P$.

Before studying the properties of ${\tilde R}_{i}^\text{DF}$, we first present the following useful lemma.
\begin{lemma}
The functions $g_1(x) = \log_2\left(1 + \frac{a_1x}{b_1x+c_1}\right)$ and $g_2(x) = \log_2\left(1 + \frac{a_2\left(d_2 - x\right)}{b_2\left(d_2-x\right)+c_2}\right)$ are all strictly concave with respect to $x$.
\end{lemma}
\proof This claim can be easily verified by showing that $g_1''(x) < 0$ and $g_2''(x) < 0$ when $a_1,b_1,c_1,a_2,b_2,c_2,d_2>0$.\endproof

Now, focusing on ${\tilde R}_{i}^\text{DF}$ and substituting $2N p_u + p_r = P$ into ${\tilde R}_{i}^\text{DF}$, it is easy to show that ${\tilde R}_{1,i}^\text{DF}$, ${\tilde R}_{AR,i}^\text{DF}$, and ${\tilde R}_{BR,i}^\text{DF}$ can be reformulated as $g_1(p_u)$, while ${\tilde R}_{RA,i}^\text{DF}$, and ${\tilde R}_{RB,i}^\text{DF}$ can be reformulated as $g_2(p_u)$, hence, ${\tilde R}_{1,i}^\text{DF}$, ${\tilde R}_{AR,i}^\text{DF}$, ${\tilde R}_{BR,i}^\text{DF}$, ${\tilde R}_{RA,i}^\text{DF}$, and ${\tilde R}_{RB,i}^\text{DF}$ are all concave functions with respect to $p_u$.

Due to the convexity preservation property of pointwise maximum and nonnegative weighted sums operations \cite{S.Byod}, ${\tilde R}_{2,i}^{\text{DF}} = \min\left({\tilde R}_{AR,i}^\text{DF}, {\tilde R}_{RB,i}^\text{DF}\right) +  \min\left({\tilde R}_{BR,i}^\text{DF}, {\tilde R}_{RA,i}^\text{DF}\right)$ is also a concave function with respect to $p_u$. Therefore, $\min\left({\tilde R}_{1,i}^{\text{DF}}, {\tilde R}_{2,i}^{\text{DF}} \right)$ is a concave function, which completes the proof.

%

\bibliographystyle{IEEE}

\begin{thebibliography}{10}

 \bibitem{H.Q.Ngo1}
H. Q. Ngo, H. A. Suraweera, M. Matthaiou, and E. G. Larsson, ``Multipair full-duplex relaying with massive arrays and linear processing,'' {\em IEEE J. Sel. Areas Commun.}, vol. 32, no. 9, pp. 1721--1737, Oct. 2014.

\bibitem{H.A.Suraweera}
H. A. Suraweera, H. Q. Ngo, T. Q. Duong, C. Yuen, and E. G.  Larsson, ``Multi-pair amplify-and-forward relaying with very large antenna arrays,'' in {\it Proc.} {\em IEEE ICC}, June 2013, pp. 4635--4640.

\bibitem{G.Kramer}
G. Kramer, M. Gastpar, and P. Gupta, ``Cooperative strategies and capacity theorems for relay networks,'' {\em IEEE Trans. Inf. Theory}, vol. 51, no. 9, pp. 3037--3063, Sep. 2005.

\bibitem{Q.Wang}
Q. Wang and Y. Jing, ``Performance analysis and scaling law of MRC/MRT relaying with CSI error in massive MIMO systems'', June 2016. [Online]. Available: http://arxiv.org/pdf/1606.07480v1.pdf

\bibitem{R.Zhang}
R. Zhang, Y-C. Liang, C. C. Chai, and S. Cui ``Optimal beamforming for two-way multi-antenna relay channel with analogue network coding,'' {\em IEEE J. Sel. Areas Commun.}, vol. 27, no. 5, pp. 699--712, Jun. 2009.

\bibitem{K.-J.Lee}
K.-J. Lee, H. Sung, E. Park, and I. Lee, ``Joint optimization for one and two-way MIMO AF multiple-relay systems,'' {\em IEEE Trans. Wireless Commun.}, vol. 9, no. 12, pp. 3671--3681, Dec. 2010.

\bibitem{G.Amarasuriya}
G. Amarasuriya, C. Tellambura, and M. Ardakani, ``Two-way amplify-and-forward multiple-input multiple-output relay networks with antenna selection,'' {\em IEEE J. Sel. Areas Commun.}, vol. 30, no. 8, pp. 1513--1529, Sep. 2012.

\bibitem{R.Vaze}
R. Vaze and R.W. Heath Jr., ``On the capacity and diversity-multiplexing tradeoff of the two-way relay channel,'' {\em IEEE Trans. Inf. Theory}, vol. 57, no. 7, pp. 4219--4234, July 2011.

\bibitem{S.Jin}
S. Jin, X. Liang, K.-K. Wong, X. Gao, and Q. Zhu, ``Ergodic rate analysis for multipair massive MIMO two-way relay networks,'' {\em IEEE Trans. Wireless Commun.}, vol. 14, no. 3, pp. 1480--1491, Mar. 2015.

\bibitem{H.Cui}
H. Cui, L. Song, and B. Jiao, ``Multi-pair two-way amplify-and-forward relaying with very large number of relay antennas,'' {\em IEEE Trans. Wireless Commun.}, vol. 13, no. 5, pp. 2636--2645, May 2014.

\bibitem{M.Tao}
M. Tao and R. Wang, ``Linear precoding for multi-pair two-way MIMO relay systems with max-min fairness,'' {\em IEEE Trans. Signal Process.}, vol. 60, no. 10, pp. 5361--5370, Oct. 2012.

\bibitem{S.Sima}
S. Sima and W. Chen, ``Joint network and dirty-paper coding for multi-way relay networks with pairwise information exchange,'' in {\it Proc.} {\em IEEE GLOBECOM}, Dec. 2014, pp. 1565--1570.

\bibitem{R.S.Ganesan}
R. S. Ganesan, H. Al-Shatri, A. Kuehne, T. Weber, and A. Klein, ``Pair-aware interference alignment in multi-user two-way relay networks,'' {\em IEEE Trans. Wireless Commun.}, vol. 12, no. 8, pp. 3662--3671, Aug. 2013.

\bibitem{T.L.Marzetta}
T. L. Marzetta, ``Noncooperative cellular wireless with unlimited numbers of base station antennas,'' {\em IEEE Trans. Wireless Commun.}, vol. 9, no. 11, pp. 3590--3600, Nov. 2010.

\bibitem{E.G.Larsson}
E. G. Larsson, O. Edfors, F. Tufvesson, and T. L. Marzetta, ``Massive MIMO for next generation wireless systems,'' {\em IEEE Commun. Mag.}, vol. 52, no. 2, pp. 186--195, Feb. 2014.

\bibitem{F.Rusek}
F. Rusek, D. Persson, B. K. Lau, E. G. Larsson, T. L. Marzetta, O. Edfors, and F. Tufvesson, ``Scaling up MIMO: Opportunities and challenges with very large arrays,'' {\em IEEE Signal Process. Mag.}, vol. 30, no. 1, pp. 40--60, Jan. 2013.

\bibitem{H.Q.Ngo3}
H. Q. Ngo and E. G. Larsson, ``Spectral efficiency of the multipair two-way relay channel with massive arrays,'' in {\it Proc.} {\em IEEE ACSSC}, Nov. 2013, pp. 275--279.

\bibitem{H.Q.Ngo2}
H. Q. Ngo, E. G. Larsson, and T. L. Marzetta, ``Energy and spectral efficiency of very large multiuser MIMO systems,'' {\em IEEE Trans. Commun.}, vol. 61, no. 4, pp. 1436--1449, Apr. 2013.

\bibitem{F.Gao}
F. Gao, R. Zhang, and Y.-C. Liang, ``Optimal channel estimation and training design for two-way relay networks,'' {\em IEEE Trans. Commun.}, vol. 57, no. 10, pp. 3024--2033, Oct. 2009.

\bibitem{C.Wang}
C. Wang, T. C.-K. Liu, and X. Dong, ``Impact of channel estimation error on the performance of amplify-and-forward two-way relaying,'' {\em IEEE Trans. Veh. Technol.}, vol. 61, no. 3, pp. 1197--1207, Mar. 2012.

\bibitem{J.Hoydis}
J. Hoydis, S. ten Brink, and M. Debbah, ``Massive MIMO in the UL/DL of cellular networks: How many antennas do we need?,'' {\em IEEE J. Sel. Areas Commun.}, vol. 31, no. 2, pp. 160--171, Feb. 2013.

\bibitem{B.Rankov}
B. Rankov and A. Wittneben, ``Spectral efficient protocols for half-duplex fading relay channels,'' {\em IEEE J. Sel. Areas Commun.}, vol. 25, no. 2, pp. 379--389, Feb. 2007.

\bibitem{J.Gao1}
J. Gao, S. A. Vorobyov, H. Jiang, J. Zhang, and M. Haardt, ``Sum-rate maximization with minimum power consumption for MIMO DF two-way relaying--part I: Relay optimization,'' {\em IEEE Trans. Signal Process.}, vol. 61, no. 14, pp. 3563--3577, July 2013.

\bibitem{J.Gao2}
---, ``Sum-rate maximization with minimum power consumption for MIMO DF two-way relaying--part II: Network optimization,'' {\em IEEE Trans. Signal Process.}, vol. 61, no. 14, pp. 3578--3591, July 2013.

\bibitem{J.Gao3}
J. Gao, J. Zhang, S. A. Vorobyov, H. Jiang, and M. Haardt, ``Power allocation/beamforming for DF MIMO two-way relaying: Relay and network optimization,'' in {\it Proc.} {\em IEEE GLOBECOM}, Dec. 2012, pp. 5657--5662.

\bibitem{I.Hammerstrom}
I. Hammerstr\"{o}m, M. Kuhn, C. E\c{s}li, J. Zhao, A. Wittneben, and G. Bauch, ``MIMO two-way relaying with transmit CSI at the relay'', in {\it Proc.} {\em IEEE SPAWC}, June 2007.

\bibitem{A.Alsharoa}
A. Alsharoa, F. Bader, and M.-S. Alouini, ``Relay selection and resource allocation for two-way DF-AF cognitive radio netwroks,'' {\em IEEE Wireless Commun. Lett.}, vol. 2, no. 4, pp. 427--430, Aug. 2013.

\bibitem{M.Avriel}
M. Avriel and A. C. Williams, ``Complementary geometric programming,'' {\em SIAM J. Appl. Math.}, vol. 19, no. 1, pp. 125--141, July 1970.

\bibitem{M.Grant}
M. Grant and S. Boyd, {\em CVX: Matlab software for disciplined convex programming}, June 2015. [Online]. Available: http://cvxr.com/cvx/

\bibitem{A.Mutapcic}
A. Mutapcic, K. Koh, S. Kim, and S. Boyd, {\em A Matlab Toolbox for Geometric Programming}, May 2006. [Online]. Available: http://www.stanford.edu/~boyd/ggplab/ggplab.pdf

\bibitem{P.C.Weeraddana}
P. C. Weeraddana, M. Codreanu, M. Latva-aho, and A. Ephremides, ``Resource allocation for cross-layer utility maximization in wireless networks,'' {\em IEEE Trans. Veh. Technol.}, vol. 60, no. 6, pp. 2790--2809, July 2011.

\bibitem{B.R.Marks}
B. R. Marks and G. P. Wright, ``A general inner approximation algorithm for nonconvex mathematical programs,'' {\em Oper. Res.}, vol. 26, no. 4, pp. 681--683, Jul./Aug. 1978.

\bibitem{S.Boyd1}
S. Boyd, S. J. Kim, L. Vandenberghe, and A. Hassibi, ``A tutorial on geometric programming,'' {\em Optim. Eng.}, vol. 8, no. 1, pp. 67--127, Apr. 2007.

\bibitem{S.Boyd2}
S. Boyd, {\em Sequential Convex Programming}, 2007. [Online]. Available: http://www.stanford.edu/class/ee364b/lectures/seq\_slides.pdf

\bibitem{C.He}
C. He, G. Y. Li, F.-C. Zheng, and X. You, ``Power allocation criteria for distributed antenna systems,'' {\em IEEE Trans. Veh. Technol.}, vol. 64, no. 11, pp. 5083--5090, Nov. 2015.

\bibitem{H.Q.Ngo4}
H. Q. Ngo, M. Matthaiou, and E. G. Larsson, ``Massive MIMO with optimal power and training duration allocation,'' {\em IEEE Wireless Commun. Lett.}, vol, 3, no. 6, pp. 605--608, Dec. 2014.

\bibitem{S.Byod}
S. Boyd and L. Vandenberghe, {\em Convex Optimization.} Cambridge, UK: Cambridge University Press, 2004.

 \end{thebibliography}
\begin{footnotesize}
 
 \end{footnotesize}

\end{document}